\documentclass{pinchcr}   % Comment the above and uncomment this
\usepackage{amsmath,epsfig,cite,graphics}%,macros,macros_classes}
\title{Introduction to Non-perturbative Heavy Quark Effective Theory}

\author{R. Sommer}
\affiliation{NIC, DESY, Platanenallee 6, 15738 Zeuthen, Germany}
\authors{1}

\usepackage{amssymb,bbm}
\usepackage{mathrsfs}  % for \mathscr

\newcommand{\cal}{\mathcal}

\usepackage{macros_lh}

\newcommand{\Lambdaqcd}{\Lambda_{\rm QCD}}

\renewcommand{\fb}{f_{\rm B}}
\renewcommand{\fbs}{f_{\rm B_\mrm{s}}}

\renewcommand{\fd}{f_{\rm D}}

\renewcommand{\cite}{\shortcite}

\begin{document}
\maketitle
\preface
%\chapter{Summary} \label{s:summ}
%
My lectures on the effective field theory for heavy quarks,
an expansion around the static limit, concentrate on
the motivation and formulation of HQET, its renormalization and
discretization.
This provides the basis for understanding 
that and how this effective theory can be formulated fully non-perturbatively
in the QCD coupling, while by the very nature of an effective field theory,
it is perturbative in the expansion
parameter $1/m$. After the couplings in the effective theory 
have been determined,
the result at a certain order in  
$1/m$ is unique up to higher order terms in $1/m$. In particular the 
continuum limit of the lattice regularized theory exists and leaves
no trace of how it was regularized. In other words, the theory yields 
an asymptotic expansion of the QCD observables in $1/m$ --
as usual in a quantum field theory modified by
powers of logarithms. 
None of these properties has been shown rigorously 
(e.g. to all orders in perturbation theory) 
but perturbative computations and recently also non-perturbative 
lattice results give strong support to this ``standard wisdom''. 
 
A subtle issue is that a theoretically consistent formulation of the theory
is only possible through a non-perturbative matching of its parameters 
with QCD at finite values of $1/m$ (\sect{s:need}). As a consequence one finds 
immediately that the splitting of a result for a certain observable
into, for example, lowest order and first order is ambiguous. Depending on how 
the matching between effective theory and QCD is done, a first order
contribution may vanish and appear instead in the lowest order. 
For example, the often cited phenomenological HQET parameters 
$\bar\Lambda$ and $\lambda_1$ lack a unique
non-perturbative definition. But this does not affect the precision
of the asymptotic expansion in $1/m$. The final result for an
observable is correct up to order $(1/m)^{n+1}$ if the theory was
treated including $(1/m)^{n}$ terms. 

Clearly, the weakest point of HQET is that it intrinsically is an expansion. In practise,
carrying it out non-perturbatively beyond the order $1/m$ will be very difficult. In 
this context two observations are relevant.  First, the expansion parameter 
for HQET applied to B-physics is
$\Lambda_\mrm{QCD}/\mbeauty \sim 1/(r_0\mbeauty)=1/10$ and indeed recent computations
of $1/\mbeauty$ corrections showed them to be very small. Second, since HQET yields
the asymptotic expansion of QCD, it becomes more and more accurate the
larger the mass is. It can therefore be used to constrain the large mass
behavior of QCD computations done at finite, varying, quark masses. At some point,
computers and computational strategies will be sufficient to simulate
with lattice spacings which are small enough for a relativistic b-quark.
One would then like to understand the full mass-behavior of observables
and a combination of HQET and relativistic QCD will again be most useful. 
Already now, there is a strategy \cite{romeII:fb,romeII:mb,ssm:comb1}, which 
is related to the one discussed in \sect{s:strat} and which, in its 
final version combines HQET and QCD in such a manner. For a 
short review of this aspect I refer to \cite{ichep08:nazario}.

\acknowledgements
I am thankful for the nice collaboration in the team of organizers, with 
the director of the school 
and the staff of the school. 
It is also a pleasure to thank the members of
the LGT discussion seminar at Humboldt-University and DESY, in particular 
Hubert Simma and Ulli Wolff, 
for their valuable suggestions on a first version of these lecture notes.
I am grateful for a fruitful collaboration with   
Benoit Blossier, Michele Della Morte, Patrick Fritzsch, Nicolas Garron,
Jochen Heitger, Georg von Hippel, Tereza Mendes, Mauro Papinutto and 
Hubert Simma on several of the subjects of these lectures and thank
Nicolas Garron for providing me with tables and figures. 
Most of all I would like to thank Dorothy for her patience with me spending
much time on these lectures and the school.  
\tableofcontents

\maintext

\chapter{Introduction}
\section{Conventions}
Our conventions for gauge fields, lattice derivatives etc.
are summarized in the appendix. 
\section{The r\^ole of HQET}
\label{s:moti}
This school focuses on lattice gauge theories. How does 
heavy quark effective theory (HQET) fit into it? The first 
part of the answer is that HQET is expected to provide the true
asymptotic expansion of quantities in powers (accompanied by logarithms)
of $1/m$, the mass of the heavy quark, 
with all other scales held fixed. The accessible
quantities are energies, matrix elements and 
Euclidean correlation functions with a single
heavy (valence) quark, while all other quarks are light. 
A full understanding of QCD should
contain this kinematical region. 

The second part of the answer has to do with the 
challenge we are facing when we perform a Monte Carlo (MC) 
evaluation of the QCD path integral.
This becomes apparent by considering the scales which are 
relevant for QCD.  For low energy QCD and flavor physics
excluding the top-quark, they range from \\[0.5ex]
\centerline{$\mpi\approx 140\,\MeV$ over
$\md=2\,\GeV$ to $\mb=5\,\GeV$.} \\[0.5ex]
In addition, the
ultraviolet cutoff of $\Lambda_{\rm UV} = a^{-1}$
of the discretized theory has to be large compared to all
physical energy scales if the  theory discretized with a 
lattice spacing $a$ is to
be an approximation to a continuum.
Finally, the linear extent of space time has to
be restricted to a finite value $L$
in a numerical treatment:
there is an infrared cutoff $L^{-1}$.
Together the following constraints have to
be satisfied.
\bes
     \Lambda_{\rm IR} \;=\;  L^{-1} & \ll\;m_\pi\,,\;\ldots\;,\md\,,\mb\;\ll &
     a^{-1} \;=\; \Lambda_{\rm UV} 
\ees
The infrared and the ultraviolet effects are systematic errors
which have to be controlled.
Infrared effects behave as~\cite{FSE:martin2} 
$\rmO(\rme^{-L\mpi})$ and are known from
chiral perturbation theory~\cite{chpt:fse3} 
to be at the percent level when $L \gtrsim 4/\mpi \approx 6\,\fm$,
while the UV, discretization, errors are $\rmO((a\, m_\mrm{quark})^2)$  
in $\rmO(a)$-improved theories.\footnote{See Peter Weisz' lectures 
for the general discussion of discretization errors and improvement 
of lattice gauge theories.} 
With a charm quark mass of around
$1\,\GeV$ we have a requirement of 
$a \lesssim 1/(2\mcharm)\;\ldots\; 1/(4\mcharm)\approx 
0.1\;\ldots\;0.05\,\fm$ \cite{zastat:pap2} 
and thus
\bes
  \label{e:size}
  L/a \approx 60 \;\ldots\; 120\,.
\ees
Including b-quarks would increase the
already rather intimidating estimate of $L/a$ by 
a factor 4. 
It is thus mandatory to resort to an effective theory where 
degrees of freedom with energy scales around the b-quark mass
and higher are summarized in the coefficients
of terms in the effective Lagrangian. 
A {\em precise} treatment of this theory 
has become very relevant because the search for 
physics beyond the Standard Model 
in the impressive first generation
of B-physics flavor experiments
has been unsuccessful so far. New physics contributions are very small
and even higher precision is needed both in 
experiment and in theory to possibly reveal them. HQET is a very 
important ingredient in this effort. 

Before we focus on our topic let us note that a factor two or so
in $L/a$ may be saved by working at somewhat higher pion mass and
extrapolating with chiral perturbation theory, see M.~Golterman's
lectures.

\section{On continuum HQET}
\label{s:intro}

\subsection{Idea \label{s:idea}}
We consider hadrons with a single very heavy quark,
e.g. a B-meson. Physical intuition tells us that these will be similar
to a
hydrogen atom with the analogy

\begin{tabular}{lclcll}
hydrogen atom &:& heavy proton &+& light electron \\
B-meson &:& heavy b-quark &+& light anti-quark & \todo{figure}\\
b-baryons &:& heavy b-quark &+& two light quarks \\
and so on.
\end{tabular} 

When we take the limit $m=\mbeauty\to\infty$ (``static'')
the b-quark is at rest 
in the {\em rest-frame of the b-hadron} (B, $\Lambda_\beauty$, \ldots).  
In this situation, we should be able to find an effective Lagrangian 
describing
the dynamics of the light quarks and glue with the heavy quark
just representing a color source. 
Corrections in $\minv$ 
should be systematically included in a series
expansion in that variable.
The Lagrangian  
is then expected to be given as a series in $D_k/m$ where 
the covariant derivatives act on the heavy quark field and
correspond to its spatial momenta 
in the rest-frame of the heavy hadron.

Before proceeding to a heuristic derivation of the 
effective field theory, let us note some general properties
of what we are actually seeking, comparing to other familiar 
effective field theories. In contrast to the low energy effective 
field theory for electroweak interactions, where the heavy particles
(W- and Z-boson, top quark) are removed completely from the Lagrangian
we here want to consider processes with b-quarks in initial and/or
final states. The b-quark field is thus contained in the Lagrangian
and we have to find its relevant modes to be kept.\footnote{However, 
when one carries
out the expansion to include $\minv^2$ terms, also a whole set of terms
generated by b-quark loops in QCD which do not contain the b-quark field
in the effective theory have to be taken into account. 
An example are 4-fermion operators
made of the light quarks, just as they appear when
one ``integrates out'' the W and Z-bosons in the Standard Model.}

Another important effective field theory to compare to is the chiral 
effective theory, covered here by Maarten Golterman. Main differences 
are that this is a fully relativistic theory with loops of the 
(pseudo-) Goldstone bosons and that the interaction of the fields in 
the effective Lagrangian disappears for zero momentum. The theory 
can therefore be evaluated perturbatively. It is also called chiral 
perturbation theory. In contrast, the b-quarks in HQET still interact 
non-perturbatively
with the light quarks and gluons. This effective field theory therefore 
needs a lattice implementation in order 
to come to predictions beyond those that 
can be read off from its symmetries.

\subsection{Derivation of the form of the 
effective field theory: FTW  trafo \label{s:ftw}}
\def\psibar{\overline{\psi}}
\def\rme{{\rm e}}
\def\LD{\lag{}}
\def\Dop{{\cal D}}
\def\Obkin{{\bar{\mathcal{O}}_\mrm{kin}}}
\def\Obspin{{\bar{\mathcal{O}}_\mrm{spin}}}
\def\nab#1{{\nabla_{#1}}}
\def\lnabstar#1{\overleftarrow{\nabla}\kern-0.5pt\smash
             {\raise 4.5pt\hbox{$\ast$}}\kern-4.5pt_{#1}}
\def\nabstar#1{\nabla\kern-0.5pt\smash{\raise 4.5pt\hbox{$\ast$}}
               \kern-4.5pt_{#1}}
\def\vecD{{\bf D}}
\def\vecB{{\bf B}}
\def\vecsig{\boldsymbol{\sigma}}
\newcommand{\vecg}{\boldsymbol{\gamma}}
\def\Dg{D_k\gamma_k}
\def\hub{\psibar_{\mrm{h},u}}
\def\hu{\psi_{\mrm{h},u}}
\def\ahub{\psibar_{\bar{\mrm{h}},{u}}}
\def\ahu{\psi_{\bar{\mrm{h}},{u}}}

%\subsubsection
{\em Strategy}\\[0.5ex]
Our strategy is to carry out the following steps, which we discuss in more detail
below. 
\bi
\item
We start from a Euclidean action. 
\item
We identify the dominant degrees of freedom for the kinematical
situation we are interested in: the ``large'' components of the
b-quark field for the quark and the ``small'' components for the 
anti-quark.
\item
We decouple large components and small components, order by order
in $D_k/m$ \\ ~[  $\heavyb D_k/m\; \heavy \ll  \heavyb \heavy$ ].
This assumes smooth gauge (and other) fields. 
It is thus essentially a classical derivation. 
The decoupling is achieved by a sequence of 
Fouldy Wouthuysen-Tani (FTW) transformations (see e.g.~\cite{books:IZ}),
following essentially \cite{hqet:cont5}.
\item
The irrelevant modes are dropped from the theory (often
it is said they are integrated out). Their  
effects are {\em not} expected to change the 
form of the local Lagrangian, but just to renormalize its parameters. 
Still it could be that local terms allowed by the symmetries 
happen to vanish in the classical theory. Thus 
the symmetries have to be considered and 
all terms of the proper dimension compatible with the 
symmetries have to be taken into account.
\item 
At tree level the values of the parameters in the effective Lagrangian 
are given by
the FTW transformation. 
In general (i.e. for any value of the QCD coupling) they have to be
determined by matching to QCD:
one expands QCD correlation functions in $\minv$ and compares to HQET.
This part of the strategy will be discussed in detail in later sections.
\ei
%\subsubsection
{\em Identifying the degrees of freedom}
\\[0.5ex]
We consider the {\em free} propagator of a Dirac-fermion
in Euclidean space, in the  time$\,/\,$space-momentum 
representation\footnote{The expectation value
$\langle . \rangle$ refers to the Euclidean path integral,
here with the free Dirac action.\\
We suggest to verify these formulae as an exercise.}:
\bes
 S(x_0;\veck) &=& \int{\rmd^3 \vecx}\,\rme^{-i\veck \vecx}  
              \langle \psi(x) \psibar(0) \rangle  
              = \int\,{\rmd k_0 \over (2\pi)}  \rme^{ik_0x_0}\, 
                  \left[i k_\mu \gamma_\mu +m\right]^{-1}
 \nonumber \\[-0.5ex] \\[-0.5ex] \nonumber
 &=& S_+(x_0;\veck) + S_-(x_0;\veck) \,,
\ees
with
\bes
  S_+(x_0;\vecp) &=& \theta(x_0) {m\over E(\vecp)} \rme^{-E(\vecp)x_0} P_+(u)  \,, \qquad
                      P_+(u) = {1-iu_\mu\gamma_\mu \over 2}\,, 
                      \; u_\mu=p_\mu/m\,,\; 
 \nonumber \\[-0.5ex] \label{e:propmixed}\\ \nonumber
  S_-(x_0;\vecp) &=& \theta(-x_0)  {m\over E(\vecp)} \rme^{E(\vecp)x_0} P_-(u)\,, \qquad 
                   P_-(u) = {1+iu_\mu\gamma_\mu \over 2}\,,
\ees
where $p_\mu$ is the on-shell momentum, i.e.
\bes
  p_0&=&{iE(\vecp)}=i\sqrt{m^2+\vecp^2} \,. 
\ees
Here $S_+(x_0;\vecp)$ describes the propagation of a quark 
from time $t=0$ to $t=x_0$ and $S_-(x_0;\vecp)$ describes the propagation of an 
anti-quark 
from $t=-x_0$ to $t=0$. Since the Euclidean  4-velocity vector $u$
satisfies 
$u^2=u_\mu u_\mu =-1$, the matrices $P\in\{P_+,P_-\}$ 
are projection operators,
\bes
 [P(u)]^2=P(u)\,, \; P_+(u)P_-(u)=0\,, \;
  P_+(u)+P_-(u)=1\,.
\ees
They allow us to project onto the on-shell components of a
quark with velocity $\vecu$. 

The ``large'' field components corresponding to the quark are given by
the projection
% \footnote{The Dirac equation is simply 
% $2mP(-u)\hu(x) = 0 = (m + ip_\mu\gamma_\mu)\hu(x)$}
\bes
  \label{e:hu}
  \hu(x) &=& P(u) \psi(x)\,,\; \hub(x) = \psibar(x) P(u)
\ees
and the ``small'' ones, the anti-quark field, are
\bes
  \label{e:ahu}
  \ahu(x) = P(-u) \psi(x)\,,\; \ahub(x) = \psibar(x) P(-u)\,,
\ees
such that for free quarks
\bes
\int{\rmd^3 \vecx}\,\rme^{-i\vecp \vecx}  
              \langle \hu(x) \hub(0) \rangle = S_+(x_0;\vecp)
\ees
and similarly for the anti-quark.\footnote{The terms ``large'' and ``small'' 
components are commonly used when discussing the non-relativistic limit
of the Dirac equation for bound states, see e.g. \cite{books:IZ}.}

For a b-hadron with velocity $\vecu$,
the fields
$\hu(x),\, \hub(x)$ are expected to be the relevant ones with the other 
field-components giving subdominant contributions in the 
path integral representation of correlation functions (or
scattering amplitudes in Minkowski space), while for 
a $\overline{\rm b}$-hadron $\ahu(x),\, \ahub(x)$ are expected to dominate.
% We note the special case of velocity $u_k=0$. Here the upper components
% in the Dirac representation are those of the quark and the lower ones
% those of the anti-quark ($P(u)=\frac12(1+\gamma_0)\,,\; P(-u)=\frac12(1-\gamma_0)$.
\\[1ex]
{\em In the presence of a gauge field} 
\\[0.5ex]
When a gauge field is present,
we therefore expect an effective Lagrangian for the b-hadrons in terms
of $\hu, \hub$ plus a term for the anti-quark. When we rewrite the 
Dirac Lagrangian in terms of these fields, 
\bes
 \lag{} &=& \psibar(m+\Dop)\psi \\
      &=& \hub(m+\Dop_\parallel) \hu 
          +  \ahub(m+\Dop_\parallel) \ahu
       + \hub \Dop_\perp \ahu + \ahub \Dop_\perp \hu \,,
      \nonumber
\ees
there are mixed contributions which involve 
\bes
  \Dop_\perp =  \gamma_\mu D_\mu^\perp\,,\quad
   D_\mu^\perp = 
       (\delta_{\mu\nu} + u_\mu u_\nu)\, D_\nu \,,
\ees
where the derivative is projected orthogonal to $u_\mu$. Analogously
we have
\bes
  \Dop_\parallel =  \gamma_\mu D_\mu^\parallel\,, \quad
  D_\mu^\parallel = -u_\mu\, D_\nu u_\nu\,.
\ees
From our general consideration of the kinematical
situation that we want to describe, $D_\mu^\perp$ acting on
the heavy quark field is to be considered small (compared to
$m$). In contrast, $D_\mu^\parallel$ applied to the field 
will yield approximately $p_\mu = u_\mu m$. 
We therefore carry out an expansion 
with 
\bes
   \Dop_\parallel \psi &=& \rmO(m)\,\psi\,, 
   \nonumber \\[-1ex] \\[-1ex] \nonumber
   \Dop_\perp \psi &=& \rmO(1)\,\psi\,   
\ees
and all other fields, such as $F_{\mu\nu}$, treated as order one.
This is often called the power counting scheme. 
\\[1ex]
%\subsubsection
{\em FTW trafo and Lagrangian at zero velocity}
\\[0.5ex]
Having identified the expansion, 
we perform a field rotation (FTW transformation) to decouple 
large  and small components order by order in $1/m$. 
First we consider the special case of
zero velocity,
\bes
  u_k=0 &:\quad& \Dop_\parallel=D_0\gamma_0\,,\quad 
                       \Dop_\perp = D_k \gamma_k  \,,
  \nonumber\\[-1ex] \\[-1ex] \nonumber
  &&
            P(u) = P_+={1+\gamma_0 \over 2}\,,\quad  
           P(-u) = P_-={1-\gamma_0 \over 2} \, .
\ees
The FTW transformation is 
\bes
  \psi &\to& \psi'=\rme^{S} \psi \,,\quad 
     S=\frac{1}{2 m} \Dg = - S^\dagger\,, 
   \nonumber \\[-1ex] \label{e:ftw1} \\[-1ex] \nonumber
  \psibar &\to& \psibar'=\psibar \rme^{-\ola{S}} 
            = \psibar \rme^{-\ola{D}_k \gamma_k /(2m)}\,. 
\ees
Its Jacobian is one.
% $S=\frac{1}{2 m} \Dg = - S^\dagger$. 
% ($\Dop_\perp=-\Dop_\perp^\dagger$ is just inherited from 
% $D_\mu=-D_\mu^\dagger$ and $\gamma_\mu=\gamma_\mu^\dagger$.)
The Lagrangian written in terms of the transformed fields,
\bes
  \LD &=&  
           \psibar' (\Dop'+m) \psi'\,,\quad 
\ees
yields a Dirac operator (note that $S$ acts to the right everywhere)
\bes
  \Dop'+m=\rme^{-S} (\Dop+m) \rme^{-S} \,.
\ees
Expanding 
$
  \rme^{-S} = 1 - S + \frac12 S^2 - \ldots
$
in $S =\rmO(1/m)$ yields 
\bes
  \Dop'+m = \underbrace{\Dop+m}_{\rmO(m)} 
         +\underbrace{\{-S,\Dop+m\}}_{\rmO(1)} 
         + \frac12 \underbrace{\{-S, \{-S,\Dop+m\}\}}_{\rmO(1/m)}
         +
        \ldots  % \frac1{k!} \{-S,\ldots \{-S,\Dop+m\}\} \ldots
       \label{e:dopp}
\ees
\renewcommand{\minv}{{1/m}}
In the evaluation of the different terms we count all fields and derivatives
of fields 
(e.g. $F_{\mu\nu}$) as order one except for $D_0$ 
{\em acting onto the heavy quark field}. We work out the expansion 
up to order $\minv$. 
A little algebra yields 
\bes
  \Dop+m+ \{-S,\Dop+m\}  &=&
  D_0\gamma_0 -  {1\over 2m} 
        [ \gamma_k\gamma_0 F_{k0} 
        + {1\over i} \sigma_{kl}\, F_{kl}+2D_k D_k] 
\ees
with $
     \sigma_{\mu\nu}\!=\!{i\over2}[\gamma_\mu,\gamma_\nu] \,, 
F_{kl}=[D_k,D_l]\,$
and
\bes
  {1\over 2} \{-S, \underbrace{\{-S,\Dop+m\}}_{-\Dg+\rmO(\minv)}\} &=&  
    {1\over 4m} [ {1\over i} \sigma_{kl}\, F_{kl}+2D_k D_k] \,,
\ees
such that 
\bes
  \Dop' &=& D_0\gamma_0 -  {1\over 2m} 
        [\underbrace{\gamma_k\gamma_0 F_{k0}}_{\mbox{off-diagonal}} 
        + {1\over 2i} \sigma_{kl}\, F_{kl}+D_k D_k]+\rmO(1/m^2)\,.
\ees
In the static part, $D_0\gamma_0$, the large and small components are 
decoupled, but one of the $1/m$ terms, $\gamma_k\gamma_0 F_{k0}$, 
is off-diagonal with respect to this split.
We therefore seek a
second transformation $\psi''=\rme^{S'}\psi'$ to cancel also that term, namely 
we want
\bes
   \{-S',\Dop'+m\} 
   =  {1\over 2m} \gamma_k\gamma_0 F_{k0}+\rmO(1/m^2)\,.
\ees
The simple choice 
$
   S' = {1\over 4m^2} \gamma_0\gamma_k F_{k0} 
$
does the job.
Now we have the classical HQET Lagrangian 
\begin{eqnarray}
  \lag{} 
        &=& \Lh^{\rm stat} +   \frac{1}{ 2m}\Lh^{(1)} \;+\; \Lhb^{\rm stat} 
       + \frac{1}{ 2m} \Lhb^{(1)}
       +\rmO(\frac{1}{ m^2}) \\[1ex]
 \Lh^{\rm stat} &=&   \heavyb(m+D_0)\heavy\,,\quad 
               P_+\heavy=\heavy\,,\quad\heavyb P_+=\heavyb\,,\quad
               P_{\pm}= \frac{1\pm\gamma_0}{ 2}  \\
\Lhb^{\rm stat} &=&  \aheavyb(m-D_0)\aheavy\,,\quad 
               P_-\aheavy=\aheavy\,,\quad\aheavyb P_-=\aheavyb\,,\\[1ex]
\Lh^{(1)} &=& -(\Okin + \Ospin ) \,,
\quad \Lhb^{(1)} = -(\Obkin + \Obspin ) \,,
\end{eqnarray}
correct up to terms of order $1/m^2$. We introduced
\bes
  && \Okin(x)=\heavyb(x)\,\vecD^2\, \heavy(x)\,, \; 
     \Ospin(x)=\heavyb(x)\,\vecsig\cdot\vecB(x)\, \heavy(x)\,,
  \\
  && \Obkin(x)=\aheavyb(x)\,\vecD^2\, \aheavy(x)\,, \; 
     \Obspin(x)=\aheavyb(x)\,\vecsig\cdot\vecB(x)\, \aheavy(x)\,,
  \\
  && \sigma_k = \frac12\eps_{ijk}\sigma_{ij}\,,\;
  \label{e:sigmak}
  \quad 
  B_k = i\frac12\eps_{ijk}F_{ij}\,,\; 
\end{eqnarray}
and the heavy quark fields are the transformed ones,
i.e. we renamed $\heavy'' \to \heavy$ etc.

Depending on the process/correlation function, just the
heavy quark part or just the
heavy anti-quark part of the Lagrangian will contribute, 
but there are also processes such as 
$\rm B - \bar{B}$ oscillations where both are needed. 
%
% We will see that the mass terms can be eliminated, they correspond
% just to an energy shift.

\newcommand{\EFT}{{effective field theory }}
It is worth summarizing some issues that arose in this formal derivation.
\bi
\item
  Assuming $D_k = \rmO(1)$ means that this is a classical derivation:
  in the quantum field theory path integral we integrate over rough fields,
  i.e. there are arbitrarily large derivatives. \\
  As emphasized before we therefore take this as a classical Lagrangian. 
  Its renormalization will be discussed
  later, guided by dimensional counting.
  % That is the standard \EFT logics.
\item
  The derivation is perturbative in $1/m$, order by order. This is 
  all that we want. In this way we expect to obtain the {\em asymptotic} expansion 
  in powers of $1/m$. 
\item 
  We note that there are alternative ways to derive the form of the Lagrangian.
  One may integrate out the components $\aheavyb,\aheavy$ in a path integral 
  and then perform a formal expansion of the resulting non-local action
  for the remaining fields in terms of a series of 
  local operators \cite{hqet:cont6}. Another option is to perform a 
  hopping parameter expansion of the Wilson-Dirac lattice propagator.  
  The leading term gives the propagator of the static action; see 
  exercise \ref{ex:hopp}.
\ei

\def\hub{\psibar_{\mrm{h},u}}
\def\hu{\psi_{\mrm{h},u}}
\def\ahub{\psibar_{\bar{\mrm{h}},{u}}}
\def\ahu{\psi_{\bar{\mrm{h}},{u}}}
%\subsubsection

\vspace*{1ex}
\noindent
{\em FTW transformation and Lagrangian at finite velocity}
\\[0.5ex]
At finite velocity the transformation is given
again by \eq{e:ftw1} but with $S=D_\mu^\perp\gamma_\mu /(2m)$.
For the lowest order (static) approximation, just the 
anti-commutator 
$$
  \{\Dop^\perp, \Dop^\parallel\} = \frac12 \{D_\mu^\perp, D_\nu^\parallel\} 2 \delta_{\mu\nu}
           + \frac12 [D_\mu^\perp, D_\nu^\parallel][\gamma_\mu,\gamma_\nu] 
$$ is needed. Since $D_\mu^\perp\,D_\mu^\parallel = 0= D_\mu^\parallel D_\mu^\perp$ and
the second term just involves a commutator of derivatives, we see that 
$\{\Dop^\perp, \Dop^\parallel\} = \rmO(1)$.  Consequently we find 
\bes
 \lag{} 
      &=& \hub(m+\Dop^\parallel) \hu 
          +  \ahub(m+\Dop^\parallel) \ahu +\rmO(1/m) \nonumber\\
      &=& \hub(m-iu_\mu D_\mu ) \hu 
          +  \ahub(m+iu_\mu D_\mu ) \ahu +\rmO(1/m) \, 
\ees
with the projected fields  
\eq{e:hu} and
\eq{e:ahu}.
\footnote{
One can also obtain this Lagrangian by performing a boost of the 
zero velocity theory~\cite{nrqcd:finitevelocity}. In the quoted reference
also the next to leading order terms are found.}
\\[2ex]
Let us add a few {\em comments on the finite velocity theory}, since
we will not discuss it further.
%\\[0.5ex]
\bi
\item 
O(4) (or Lorentz) invariance is broken. One therefore has to 
expect a different renormalization of $D_0$ and $D_k$ (or as is usually
said, a renormalization of $\vecu$ \cite{hqet:CDM,hqet:MO}).
\item 
The operator 
$
 -i D_k u_k 
$
is unbounded from below. Since it enters the Hamiltonian the
theory seems to contain states with arbitrarily large negative
energies. 
Resulting problems in the Euclidean formulation 
of the theory have been discussed in the 
literature~\cite{Aglietti:1992in,Aglietti:1993hf}, but
a compelling formulation of the theory seems not to have been found. 
There are also no
modern applications of the finite velocity theory on the lattice.  
We will therefore 
concentrate entirely on zero velocity HQET from now on.

\todo{Since it enters the Hamiltonian, one might think
the correlation functions of the theory do not exist. There is a number
of papers on this [References to be added]. However, in a regularized 
form, it is bounded and the positive, arbitrarily large $m$
removes the problem completely.  
(when one wants to discuss the theory with the mass term removed, 
one may add a sufficiently large counter-term).  
to do this properly I should know the transfer matrix ...}
\ei
\subsection{Propagator and Symmetries \label{s:symm}} 
\def\Sh{G_\mrm{h}}
\def\Sah{G_\mrm{\bar h}}

\mysubsubsection{The continuum propagator.}
We consider the static approximation
at zero velocity and the latter always from now on.
The static Dirac operator for the quark is just
$
  D_0+m 
$ 
so its Green function, $\Sh$, (the propagator) in a gauge field $A_\mu(x)$
% is given by 
% \bes
%   \Sh(x,y) = {1 \over Z[A]} \int \rmD[\heavy]\rmD[\heavyb] 
%   \exp\left\{-\int\rmd^4z \,\aheavy(z)(D_0+m)\heavy(z)\right\} 
%   \,\heavy(x)\heavyb(y)\,. \nonumber \\[-0.5ex]
% \ees
% The fields $\heavyb,\heavy$ are taken as Grassmann variables and
% the integration is only over the non-vanishing components. 
% A proper definition requires a regularization and in the lattice 
% regularization which we will discuss shortly, one sees immediately
% that $Z[A] = \det(D_0 +m) =$ const. For now we want to continue formally in
% the continuum.  The propagator 
then satisfies 
\bes
  \label{e:defprop}
  (\partial_{x_0}+A_0(x)+m) \Sh(x,y) = \delta(x-y)\,P_+\,.
\ees
The solution of this equation is simply 
\bes
   \Sh(x,y) =\theta(x_0-y_0) \, \exp(-m\,(x_0-y_0)) \,{\cal P} 
    \exp\left\{-\int_{y_0}^{x_0}\rmd z_0 A_0(z_0,\vecx)\right\}
    \delta(\vecx-\vecy)\,P_+\,,  \nonumber \\[-0.5ex]
\ees
were $\cal P$ denotes path ordering (fields at the end of 
the integration path to the left).
In the same way the propagator for the anti-quark
is~\footnote{Note 
${\cal P} \exp\left\{-\int_{y_0}^{x_0}\rmd z_0 A_0(z_0,\vecx)\right\} =
 {\cal P} \exp\left\{-\int_{x_0}^{y_0}\rmd z_0 A_0(z_0,\vecx)\right\}^\dagger$. 
}
\bes
   &&\Sah(x,y) =\theta(y_0-x_0) \, \exp(-m\,(y_0-x_0)) \,{\cal P} 
    \exp\left\{-\int_{y_0}^{x_0}\rmd z_0 A_0(z_0,\vecx)\right\}
    \delta(\vecx-\vecy)\,P_-\,,  \nonumber \\
  &&(-\partial_{x_0}-A_0(x)+m) \Sah(x,y) = \delta(x-y)\,P_-\,.
  \label{e:defaprop}
\ees
\\
The mass appears in a trivial way, with an explicit factor 
$\exp(-m\,|x_0-y_0|)$ for any gauge field
$A_\mu$. This exponential
decay is then present also after path integration over the
gauge fields in 
any 2-point function with a heavy quark,
\bes
  C_\mrm{h}(x,y;m) = C_\mrm{h}(x,y;0)\,\exp(-m\,(x_0-y_0)) \,.
  \label{e:massiveC}
\ees 
An explicit example is
\bes
   C_\mrm{h}^{\rm PP}(x,y;m) = \langle \lightb(x)\gamma_5\heavy(x)\;
        \heavyb(y)\gamma_5\light(y) \rangle\,,
\ees
with $\light(x)$ a light-quark fermion field. 
\Eq{e:massiveC} means that $m$ shifts {\em all} energies in the sector
of the Hilbert space with a single heavy quark (or anti-quark).
We may remove $m$ from the effective Lagrangian and add it to the energies
later.
We only have to be careful that $m \geq 0$ in \eq{e:defprop}, \eq{e:defaprop} 
selects the 
forward/backward propagation. Therefore we set
\bes
 \Lh^{\rm stat} =   \heavyb (D_0+\eps)\heavy\,,\quad
\Lhb^{\rm stat} =  \aheavyb (-D_0+\eps)\aheavy\,,\quad  
E_{\rm h/\bar h}^\mrm{QCD} = E_{\rm h/\bar h}^\mrm{stat}+m\,,\label{e:lstat2}
\ees
where the limit $\eps\to0_+$ is to be understood.

We note that after performing this shift of the energies, 
there is no difference 
in the Lagrangian of a charm or a b-quark if both are treated at
the lowest order in 
this expansion. We turn to discussing this as well as other 
symmetries of the static theory.
 
\mysubsubsection{Symmetries}
1. Flavor\\[1ex]
If there are $F$ heavy quarks, we just add a corresponding 
flavor index and use a notation
\bes
  \heavy &\to& \heavy=({\heavy}_1,\ldots,{\heavy}_F)^T \,, 
  \quad
  \heavyb \to \heavyb=({\heavyb}_1,\ldots,{\heavyb}_F)
  \\
  \Lh^{\rm stat}&=&\heavyb (D_0+\eps)\heavy \,.  
\ees
Then we obviously have the symmetry
\bes
  \heavy(x) &\to& V \, \heavy(x) \,,\quad
  \heavyb(x) \to \heavyb(x) V^\dagger\,,\qquad V\in {\rm SU}(F)
   \label{e:flavsymm}
\ees 
and the same for the anti-quarks. Note that this symmetry emerges
in the large mass limit irrespective 
of how the limit is taken. For example we may take ($F=2$ with the
first heavy flavor identified with charm and the second with
beauty)
\bes
 \mbeauty - \mcharm = c\times \Lambdaqcd\,, \quad \mbox{or} \quad
 \mbeauty / \mcharm = c'\,,\quad \mbeauty\to\infty
\ees
with either $c$ or $c'$ fixed when taking $\mbeauty\to\infty$.
\\[2ex]%%%%%%%%%%%%%%%%%%%%%%%%%%%%%%%%%%%%%%%%%%%%%%%%%%%%%%%%%%%%%%%
2. Spin\\[1ex]
We further note that for each field 
there are also the two spin components but the Lagrangian 
contains no spin-dependent interaction. The associated
${\rm SU}(2)$ rotations  are generated by the spin 
matrices \eq{e:sigmak}
(remember that $\heavy\,,\;\heavyb$ are kept as 4-component fields
with 2 components vanishing)
\bes
  \label{e:sigmamat}
  \sigma_k   = 
  {1\over 2} \eps_{ijk} \sigma_{ij} \equiv \pmat{\sigma_k & 0 \\ 0 &\sigma_k} \,,
\ees
where the symbol $\sigma_k$ is used at the same time for
the Pauli matrices and the $4\times4$ matrix. We here are in the Dirac 
representation where 
\bes
  \gamma_0= \pmat{1 & 0 \\ 0 &-1 }\,,\;
  P_+= \pmat{1 & 0 \\ 0 & 0 }\,,\;
  P_-= \pmat{0 & 0 \\ 0 & 1 }\,.
\ees
The spin rotation is then
\bes
  \label{e:spinsymm}
  \heavy(x) \to \rme^{i\alpha_k\sigma_k} \, \heavy(x) \,, \qquad
  \heavyb(x) \to \heavyb(x) \rme^{-i\alpha_k\sigma_k}\,,
\ees 
with arbitrary real parameters $\alpha_k$.
It acts on each flavor component of the field. 
Obviously, the symmetry 
is even bigger. We can take  
$
  V\in {\rm SU}(2F)\,
$
in \eq{e:flavsymm}.
This plays a r\^ole in heavy meson ChPT \cite{chpt:b1,chpt:b2,chpt:b3}. 
\\[4ex]%%%%%%%%%%%%%%%%%%%%%%%%%%%%%%%%%%%%%%%%%%%%%%%%%%%%%%%%%%%%%%%
3. Local Flavor-number\\[1ex]
The static Lagrangian contains no space derivative.  The transformation
\bes
  \heavy(x) \to \rme^{i\eta(\vecx)} \, \heavy(x) \,, \qquad
  \heavyb(x) \to \heavyb(x) \rme^{-i\eta(\vecx)}\,, \label{e:locflav}
\ees
is therefore a symmetry for any local phase $\eta(\vecx)$. 
For every point $\vecx$  there is a corresponding Noether charge
\bes
  Q_\mrm{h}(x)= \heavyb(x)\heavy(x)\,[\,=\heavyb(x)\gamma_0\heavy(x)\,] \,
\ees
which we call local quark number. It is conserved,
\bes
   \partial_0  Q_\mrm{h}(x) = 0 \;\forall x \,.
\ees

\subsection{Renormalizability of the static theory \label{s:ren1}}
Our effective field theory is in the category of 
local field theories with a Lagrangian made up from
local fields. In $d$ space-time dimensions, 
standard wisdom says that such theories are renormalizable
if the mass-dimension of the fields in the Lagrangian
does not exceed $d$. Ultraviolet divergences
can then be absorbed by adding a complete set of (composite) local fields 
with mass dimension smaller or equal to $d$
to the Lagrangian.

According to this (unproven\footnote{Power counting 
as discussed by Peter Weisz at this school is not applicable here, since
the propagator does not fall off with {all} momentum components.})  
rule, the static theory is renormalizable. The possible counter-terms 
have to share the symmetries of the bare Lagrangian. They are
easily found. 
From the kinetic term in the Lagrangian \eq{e:lstat2}
we see that the dimension of the fields is $[\heavy] = 3/2$.
Only 2-fermion terms with up to one derivative are then possible.
Space-derivatives are excluded by the local phase invariance
\eq{e:locflav}.   
We then have the total quantum Lagrangian
\bes
     \label{e:lstat3}
     \Lh(x) &=& c_1 \op1(x) + c_2  \op2(x) \\ 
     \op1(x) &=& \heavyb(x) \heavy(x)\,,\quad \op2(x) = \heavyb(x) D_0\heavy(x)\,, 
\ees
where the convention $c_2=1$ can be chosen since it
only fixes the unphysical field normalization,  and
$c_1=\dmstat$ has mass dimension $[\dmstat] = 1$ and
corresponds to an additive mass renormalization. From dimensional 
analysis and neglecting for simplicity the masses of the light quarks,
it can be written as
$\dmstat = (e_1 g_0^2 + e_2 g_0^4 + \ldots)\,\Lambda_\mrm{cut}$ in
terms of the bare gauge coupling $g_0$ and a cutoff $\Lambda_\mrm{cut}$,
which in lattice regularization is $\Lambda_\mrm{cut}=1/a$. 
For a static quark there is of course no chiral symmetry to 
forbid additive mass renormalization.

This is the complete static Lagrangian. After the standard QCD
renormalization of coupling and light quark masses, all divergences 
can be absorbed
in $\dmstat$, i.e. an energy shift. Flavor symmetry
tells us that with several heavy flavors, $\dmstat$ is proportional
to the unit matrix in flavor space.
Energies of {\em any state} are then 
\bes
      E_{\rm h/\bar h}^\mrm{QCD} &=& 
      \left.E_{\rm h/\bar h}^\mrm{stat}\right|_{\dmstat=0}+\dmstat+m = 
      \left.E_{\rm h/\bar h}^\mrm{stat}\right|_{\dmstat=0}+\mhbare\,.
\ees 
Here $\mhbare$ and $\dmstat$ compensate the linear divergence 
(self energy) of the static theory, while $m$ is finite. Note
that there is no symmetry which would suggest a natural way of splitting 
$\mhbare$ into $\dmstat$ and $m$. This split is arbitrary and 
convention dependent. The quantity $\dmstat$ is often called the residual mass.

A rigorous proof of {\em renormalizability} to all orders in 
perturbation theory has not been given but we note the following.
\bi 
\item
Perturbative computations 
have confirmed
the standard wisdom.
These computations reach up to three loops in dimensional 
regularization \cite{ChetGrozin,hqet:gammamag3}, while in various 
different lattice regularizations 1-loop computations have
been carried out 
\cite{stat:eichhill1,stat:eichhill2,stat:eichhill_za,stat:boucaud_za,stat:boucaud_rig,stat:4ferm_flynn,BorrPitt,zastat:pap1,zastat:pap2,stat:actpaper,zvstat:filippo,hqet:zospin,castat:filippo,stat:zbb_pert,stat:derop} 
\item
We will see non-perturbative results which again yield a rather
strong confirmation.
\item
Nevertheless a proof of renormalizability would be very desirable.
\ei

\subsection{Normalization of states, scaling of decay constants}

For the discussion of the mass-dependence of
matrix elements we have to think about the 
normalization of states. Standard, relativistic invariant, normalization
of bosonic one-particle states is
\bes
  \langle \vecp |  \vecp' \rangle_\mrm{rel} &=&
  (2\pi)^3\,2E(\vecp)\,\delta(\vecp-\vecp')\,.
\ees
The states have a mass-dimension
$[\,|  \vecp \rangle_\mrm{rel}\,] = -1$. 
The factor $ E(\vecp)$ introduces 
a spurious mass-dependence.
In the large mass limit, relativistic invariance plays no r\^ole 
and we should choose a mass-independent normalization instead.
The standard convention for such a non-relativistic normalization is
\bes
  \langle \vecp |  \vecp' \rangle_\mrm{NR} &\equiv& 
  \langle \vecp |  \vecp' \rangle = 2\,(2\pi)^3\,\delta(\vecp-\vecp')
\ees
with $[\,|  \vecp \rangle\,] =-3/2$ and
\bes
  |\vecp \rangle_\mrm{rel} &=& \sqrt{E(\vecp)}\,|  \vecp \rangle\,.
  \label{e:nrnorm}
\ees
\mynotes{The normalization 2 is e.g. in agreement with 
 the Neubert review \cite{reviews:neubert}.\\}

Consider as an example where the normalization of states plays a role,
the leptonic decay of a B-meson, 
$  B^-\to \tau^- \bar\nu_\tau$.  The transition amplitude $\cal A$
for this decay is given to a good approximation in terms of the 
effective weak Hamiltonian. It factorizes into a leptonic and
a hadronic part as
\bes
{\cal A} \propto  
  \langle \tau\; \bar \nu | \tau(x)\gamma_\mu(1-\gamma_5) \bar \nu_\tau(x) | 0
  \rangle\,
  \langle 0| \bar u(x) \gamma_\mu(1-\gamma_5) b(x)  | B^- \rangle \,.
\ees
Using parity and Lorentz invariance, the hadronic part is
\bes
  \langle 0| \bar u(x) \gamma_\mu(1-\gamma_5) b(x)  | B^-(\vecp) \rangle
  = 
    \langle 0| A_\mu(x)  | B^-(\vecp) \rangle = p_\mu \fb \rme^{ipx} 
\ees
in terms of the flavored axial current
\bes
     A_\mu(x) = \bar u(x) \gamma_\mu\gamma_5 b(x)\,.
\ees
There is a single hadronic parameter $\fb$ (matrix element) parameterizing
the bound state dynamics in this decay.
We note that it is very relevant for the phenomenological 
analysis of the CKM matrix \cite{CKM08}.

We may now use HQET to find the asymptotic mass-dependence 
of $\fb$ for large $m=\mbeauty$. Since to lowest order in $\minv$ the FTW
transformation is trivial, the HQET current is just
\be
A_0^\mrm{HQET}(x) =  A_0^\mrm{stat}(x) + \rmO(\minv)\,,\;
A_0^\mrm{stat}(x) = \bar u(x) \gamma_0\gamma_5 \heavy(x)\,.
\ee
The static current $A_0^\mrm{stat}$ has no explicit mass dependence.
In static approximation we then have   
\bes
  \label{e:Phistat}
  \langle 0| A_0^\mrm{stat}(0)  | B^-(\vecp=0) \rangle = \Phistat\,,
\ees
with a mass-independent $\Phistat$. Its relation to 
$\fb$,
\bes
  \Phistat = \mb^{-1/2} \,p_0\,\fb = \mb^{1/2} \fb \,,
\ees
takes \eq{e:nrnorm} into account ($p_0=E(\mathbf{0})= \mb$). We arrive at the prediction
\bes
  \fb = {\Phistat \over \sqrt{\mb}} +\rmO(1/\mbeauty)\,, 
  \quad {\fb \over \fd} = {\sqrt{\md} \over \sqrt{\mb}} + \rmO(1/\mcharm)\,.
\ees
The latter use of course assumes  $\Lambda_\mrm{QCD}/\mc \ll 1$. 
We will see later
that these predictions are modified by the renormalization 
of the effective theory.
 
\subsection{HQET and phenomenology \label{s:phen}}

Heavy quark spin/flavor symmetry 
is very useful to classify the spectrum in terms of
a few non-perturbative parameters or predict relations 
between different masses, e.g.
\bes
       \mBstar^2-\mB^2 &\approx& \mDstar^2-\mD^2\,, \\
       \mBp-\mB &\approx& \mDp-\mD\,,
\ees
where $\mBstar,\,\mDstar$ are the vector meson masses and with
$\mBp,\,\mDp$ we indicate the first excitation in the pseudo-scalar sector. 
The first of these relations
has been seen to be approximately realized in nature.

More detailed statements about semi-leptonic transitions
$B\to D l \nu $, $B^\star \to D^\star l \nu $ are possible. 
In the heavy quark limit for both beauty and charm these are described 
by a single form factor, the Isgur Wise function, instead of several 
\cite{hqet:cont1,hqet:cont2}.
These topics and many others
are discussed in many reviews, e.g. \cite{HQET:neubert}. 
We here concentrate on
lattice HQET and where HQET helps to understand lattice results for
states with a b-quark.

\exercis{Static quarks from the 
hopping parameter expansion \label{ex:hopp}}
{Consider a Wilson quark propagator in a gauge background field. 
Evaluate the leading non-vanishing term in the hopping parameter expansion 
(with  non-zero
time-separation). Check that it is the continuum HQET propagator 
(restricted to the lattice points)
up to an energy shift. Even if this is a nice piece of confirmation,
note that one here takes the limit $\kappa\to0$ corresponding
to $ma \to \infty$, while the true limit for relating
QCD observables $\Phiqcd$ to those of HQET is
$$
  \Phi^\mrm{HQET} \sim \lim_{m\to\infty} \lim_{a\to0} \Phi^\mrm{QCD},
$$ 
in that order!
}

\chapter{Lattice formulation}
\label{s:latt}
We start with the static approximation. The $\minv$ terms will be added
after a discussion of the renormalization of the static theory.
\section{Lattice action} \label{s:act}
For a static quark there is no chiral symmetry. Since 
we want to avoid doublers, 
we discretize \`a la Wilson (with $r=1$). The continuum 
$D_0\,\heavy(x)$ is transcribed to the lattice as
\bes
   D_0\gamma_0 &\to& \frac12 \{
     (\nab{0}+\nabstar{0})\gamma_0 - a \nabstar{0}\nab{0} 
     \} \,,
\ees
and with $P_+\heavy=\heavy\,,\; P_-\aheavy=\aheavy$, we have the 
lattice identities
\bes
   D_0\,\heavy(x) =   \nabstar{0} \heavy(x)\,, \quad
   D_0\,\aheavy(x) =   \nab{0} \aheavy(x)\,.
\ees
For later convenience we insert a specific normalization factor,
defining the static lattice Lagrangians
\bes
   \label{e:llstat}
   \Lh &=& {1 \over 1+a\dmstat} \heavyb(x) [\nabstar{0}+\dmstat]
   \heavy(x)\,,\\
   \label{e:llstata}
   \Lhb &=& {1 \over 1+a\dmstat} \aheavyb(x) [-\nab{0}+\dmstat] \aheavy(x) \,.
\ees
The following points are worth noting.
\bi
\item Formally, this is just a one-dimensional Wilson fermion replicated 
      for all space points $\vecx$, see also exercise \ref{ex:hopp}.
\item As a consequence there are no doubler modes.
\item The construction of a positive hermitian transfer matrix
      for Wilson fermions \cite{Luscher:TM,books:MM} can just be taken over.
\item The choice of the backward derivative for the quark and
      the forward derivative for the anti-quark is selected by the Wilson term.
      We will see that this selects forward/backward propagation and
      an $\eps$-prescription as in \eq{e:lstat2} is not needed.
\item The form of this Lagrangian was first written down by Eichten and
      Hill
      \cite{stat:eichhill1}.
\item The lattice action preserves all the 
      continuum heavy quark symmetries discussed in the previous section.
\ei

\section{Propagator} \label{s:prop}

From the Lagrangian \eq{e:llstat} we have the 
defining equation for the  propagator
\bes
  &&{{1}\over{1+a\; \dmstat }} (\nabstar{0} + \dmstat ) G_\mrm{h}(x,y)=
  \delta(x-y)P_+ 
  \equiv a^{-4} \prod_\mu \delta_{\frac{x_\mu}a \frac{y_\mu}a} P_+ \,.
\ees
Obviously $G_\mrm{h}(x,y)$ is proportional to $\delta(\vecx-\vecy)$.
Writing 
$  G_\mrm{h}(x,y) = g(n_0,k_0;\vecx) \delta(\vecx-\vecy) P_+ $ with
$x_0=an_0\,,\;y_0=ak_0$, the above 
equation yields a simple recursion for $g(n_0+1,k_0;\vecx)$ in terms of 
$g(n_0,k_0;\vecx)$ which is solved by 
\bes
  g(n_0,k_0;\vecx) &=&  \theta(n_0-k_0) 
  (1+a\dmstat)^{-(n_0-k_0)}{\cal{P}}(y,x;0)^{\dagger}\,, \\
{\cal{P}}(x,x;0)&=&1 \; ,\quad
{\cal{P}}(x,y+a\hat0;0) = {\cal{P}}(x,y;0)U(y,0)\,,  
\ees
where 
\bes
 \theta(n_0-k_0) = 
 \begin{cases}  0 &  n_0<k_0 \\ 
                1 &  n_0\geq k_0  \,.
 \end{cases}
\ees
The static propagator reads
\bes
G_{\rm h}(x,y) &=& \theta(x_0-y_0) \;
 \delta({\vecx-\vecy}) \;\exp\big(- \dmstathat\,(x_0-y_0)\big)\;
\Ptrans(y,x;0)^{\dagger} \; P_+ \; ,
 \label{e:stat_prop}
\\ && \dmstathat=\frac1a \ln(1+a\dmstat)\,.
 \label{e:dmstathat}
\ees
The object $\Ptrans(x,y;0)$ parallel transports fields in the fundamental 
representation from $y$ to $x$ along a time-like path. 
Note that the derivation fixes 
$ 
\theta(0)=1
$
for the lattice $\theta$-function. As in the continuum, 
the mass counter term $\dmstat$
just yields an energy shift;  now, on the lattice, the shift is
\bes
      E_{\rm h/\bar h}^\mrm{QCD} &=&
      \left.E_{\rm h/\bar h}^\mrm{stat}\right|_{\dmstat=0}+\mhbare\,,
      \quad \mhbare = \dmstathat + m\,.
\ees
It is valid for all energies of states with a single heavy quark 
or anti-quark. As in the continuum the split between
$\dmstat$ and the finite $m$ is convention dependent.

In complete analogy the anti-quark propagator is given by
\bes
G_{\bar{\mrm{h}}}(x,y) &=& \theta(y_0-x_0) \;
 \delta({\vecx-\vecy}) \;\exp\big(- \dmstathat\,(y_0-x_0)\big)\;
\Ptrans(x,y;0) \; P_- \;.
 \label{e:stat_aprop}
\ees

\section{Symmetries}

All HQET symmetries are preserved on the lattice,
in particular the U($2F$) spin-flavor symmetry 
and the local flavor-number conservation. The symmetry
transformations can literally be carried over
from the continuum, e.g. \eq{e:locflav}. One just replaces
the continuum fields by the lattice ones.

Note that these HQET symmetries are defined in terms
of transformations of the heavy quark fields while
the light quark fields do not change (unlike e.g. standard parity). 
Integrating out just
the quark fields in the path integral while leaving
the integral over the gauge fields,
they thus yield identities for the integrand or one may say for
``correlation functions in any fixed gauge background field''.

\section{Symanzik analysis of cutoff effects} \label{s:symanzik}

According to the --- by now well tested\footnote{See Peter Weisz' lectures 
for a theoretical discussion and chapter I of \cite{nara:rainer}
for an overview of tests. Finally~\cite{impr:sigma_BNW,impr:sigma_BNWl} 
represents the most advanced understanding of the subject.} ---
Symanzik conjecture, the cutoff effects of a lattice theory
can be described in terms
of an effective {\em continuum} theory~\cite{impr:Sym1,impr:Sym2,impr:pap1}. 
Once the terms in Symanzik's effective Lagrangian are known,
the cutoff effects can be canceled by adding terms of the same form
to the lattice action,
resulting in an improved action.

For a static quark, Symanzik's effective action is~\cite{zastat:pap1}
\bes
  S_{\rm eff} &=& S_0 + a S_1 +\ldots\,,\quad S_i = \int\rmd^4x \,\lag{i}(x)
\ees
where $\lag{0}(x)=\Lh^{\rm stat}(x)$ is the continuum static 
Lagrangian of \eq{e:lstat3} and 
\bes
  \lag{1}(x)&=&\sum_{i=3}^5  c_i\,\op{i}(x)\,,
\ees
is given in terms of local fields with mass dimension  $[\op{i}(x)] =5$. 
Their coefficients $c_i$ are functions of the bare gauge coupling. 
Assuming for simplicity 
mass-degenerate light quarks with a mass $m_\mrm{l}$, the set of possible
dimension five fields, which share the symmetries of the lattice theory,
is 
\bes
  \op3 & = & \heavyb D_0D_0\heavy\,,
  \quad
  \op4  =   m_\mrm{l}\; \heavyb D_0 \heavy\,,
  \quad  
  \op5  =   m_\mrm{l}^2\; \heavyb \heavy\,.
\ees
Note that $P_{+}\sigma_{0j} P_{+}=0$ means there is no term
$\heavyb\sigma_{0j}F_{0j}\heavy$, and 
$\heavyb D_jD_j\heavy$ can't occur because it violates the local phase 
invariance \eq{e:locflav}. Finally $\heavyb\sigma_{jk}F_{jk}\heavy$
is not invariant under the spin rotations \eq{e:spinsymm}. 

Furthermore, we are only interested in on-shell correlation functions 
and energies. For this class of observables 
$\op3$, $\op4$  do not contribute~\cite{impr:onshell,impr:pap1}
because they vanish by the equation of motion\footnote{The equations of motion
follow just from a change of variable in the path integral.  Contact terms
are re-absorbed into the free coefficients $c_i$. We refer to Peter Weisz' 
lectures or
\cite{impr:pap1} for a more detailed discussion.}, 
\bes
  \label{e:eomh}
  D_0 \heavy =0 \,.
\ees
The only remaining term,
$\op5$, induces a redefinition of the mass counter-term
$\dmstat$ which therefore depends explicitly on the 
light quark mass. 

We note that 
for almost all applications, $\dmstat$ is explicitly canceled
in the relation between physical observables and one thus has
automatic on-shell $\Oa$ improvement for the static action.
No parameter has to
be tuned to guarantee this property. 
Still, the improvement of matrix elements and
correlation functions requires to also consider composite fields in
the effective theory.

\renewcommand{\light}{\psi}
\renewcommand{\lightb}{\overline{\psi}}

\exercis{
The static quark anti-quark potential.\\}{
\label{ex:pot}
A (time-local) field 
\bes
   O(t,\vecx,\vecy) &=& \heavyb(x)\, \Ptrans(x,y) \gamma_5 \,\aheavy(y)\,,
   \quad x_0=y_0=t \nonumber
\ees
with $\Ptrans(x,y)$ being a parallel transporter from $y$ to $x$ in $x_0=t$ plane, 
can be used to annihilate a quark-anti-quark pair at a separation
$\vecx-\vecy$,
while 
\bes
   \bar O(t,\vecx,\vecy) &=& -\aheavyb(y)\, \Ptrans(y,x) \gamma_5 \,\heavy(x)\,,
   \quad x_0=y_0=t 
\ees
will create a quark-anti-quark pair at a separation $\vecx-\vecy$.\\[1ex]
Show that for $t>0$
\bes
  \langle \,\bar O(t,\vecx,\vecy)\;O(0,\vecx,\vecy)\,\rangle 
  = \mrm{const.}\; \rme^{-2t\,\dmstathat} W(t,\vecx-\vecy)
\ees
where $W$ is the Wilson loop introduced in the lectures of P. Hernandez. 
Since the energy levels of HQET are finite (after inclusion
of a suitable $\dmstat$),
one can conclude that 
\bes
  V_\mrm{R}(\vecx-\vecy) = - \lim_{t\to\infty} \partial_t\ln(W(t,\vecx-\vecy))
  +2\dmstathat
\ees
is a finite quantity: the divergent constant in the bare potential is
absorbed by $\dmstathat$, i.e. by a renormalization of
the heavy quark mass.

Furthermore, from the $\Oa$ improvement of HQET, one concludes 
\cite{pot:intermed}
\bes
   V_\mrm{R}(\vecx-\vecy) =  V^\mrm{cont}_\mrm{R}(r=|\vecx-\vecy|)
   + \rmO(a^2) 
\ees
if the action for the light fields is $\rmO(a)$ improved.
\todo{
Note that we can find local fields of dimension six, compatible with the 
symmetries of the static action, which will contribute to the 
$\rmO(a^2)$ terms. A conflict with a determination of $O(a^2)$
improvement terms from the potential???; see Peter.}
}

\subsection{Renormalized and improved axial current.}
We now also have to specify the discretization of the light
quark field $\light$. We will generically think of a standard
$\rmO(a)$-improved Wilson discretization~\cite{impr:sw,impr:pap1}
but occasionally mention changes which occur when one has an 
action with exact chiral 
symmetry~\cite{exactchi:neub,exactchi:perfect,exactchi:martin}\footnote{When the 
chiral symmetry realization of domain wall fermions \cite{exactchi:shamir} 
is good enough, these fermions can of course also be considered
to have an in practice exact chiral symmetry.}
or a Wilson regularization 
with a twisted mass term\cite{tmqcd:pap1,tmqcd:pap2,tmqcd:FR1}.
As an example we study the time component of the axial current. 
In Symanzik's effective theory it is represented by
\be
 (\Astat)_{\rm eff} = \Astat + a \sum_{k=1}^4 \omega_k (\delta\Astat)_k\,,
 \quad \Astat= \lightb\gamma_0\gamma_5\heavy
\ee
with some coefficients $\omega_k$. Here the flavor index of the field
$\lightb$ is suppressed. It is considered to have some fixed but arbitrary value
for our discussion, except where we indicate this explicitly. 
A basis for the dimension four fields
$\left\{(\delta\Astat)_k\right\}$ is
\begin{eqnarray}
  \label{e_cur_basis}
  (\delta\Astat)_1 & = &  \lightb\ola{D}_j\gamma_j\gamma_5\heavy\,,
  \quad
  (\delta\Astat)_2  =  \lightb\gamma_5D_0\heavy\,, 
 \nonumber\\[-1ex] \\[-1ex] \nonumber
  (\delta\Astat)_3 & = & \lightb\ola{D}_0\gamma_5\heavy\,,
  \quad 
  (\delta\Astat)_4  =  m_\mrm{l}\,\lightb\gamma_0\gamma_5\heavy\,.
\end{eqnarray}
From \eq{e:eomh} we see that $k=2$ does not contribute, 
while the equation of motion for $\lightb$ relates $(\delta\Astat)_3$,
$(\delta\Astat)_4$
and  $(\delta\Astat)_1$. We choose to remain with $k=1$ 
(and in principle $k=4$),
but for simplicity assume\footnote{If 
the light quark action has an exact chiral symmetry or the light
quarks are discretized with a twisted mass term at full twist,
this restriction is unnecessary, since the term is excluded by the symmetry.
Note that $(\delta\Astat)_1$ is, however, not forbidden
by chiral symmetry and  $\castat$ is necessary for 
$\rmO(a)$-improvement in any case.}
$a\,m_\mrm{l} \ll 1$; we can then drop $(\delta\Astat)_4$.
So for on-shell quantities the effective theory representation
is
\bes
  (\Astat)_{\rm eff} = \Astat + a \tilde\omega_1 (\delta\Astat)_1  \,.
                              % + a \tilde\omega_4 (\delta\Astat)_4  \,.
\ees
In order to achieve
a cancellation of the $\Oa$ lattice spacing effects, we
add a corresponding combination of correction terms to the
axial current in the lattice theory and
write the improved and renormalized current in the form
\bes
 \Aren &=& \zastat(g_0,a\mu)\,% (1+\bastat(g_0) a\mq) 
           \Astatimpr 
   \,,\quad \\
   \Astatimpr  &=&  \Astat +  a\castat(g_0)\,
\lightb\gamma_j\gamma_5
       \frac12(\lnab{j}+\lnabstar{j})\heavy
   \,,\quad 
\ees
with a mass-independent renormalization constant $\zastat$ and
a dimensionless improvement coefficient, $\castat$, depending again
on $g_0$ but not on the light quark mass. 

The improvement coefficients can be determined such that
for this (time component of the) {\em improved axial current} we have
the representation 
\bes
  (\Astat)_{\rm eff} = \lightb\gamma_0\gamma_5 \heavy +\rmO(a^2) \,,
\ees
in the Symanzik effective theory. In other words
$\tilde\omega_1$ is then $\Oa$ % = \tilde\omega_4 =0$ 
and cutoff effects are $\rmO(a^2)$.

The symmetries of the static theory are strong enough to improve
all components of the flavor currents in terms of just
$\castat$ and to renormalize them by $\zastat$. Let us discuss 
how this works.

\section{The full set of flavor currents} \label{s:statfields}
The previous discussion literally carries over 
to the time component of the vector current,
\bes
  \Vstat=\lightb \gamma_0 \heavy \,.
\ees
Its improved and renormalized lattice version may be chosen as
\bes
 \Vren &=& \zvstat\,% (1+\bvstat a\mq) 
           \Vstatimpr \\
   \Vstatimpr  &=&  \lightb \gamma_0 \heavy  +
           a\cvstat \lightb\gamma_j \frac12(\lnab{j}+\lnabstar{j})\heavy
   \,.
\ees
The chiral symmetry of the continuum limit can be used to
relate $\zvstat\,,\;\cvstat$ to $\zastat\,,\;\castat$
in the following way. 
We assume $\nf\geq 2$ massless light quarks. 
Then the infinitesimal transformation
\bes
  &&\da^a\light(x)=\frac{1}{2}\tau^a\dirac{5}\light(x),
  \qquad\phantom{\dirac{5}}
  \da^a\lightb(x)=\lightb(x)\dirac{5}\frac{1}{2}\tau^a \enspace ,
  \label{e_var_quarks}
\ees
with the Pauli matrices $\tau^a$ acting on two of the flavor components
of the light
quark fields $\light,\;\lightb$, is a (non-anomalous) symmetry of the theory.
Identifying $\Vstat=\lightb_1 \gamma_0 \heavy$, where 
$\lightb_1$ is the first flavor component of $\lightb$,
the vector current transforms as
$
   \da^3 \Vstat = -\frac12\Astat\,. 
$
The same property can then be required for the renormalized 
and improved lattice fields,% (up to $\rmO(a^2)$ corrections), 
\bes
   \label{e:WIstat}
   \da^3 \Vren &=& -\frac12\Aren + \rmO(a^2) \,.
\ees
This condition can be implemented in the form of Ward identities 
relating different correlation functions, 
in particular in the {\SF}. We refer to A. Vladikas' lectures 
and \cite{reviews:leshouches2} for the principle; 
practical implementations have been studied in 
\cite{zvstat:onogi,zvstat:filippo}. Such Ward identities 
determine $\zvstat\,,\;\cvstat$ in terms of $\zastat\,,\;\castat$.

Furthermore by a finite spin-symmetry transformation (with $\sigma_k$ of
\eq{e:sigmamat})
\bes
  \heavy\to\heavy'=\rme^{-i\pi\sigma_k/2} \heavy=-i\sigma_k \heavy
        \label{e:spinsymtrafo}
        \,,\qquad
        \heavyb' = \heavyb i\sigma_k \,, \qquad
\ees
we have
\bes
  \Vstat \to \big[\Vstat\big]' = \Akstat 
          \equiv\lightb \gamma_k\gamma_5 \heavy\,, \quad
        \label{e:ddvk} 
  \big[\Astat\big]' = \Vkstat
          \equiv\lightb \gamma_k \heavy \,, \quad
\ees
and we can require the same for the correction terms,
\bes
  \big[\delta\Vstat\big]' = \delta\Akstat\,, \quad
  \big[\delta\Astat\big]' = \delta\Vkstat\,. 
\ees
We leave it as an exercise to determine the form of 
$\delta\Akstat\,,\; \delta\Vkstat$.  
The discussed transformations are valid for the bare lattice fields at any lattice
spacing. Thus renormalization and improvement of the spatial components 
is given completely in terms of the time-components once we define
the renormalized fields to transform in the same way as the
bare fields.
A last property to note before writing down the renormalized
and improved fields is that we have
\bes
    \zvstat(g_0,a\mu)  = \zavstat(g_0)\,\zastat(g_0,a\mu)\,
\ees
with a $\mu$-independent function $ \zavstat(g_0)$ and up to $\rmO(a^2)$, 
as soon as we require \eq{e:WIstat}. 
\footnote{A formal argument is as follows. 
Rewrite \eq{e:WIstat} in terms of the bare operators,
$  \zvstat(g_0,a\mu) \da^3\Vstat 
 = -\frac12\zastat(g_0,a\mu) \Astat + \rmO(a^2)$.
Since the bare, regularized,  operators $\Vstat,\Astat$ carry no $\mu$-dependence,
we see that $\zvstat(g_0,a\mu)/\zastat(g_0,a\mu)$ is a function of $g_0$ only,
apart from $\rmO(a^2)$ cutoff effects. To make the argument more rigorous 
one should rewrite the equation in the form of correlation functions
which represent a Ward identity equivalent to \eq{e:WIstat}. 
}

Let us disregard the $\rmO(a)$ improvement terms for 
simplicity.
We can then summarize what we have learnt about the renormalization of
the static-light bilinears as
\bes
 \Aren &=& \zastat(g_0,a\mu)\,\Astat\,, \\
 \Vren &=& \zastat(g_0,a\mu)\,\zavstat(g_0)\,\Vstat\,, \\
 \Vkren &=& \zastat(g_0,a\mu)\,\Vkstat\,, \\
 \Akren &=& \zastat(g_0,a\mu)\,\zavstat(g_0)\,\Akstat\,,  
\ees
where $\zavstat(g_0)$ can be determined from a chiral Ward identity
\cite{zvstat:onogi,zvstat:filippo}. Note that we 
denote the flavor currents in HQET in complete analogy to QCD. Still
they do not form 4-vectors, as 4-dimensional rotation
invariance is broken in HQET. For example $\Aren$ cannot be rotated into 
$\Akren$ by a 90 degree lattice rotation. 

The only bilinears which are missing here are scalar, pseudo-scalar
densities (and the tensor). These are equivalent to $\Astat$ and $\Vstat$ in 
static approximation, for example
\bes
   &&\lightb\gamma_5\heavy = \lightb\gamma_5\gamma_0\heavy =-\Astat\,,\quad 
   \lightb\heavy = \lightb\gamma_0\heavy =\Vstat\,.   
\ees
At this stage it is therefore unnecessary to introduce renormalized 
scalar and pseudo-scalar densities. 

We have so far written down expressions for the relevant renormalized heavy-light
quark bilinears. The $Z$-factors can be chosen such that
correlation functions of these fields 
have a continuum limit (with $\dmstat$, gauge coupling and light quark masses
properly determined). Beyond this requirement, however, also the finite
parts need to be fixed by renormalization conditions. 
We have fixed some of them such that the renormalized fields
satisfy chiral symmetry and heavy quark spin symmetry. 
Only one finite part (in $\zastat$) then remains free. 
Preserving these symmetries by the renormalization is natural,
but not absolutely required; e.g. \eq{e:ddvk} could be violated in
terms of the renormalized fields. 
As long as one just remains inside the
effective field theory these ambiguities are not fixed.
The proper conditions for the finite parts, valid for
HQET as an effective theory of QCD, 
have to be determined from QCD with finite heavy
quark masses. We will return to this later. 

We may, however, already note that for renormalization group invariant fields,
these ambiguities are not present. The renormalization group invariants are 
thus very appropriate. 
Still, relating the bare lattice fields to the renormalization group invariant
ones is a non-trivial task in practice\cite{alpha:sigma,mbar:pap1}. 
We will briefly discuss how it can  be done
(and has been done) for the static-light 
bilinears~\cite{zastat:pap1,zastat:pap3,zastat:nf2}. 
For this and other purposes we need the 
{\SF}. In the following we just give a simplified review
of it and describe
how static quarks
are incorporated.  Some more details are discussed by Peter Weisz.

\section{HQET and Schr\"odinger Functional} \label{s:SF}

The \SF \cite{SF:Sym,SF:LNWW,SF:stefan1,SF:stefan2} can just be seen as 
QCD in a finite Euclidean space-time of size $T\times L^3$,
with specific boundary conditions. It is useful as a renormalizable
probe of QCD, providing a definition of correlation functions which 
are accessible at all distances, short or long: gauge invariance 
is manifest and even at short distances (large momenta) cutoff effects
can be kept small. It will help us to perform the non-perturbative
renormalization of HQET and its matching to QCD. 
In all these applications it is advantageous to have a variety of 
kinematics at ones disposal. One element is to have 
access to finite but small momenta of the quarks (think of the free 
theory, a relevant starting point for the short distance regime).  

To this end, the spatial boundary conditions were chosen to be 
$
   \light(x+L\hat k) = \rme^{i\theta_k} \light(x)\,, \;
   \lightb(x+L\hat k) = \rme^{-i\theta_k} \lightb(x)
$ in \cite{pert:1loop}, which allows momenta 
\bes
   p_k = {2\pi l_k \over L} +{\theta_k \over L} \,, \; l_k\in \gz \,,
\ees
in particular small ones when $l_k=0$.
Performing a variable transformation 
$\psi(x)\to\rme^{i\theta_kx_k/L}\psi(x)$, 
$\psibar(x)\to\rme^{-i\theta_kx_k/L}\psi(x)$, 
for $0\leq x_k \leq L-a$,
we see that this boundary condition is equivalent to periodic
boundary conditions (without a phase) for the new fields, while the
spatial covariant derivatives contain an additional phase, for example
\bes
  \nab{k}\psi(x)&=&
  {1\over a}\bigl[\rme^{i\theta_ka/L}U(x,\mu)\psi(x+a\hat{k})-\psi(x)\bigr]\,,
   \label{e:gaugetrafotheta}
\ees
see also \sect{s:notation}. The phase $\theta_ka/L$ can be seen as a 
constant abelian 
gauge potential and the above variable transformation as a gauge transformation.
Of course, the angles $\theta_k$ which we will set all equal from now
on ($\theta_k=\theta$), are not specific to the \SF; they just have first been 
used in this context.

The standard \SF boundary conditions in time are~\cite{SF:stefan1,SF:stefan2}
\be
  P_{+}\light(x)|_{x_0=0}=0\,,\quad
  P_{-}\light(x)|_{x_0=T}=0\,,
\ee
and
\be
  \lightb(x)P_{-}|_{x_0=0}=0\,,\quad
  \lightb(x)P_{+}|_{x_0=T}=0\,.
\ee
The gauge fields are taken periodic in space and the space components
of the continuum gauge fields are set to zero at $x_0=0$ and $x_0=T$ 
(on the lattice the 
boundary links $U(x,k)$ are set to unity).\footnote{In
\cite{SF:LNWW,alpha:SU3} the definition of a renormalized coupling
uses more general boundary conditions for the gauge fields,
but these are not needed here.} 

For the static quark the components projected by $P_-$ vanish anyway, 
so there is just
\be
  P_{+}\heavy(x)|_{x_0=0}=0,\quad
  \heavyb(x)P_{+}|_{x_0=T}=0.
\ee
Defining
\be
  \heavy(x)=0\quad\hbox{if $x_0<0$ or $x_0 \geq T$},
\ee
the lattice action for the static quark with \SF boundary conditions
can be written as
\be
  \label{e_HeavySFLatt}
  S_{\rm h} =  {1 \over 1+a\dmstat}a^4\sum_x
    \heavyb(x)[\nabstar{0}+\dmstat]\heavy(x)
\ee
as before.
In general the improvement of the \SF requires to add boundary terms
to the action as a straightforward generalization of Symanzik improvement. 
These terms are dimension four composite fields located
on or at the boundaries, summed over space \cite{SF:LNWW,impr:pap1}. 
Since they are not so important 
here and are also known sufficiently well, we do not discuss them. 
We just note that no boundary improvement terms involving static fields 
are needed~\cite{zastat:pap1}, since the dimension four fields
vanish either due to the equation of motion or the 
heavy quark symmetries.

We take the same periodicity in space as for relativistic quarks,
\bes
   \heavy(x+L\hat k) = \heavy(x)\,, \quad
   \heavyb(x+L\hat k) = \heavyb(x)\,.
\ees
In the static theory this has no effect, since quarks at different 
$\vecx$ are not coupled, but it plays a r\^ole at order $\minv$ 
where $\theta$ is a useful kinematical variable.

An important feature of the \SF is that one can form
gauge invariant correlation functions of 
boundary quark fields. In particular, one can project those
quark fields to small spatial momentum, e.g. $\vecp=1/L\times(\theta,\theta,\theta)$ for the quarks and $-\vecp$ for the anti-quarks. 
For the precise definition of the boundary quark fields
we refer to 
\cite{impr:pap1} or for an alternative view we refer to \cite{SF:martin2}. 
The details
are here not so important. We only need to know that these 
boundary fields, i.e. 
fermion fields localized at the boundaries, exist. Those at $x_0=0$
are denoted by
\bes
 \zetal(\vecx)\,,\; \zetalb(\vecx)\,,\; \zetaah(\vecx)\,,\; \zetahb(\vecx)\,,
 \nonumber
\ees
and those at $x_0=T$ by
\bes
 \zetalprime(\vecx)\,,\; \zetalbprime(\vecx)\,,\; \zetahprime(\vecx)\,,\; 
 \zetaahbprime(\vecx)\,.
 \nonumber
\ees

\subsection{Renormalization}
These boundary quark fields are multiplicatively renormalized with 
factors 
$
  \zzeta\,, \zzetah \,,
$ such that $(\zetal(\vecx))_\mrm{R}=\zzeta\zetal(\vecx)$ etc.

To illustrate a first use of the \SF and the boundary
fields we introduce three correlation functions 
\bes
   \nonumber
  && \includegraphics[width=0.2\textwidth]{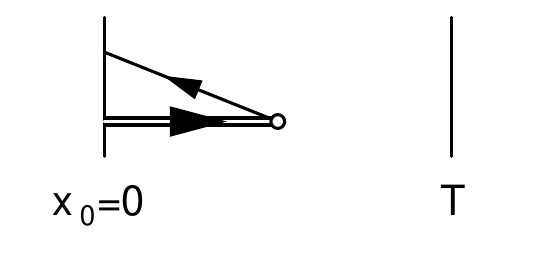} \\[-10ex]
  %\nonumber
  \fastat(x_0,\theta) = -{a^6 \over 2}\sum_{\vecy,\vecz}\,
  \left\langle
  \Astatimpr(x)\,\zetabar_{\rm h}(\vecy)\gamma_5\zeta_{\rm l}(\vecz)
  \right\rangle  &:&  
  \\[6ex]
  && 
  \nonumber \includegraphics[width=0.2\textwidth]{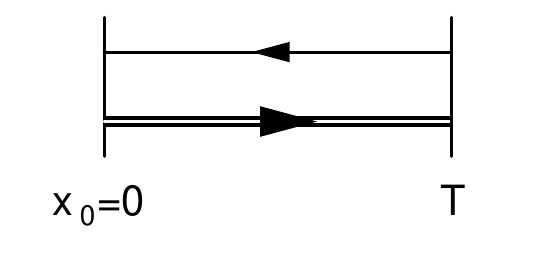} 
  \\[-10ex]
  %\nonumber
  \fonestat(\theta) = -{a^{12} \over 2L^6}\sum_{\vecu,\vecv,\vecy,\vecz}
  \left\langle
  \zetalbprime(\vecu)\gamma_5\zetahprime(\vecv)\,
  \zetabar_{\rm h}(\vecy)\gamma_5\zetal(\vecz)
  \right\rangle &:&
  \\[6ex]
  && 
  \nonumber \includegraphics[width=0.2\textwidth]{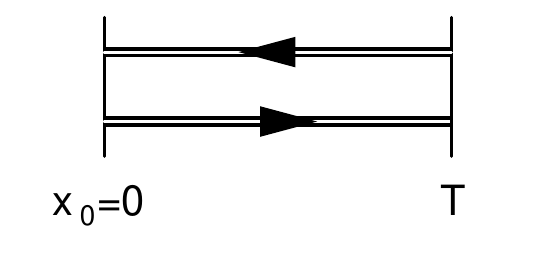} 
  \\[-10ex]
\fonehh(x_3,\theta)= -{{a^8}\over{2L^2}} \sum_{x_1,x_2,{\bf y},{\bf z}}
 \; \langle \zetaahbprime ({\bf x})
\gamma_5
\zetahprime ({\bf 0}) \; \zetahb ({\bf y}) \gamma_5
\zetaah ({\bf z}) \rangle &:&
\label{e:f1hh}
  \\[2ex]   \nonumber
\ees
In the graphs, double lines are static quark propagators. 
Note that the sum in \eq{e:f1hh} runs on $x_1$ and $x_2$ and 
therefore yields an
$x_3$-dependent correlation function.
We further point out that $\sum_\vecy$ etc. project the boundary quark fields 
onto zero (space) momentum, but together with the abelian gauge field, 
this is equivalent to a physical momentum $p_k=\theta/L$. For example 
the time-decay of a free mass-less quark propagator projected 
this way contains an energy $E(\theta/L,\theta/L,\theta/L)=\sqrt{3} \theta/L$,
cf. \eq{e:propmixed}. 

The above functions are renormalized as
\bes
  \left[\fastat\right]_\mrm{R} = \zastat \zzetah\zzeta \;\fastat
  \,,\quad
  \left[\fonestat\right]_\mrm{R} = \zzetah^2\zzeta^2 \;\fonestat
  \,, \quad
  \left[\fonehh\right]_\mrm{R} = \zzetah^4 \;\fonehh\,.
  \nonumber\\[-1ex]
  \label{e:ren_stat_correl}
\ees
We remind the reader that an additional renormalization is 
the mass counter-term of the static action.

The ratio
\bes
  \label{e:renorm_correl}
  \left[{\fastat(T/2,\theta) \over \sqrt{\fonestat(\theta)}}\right]_\mrm{R}
  = \zastat(g_0,a\mu)\,{\fastat(T/2,\theta) \over \sqrt{\fonestat(\theta)}}
  \,
\ees
renormalizes in a simple way and also needs no knowledge of $\dmstat$,
since it cancels out due to \eq{e:stat_prop}.
It is hence an attractive possibility to {\em define} the renormalization constant
$\zastat$ through this ratio. Explicitly we may choose
\bes
  \label{e:zastatdef}
  \zastat(g_0,a\mu)\, \equiv 
  {\sqrt{\fonestat(\theta)} \over \fastat(L/2,\theta)}
  \left[{\fastat(L/2,\theta) \over \sqrt{\fonestat(\theta)}}\right]_\mrm{g_0=0}
  \quad\mu=1/L\,,\quad T=L \,,\quad \theta=\frac12\,, \nonumber \\[-0.5ex]  
\ees
which defines the finite part of $\zastat$ in a so-called \SF scheme.
As usual the factor 
$\left[{\fastat(L/2,\theta) / \sqrt{\fonestat(\theta)}}\right]_\mrm{g_0=0}$
is inserted to ensure $ \zastat=1+\rmO(g_0^2)$.
The name \SF scheme just refers to the fact that the renormalization factor 
is defined in terms of correlation functions with \SF
boundary conditions. While $\zastat$ refers to a specific regularization,
the renormalization scheme is independent of that and can in principle
be applied in a continuum regularization. Many similar
\SF schemes can be defined (e.g. \cite{zastat:pap3}), 
but by the choice $T=L, \theta=0.5$ we
have made   \eq{e:zastatdef}
 unique. It is implied that the light quark masses are
set to zero. We will soon come back to the $\mu$-dependence of 
the renormalized current and its relation to the RGI current.
First let us show some numerical results which provide a 
non-perturbative test of the 
renormalizability of the static theory. 

\section{Numerical test of the renormalizability}

The above listed renormalization structure
of the \SF correlation functions is just deduced from
a simple dimensional analysis. A number of 1-loop calculations 
of the correlation functions defined above as well as of others
\cite{zastat:pap1,zastat:pap2,stat:actpaper,zvstat:filippo} 
confirm the structure
  \eq{e:ren_stat_correl} and more 
generally the renormalizability of the theory (by local counter-terms). 

\begin{figure}[ht!]
\vspace{0pt}
\centerline{\epsfig{file=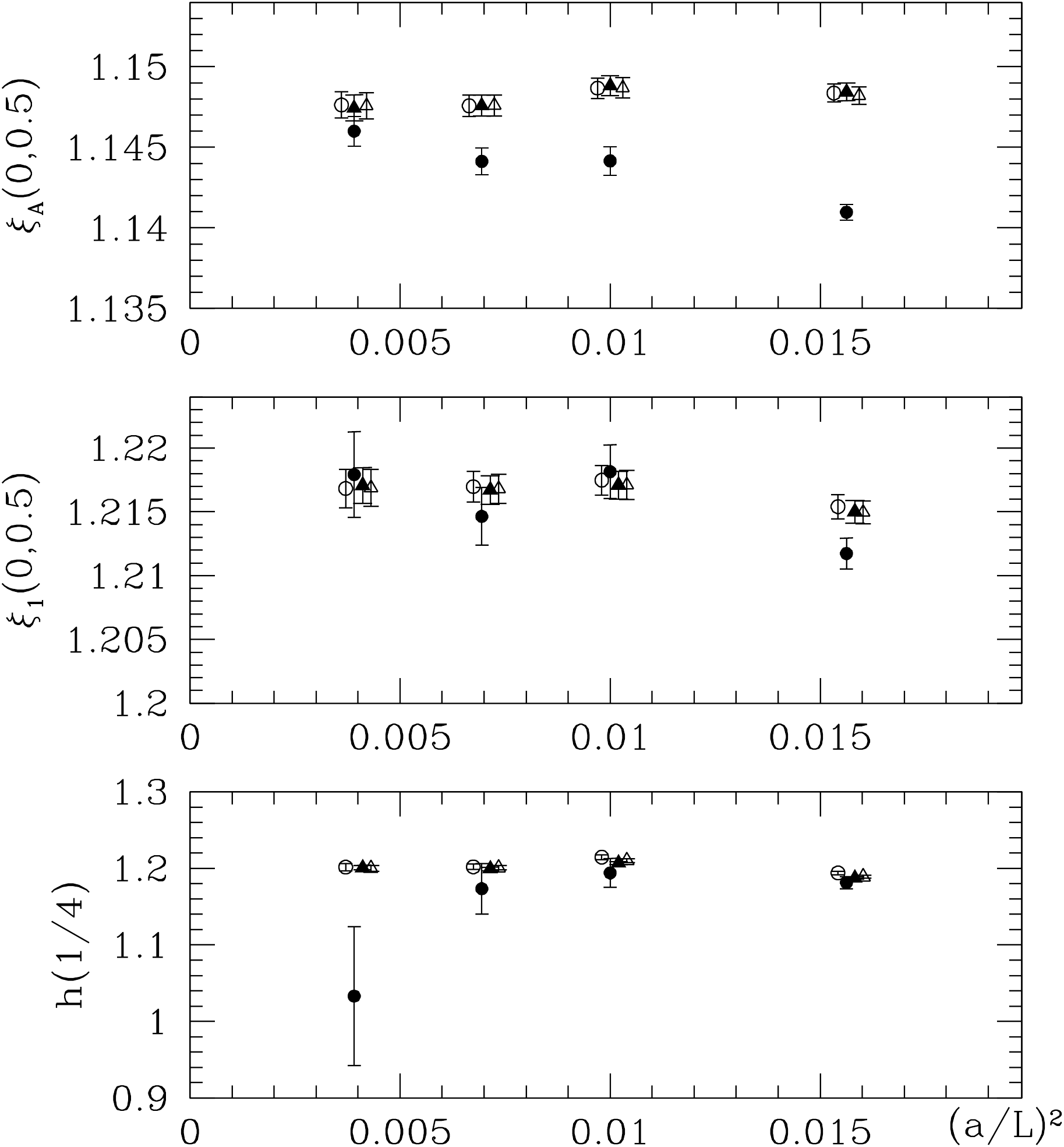,
       width=7.5cm} }
  \caption{\footnotesize 
        Lattice spacing dependence of various ratios
        of correlation functions for which Z-factors cancel. 
        Different symbols correspond to different actions.
        Computation and figure from \protect\cite{stat:actpaper}. 
        }\label{f:scal}
\end{figure}

Also non-perturbative tests exist. 
A stringent and precise one~\cite{stat:actpaper}
is based on the ratios
\begin{equation}
\xi_{\rm A}(\theta,\theta')={\fastat(T/2,\theta) \over
                           \fastat(T/2,\theta')} \; , \quad
\xi_1(\theta,\theta') = {\fonestat(\theta) \over
                           \fonestat(\theta')} \; ,
\quad h(d/L,\theta)={{f_1^{\rm hh}(d,\theta)}\over{f_1^{\rm hh}(L/2,\theta)}}\; .
\label{xi_del_h}
\end{equation}
The additional dependence on $L$ and the lattice resolution $a/L$ 
of these ratios is not indicated explicitly. With \eq{e:ren_stat_correl}, 
we see that all renormalization factors cancel in
these ratios. They should
have a finite limit $a/L\to 0$, approached asymptotically
with a rate $(a/L)^2$. This is tested in \fig{f:scal},
where $L$ is kept fixed in units of the reference length
scale $r_0$\cite{pot:r0} to $L/r_0=1.436$. This choice
corresponds to about 
$L\approx0.7\,\fm$. The same continuum limit has to be
reached for different lattice discretizations. Also this 
universality is
tested in the graphs, where four different choices of the 
covariant derivative $D_0$ in the static action are used. 
All actions defined by the different choices of $D_0$
have the symmetries discussed earlier. 

%%%%%%%%%%%%%%%%%%%%%%%%%%%%%%%%%%%%%%%%%%%%%%%%%%%%%%%%%%%%%%%%%%%%%%%%%%%%%%%%%%%%
%%%%%%%%%%%%%%%%%%%%%%%%%%%%%%%%%%%%%%%%%%%%%%%%%%%%%%%%%%%%%%%%%%%%%%%%%%%%%%%%%%%%%%%%%%

\section{Scale dependence of the axial current and the RGI current}

Let us first recapitulate the scale dependence in perturbation theory. 
At one-loop
order one has 
\bes
   &&\Aren(x) = \zastat(g_0,\mu a) \Astat(x)\,,
\\ && \zastat(g_0,\mu a) = 1+g_0^2\,[B_0-\gamma_0\ln(a\,\mu)]+\ldots
\,,\quad \gamma_0 = -{1\over 4\pi^2}\,.
\ees
In the lattice minimal subtraction scheme the $Z$--factors are polynomials
in $\ln(a\mu)$ without constant part; thus 
$B_0=0$. Instead, when the renormalization scheme is defined by 
\eq{e:zastatdef} a one-loop computation of $\fastat,\,\fonestat$
yields\cite{zastat:pap1}~\footnote{As usual in perturbation
theory, terms of order $(a\mu)^n\,,n\geq1$ are dropped.} $B_0=-0.08458$.
As usual there is the renormalization group equation (RGE)
(remember $g_0^2 =\gbar^2 + \rmO(\gbar^4)$)
\bes
  \mu  {\partial \over \partial\mu} \Aren =  \gamma(\gbar) \Aren \,,\quad
  \gamma(\gbar) =
  - \gbar^2 \left\{ \gamma_0 + {\bar g}^{2}  \gamma_1 + \ldots \right\}\,.
\ees
Combining it with the RGE for the coupling \eq{e_RG} it is
easily integrated to (see \eq{e_RGpert} for the definition of
the beta-function coefficients $b_i$)
\bes
  \Aren(\mu) &=&\Argi \; \exp \left\{\int^{\gbar(\mu)} \rmd x
           {\gamma(x) \over \beta(x)}        \right\}
           \label{e:arenrgi0}
            \\ &\equiv& 
 \Argi\; \left[\,2b_0 \gbar^2\,\right]^{\gamma_0/2b_0}
                   \exp\left\{\int_0^{\gbar} \rmd x
                     \left[\,{ \gamma(x) \over\beta(x)}
                           -{\gamma_0 \over b_0 x}\,\right]
                     \right\} \, \label{e:arenrgi} 
\ees
where \eq{e:arenrgi} provides the definition of the 
lax notation for the second factor in \eq{e:arenrgi0}. 
The integration ``constant'' is the renormalization group invariant
field. It
can also be written as
\bes
   \Argi = \lim_{\mu\to\infty} \left[\,2b_0
   \gbar^2(\mu)\,\right]^{-\gamma_0/2b_0} \Aren(\mu)\; \nonumber \,,
\ees
since the last factor in \eq{e:arenrgi} converges to one
as $\mu\to\infty$. Using also that $\gamma_0,b_0$ are independent
of the renormalization scheme, as well as 
$O_S(\mu)=O_{S'}(\mu)(1+\rmO(\gbar^2(\mu))$ (valid for any operator $O$ and
standard schemes $S,S'$),  
this representation also
shows that the renormalization group invariant operator
$\Argi$ is independent of scale and scheme.\footnote{Of course,
a trivial definition dependence due to the choice of pre-factors
in \eq{e:arenrgi} is present. Unfortunately there is no uniform 
choice for those in the literature.}

Let us now go beyond perturbation theory and start from
a non-perturbative definition of the renormalized current,
such as \eq{e:zastatdef}, together with a non-perturbative definition of 
a renormalized coupling \cite{alpha:sigma,SF:LNWW,alpha:SU3}. 
With the step scaling method discussed in more detail
by Peter Weisz, one can then determine the change 
\bes
  \Aren(\mu) = \sigmaAstat(\gbar^2(2\mu))\,  
  \Aren(2\mu) \,,\quad \mu=1/L
\ees
of the renormalized field $\Aren(\mu)$
when the renormalization scale $\mu$ is changed
by a factor of two. The so-called step scaling function
$\sigmaAstat$ is parameterized in terms of the running coupling
$\gbar(\mu)$. Its argument is $\mu=1/L$ in terms of the
linear extent, $L=T$,  of a {\SF}. 

Instead of the scale dependence of
$\Aren(\mu)$ we will often discuss 
a generic matrix element 
\bes
  \Phi(\mu) = \langle \alpha | \widehat{\Aren}(\mu) |\beta \rangle
\ees
of the associated operator $\widehat{\Astat}$ in Hilbert space.
% (since we can also divide $\Phi(\mu)/\Phi(\mu')$.

%%%%%%%%%%%%%%%%%%%%%%%%%%%%%%%%%%%%%%%%%%%%%%%%%%%%%%%%%%%%%%%%%%%%%%%%%%%%%%%%%%%%
\begin{figure}[ht!]
\vspace{0pt}
\centerline{\epsfig{file=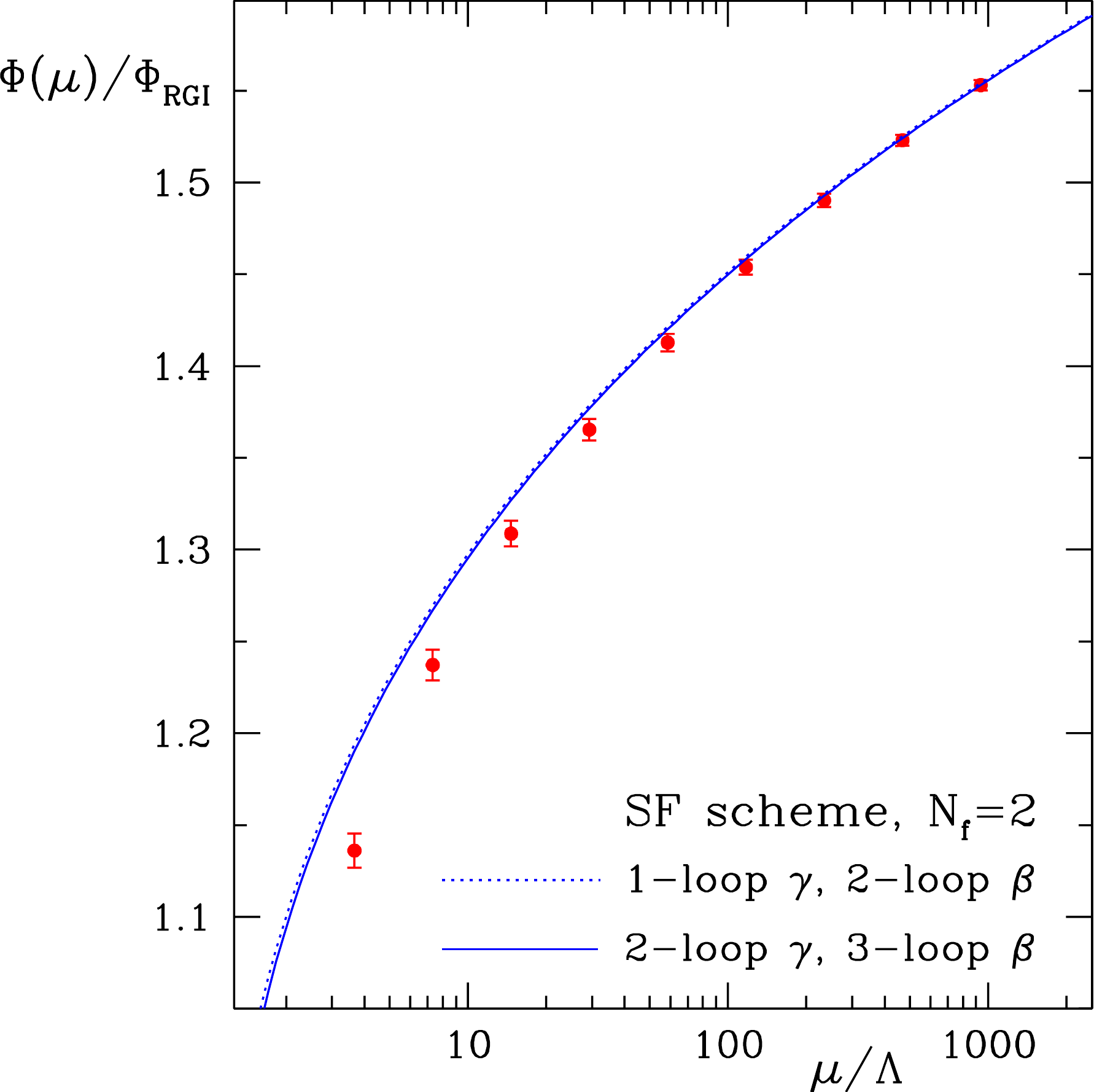,
       width=7.5cm} }
  \caption{\footnotesize Relation $\Phi(\mu)/\Phi_\mrm{RGI}$
        between RGI and matrix element at
        finite $\mu$ in a \SF scheme and for $\nf=2$. 
        The $\Lambda$-parameter in the 
        SF-scheme is around $100\,\MeV$. Everything  
        was computed non-perturbatively from continuum extrapolated
        step scaling functions \protect\cite{zastat:nf2}. 
        }\label{f:phibar}
\end{figure}
%%%%%%%%%%%%%%%%%%%%%%%%%%%%%%%%%%%%%%%%%%%%%%%%%%%%%%%%%%%%%%%%%%%%%%%%%%%%%%%%%%%%%%%%%%
In a non-perturbative calculation, the continuum $\sigmaAstat$
is obtained through a numerical extrapolation 
\bes
  \sigmaAstat(u) = \lim_{a/L\to0}\SigmaAstat(u,a/L)
\ees
of the lattice step scaling functions
\bes
   \SigmaAstat(u,a/L) = \left.\zastat(g_0,a/2L) \over \zastat(g_0,a/L)
   \right|_{\gbar^2(1/L) = u}
\ees
obtained directly from simulations. Here, $\gbar^2(1/L)$ is kept fixed 
to remain at constant 
$L$ while $L/a$ is varied in the continuum extrapolation. 

The $\mu$-dependence of $\Phi$ can then be constructed iteratively
via
\bes
  u_0 = \gbar^2(1/L_0)\,, \nonumber \quad
  L_n=2^n\, L_0 \,,\;&:&   
  \Phi(1/L_{n+1}) = \sigmaAstat(u_n)\,  \Phi(1/L_{n})  
  \\ \nonumber &&
  u_{n+1} = \sigma(u_n)\,,
\ees
where the step scaling function $\sigma$ of the running coupling enters. 
The length scale $L_0$ is chosen deep in the perturbative domain, typically
$L_0 \approx 1/100\,\GeV$ and therefore the $\mu$-dependence can be completed
perturbatively to infinite $\mu$, i.e. to the RGI using \eq{e:arenrgi}.

For $\nf=2$ the analysis has been done 
for $\mu\approx300\,\MeV ... 80\,\GeV$. After it was
verified that the steps at smallest $L$ ($L\leq L_2$)
are accurately described by perturbation theory (see \fig{f:phibar}), 
the two-loop anomalous dimension was used in \eq{e:arenrgi}
with $\mu=1/L_p$ to connect to the RGI current.
The result \cite{zastat:pap1,zastat:pap3,zastat:nf2}
is conveniently written as
\bes
   \label{e:zastatrgi}
   \zastatrgi(g_0) &=& 
   \left.{\Phi_\mrm{RGI} \over \Phi(\mu)} \times \zastat(g_0,a\mu)
        \right|_{\mu=1/(2\lmax)}\,,   
\ees
where only the second factor depends on the lattice action, and
$\gbar^2(1/\lmax)=u_\mrm{max}$ is a convenient value covered by the 
non-perturbative results for the above recursion.

We show the result for the first factor in \fig{f:phibar}
for a series of $\mu$ with $\nf=2$ dynamical quarks. The different
points in the graph correspond to different $n$ in the recursion. 
Note that the two-loop running becomes accurate only at rather 
small $L$. There is an about 5\% difference
in $\Phi_\mrm{RGI} \over \Phi(\mu)$ between a two-loop result and the 
non-perturbative one at the smallest $\mu$. 

%%%%%%%%%%%%%%%%%%%%%%%%%%%%%%%%%%%%%%%%%%%%%%%%%%%%%%%%%%%%%%%%%%%%%%%%%%%%%%%%%%%%
\begin{figure}[ht!]
\vspace{0pt}
\centerline{\epsfig{file=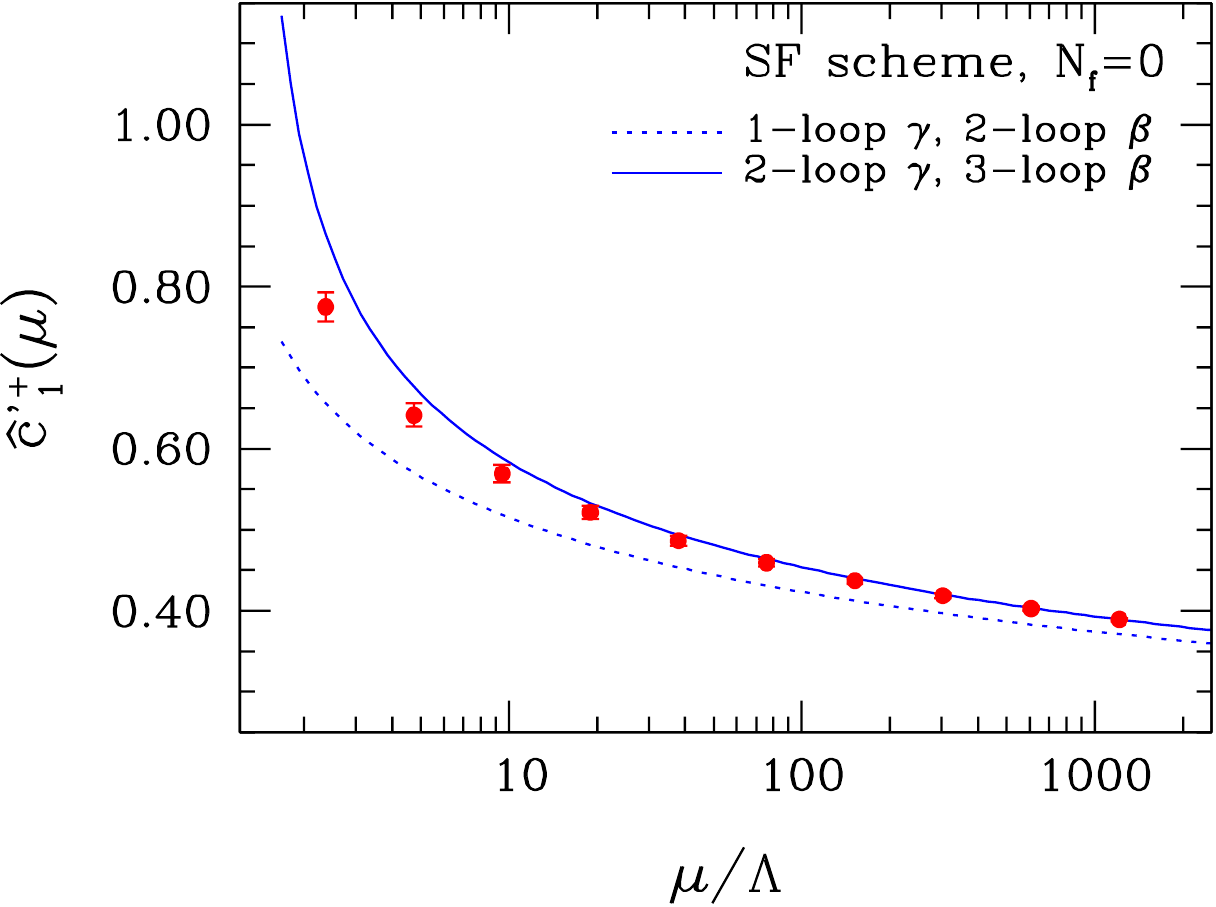,
       width=7.5cm} }
  \caption{\footnotesize  
        Relation $\hat c=\Phirgi/\Phi(\mu)$ of the RGI matrix element $\Phirgi$
        and the matrix element at
        finite $\mu$ of the ``$VA+AV$'' four--fermion operator in a \SF scheme. 
        It was computed non-perturbatively from continuum extrapolated
        step scaling functions \protect\cite{hqet:4ferm_nf2}.  
        }\label{f:4ferm}
\end{figure}
%%%%%%%%%%%%%%%%%%%%%%%%%%%%%%%%%%%%%%%%%%%%%%%%%%%%%%%%%%%%%%%%%%%%%%%%%%%%%%%%%%%%%%%%%%
The details of this calculation and strategy is not that important
for the following.
We have mainly discussed it since\\
-- first the RGI matrix elements play a 
prominent role in HQET in static approximation and 
it is relevant to understand that they can be obtained completely 
non-perturbatively  and \\
-- second we also want to later emphasize the 
difference between the here used -- by now more classic -- renormalization
of the static theory and the strategy discussed in \sect{s:strat}.

Let us further note that the strategy above has been extended to
four-fermion operators relevant for $B-\bar B$ oscillations in
\cite{stat:zbb_pert,hqet:4ferm_nf2}. Since more than one 
operator is involved and twisted mass QCD is used
in order to avoid mixing with operators of wrong chirality, 
the strategy and computation are somewhat more involved. In particular
two different four-fermion operators contribute at leading order in
$\minv$. As an example, 
we just show in \fig{f:4ferm} the result for
the four--fermion operator which dominates in the physical process.

%\neben{
\section{Eigen-states of the Hamiltonian}

The eigenstates of the static Hamiltonian can be diagonalized
simultaneously with the local heavy flavor number operator 
(remember  $Q_\mrm{h}(x)= \heavyb(x)\heavy(x)$ ). 
We consider a finite volume with periodic boundary conditions.
Since the theory is translation invariant, there is a $k\times L^3/a^3$ -- fold
degeneracy of states with a single heavy quark~\footnote{In certain types of quark smearing,
this has to be properly taken into account\cite{stat:degeneracy}.},
where $k$ arises from degeneracies on top of the 
translation invariance discussed here. 
For the lowest energy level one can choose 
a basis of eigenstates of the Hamiltonian as 
\bes
    | \tilde B(\vecx) \rangle \,,  \quad
    \langle \tilde B(\vecx) | \tilde B(\vecy) \rangle = 
    2\delta(\vecx-\vecy) \,,\quad \hat Q_\mrm{h}(\vecy)  | \tilde B(\vecx) \rangle
    = \delta(\vecx-\vecy)
\ees 
or their Fourier transformed
\bes
   &&| B(\vecp) \rangle = a^3\sum_{\vecx} \rme^{-i\vecp\vecx}
   |\tilde B(\vecx) \rangle \,,\quad 
   \langle\tilde B(\vecp') |\tilde B(\vecp) \rangle = 2(2\pi)^3 \delta(\vecp-\vecp')\\
   && \delta(\vecp-\vecp') = (L/(2\pi))^3 \prod_i 
   \delta_{l_i l_i'} \,,\quad k_i= {2\pi l_i \over L}\,, l_i\in\gz.
\ees
(Here we set $\theta=0$.)
We usually work with the zero momentum eigenstate, denoted for short
by $| B \rangle = | B(\vecp=0) \rangle$,
as this is related to an eigenstate of the finite mass QCD Hamiltonian,
which in finite volume has normalization $\langle B | B \rangle = 2L^3$.
%}

\chapter{Mass dependence at leading order  in 1/m:
 Matching} \label{s:mass}

We now discuss the ``matching'' of HQET to QCD 
using the example of a simple correlation function. 
As mentioned before, the issue is to fix the 
finite parts of renormalization constants such that the
effective theory describes the underlying theory QCD. 
Throughout this section we remain in static approximation. 
Matching including  $1/m$ terms will be discussed in the next
section. 

\section{A correlation function in QCD}  
\label{s:corrqcd}

We start from a simple QCD correlation
function, which we write down in the lattice regularization,
\bes \label{e:caa}
  \caar(x_0) = \za^2\, a^3\sum_{\vecx} \Big\langle A_0(x)  A_0^{\dagger}(0)
              \Big\rangle_\mrm{QCD}
\ees
with the bare heavy-light axial current in QCD,
$A_\mu=\lightb\gamma_\mu\gamma_5\psi_\beauty$, and
$A_\mu^{\dagger}=\psibar_\beauty\gamma_\mu\gamma_5\light$. 
The current is formed with
the relativistic b-quark field $\psi_\beauty$. 
In QCD, the renormalization
factor, $\za(g_0)$, is fixed by chiral Ward identities~\cite{Boch,impr:pap4}.
It therefore does not depend on a renormalization scale.

One reson to consider this correlation function is that
at large time the B-meson state dominates its spectral
representation via
\bes 
   \caar(x_0) &=& \za^2 a^3\sum_{\vecx} 
              \langle0| A_0^{\dagger}(x) |B\rangle{1\over 2L^3}
              \langle B|A_0(0)|0 \rangle \rme^{-x_0 \mB} 
              \left[1+\rmO(\rme^{-x_0\Delta})\right] \nonumber
   \\
   &=& \za^2 \frac12 
              \langle0| A_0^{\dagger}(0) |B\rangle
              \langle B|A_0(0)|0 \rangle \rme^{-x_0 \mB} 
              \left[1+\rmO(\rme^{-x_0\Delta})\right]
\ees
and the B-meson mass and its decay constant can  be obtained 
from\footnote{It is technically of advantage
to consider so-called smeared-smeared and local-smeared correlation functions,
but this is irrelevant in the present discussion.}
\bes
   \label{e:meff}
    \meff_\mrm{AA}(x_0) &=& -\dzero\, \ln(\caa(x_0))\,
    = \mB + \rmO(\rme^{-x_0\Delta})
   \\ \label{e:fb}
   \quad
    \big[\Phiqcd\big]^2 &\equiv& \fb^2\,\mB  \\ \nonumber
        &=& \big|\langle B | \za A_0 |0 \rangle\big|^2 =
         2\lim_{x_0\to\infty} \exp(x_0\,\meff_\mrm{AA}(x_0)) \caa(x_0)\,.
\ees
Note that we use the normalization \eq{e:nrnorm} (in finite volume
$\langle B | B \rangle = 2L^3$) for the zero momentum state
$|B\rangle$. The gap $\Delta$ is the energy difference 
between the second energy level and the first energy level
in the zero momentum (flavored) sector of the Hilbert space
of the finite volume lattice theory. 

\section{The correlation function in static approximation} 
In the static approximation we replace 
$\za A_0 \to  \zastat(g_0,\mu a) \Astat$ and define
\bes
 \label{e:caastat}
  \caarstat(x_0) &=& (\zastat)^2\, \caastat(x_0) = 
  (\zastat)^2\,a^3\sum_{\vecx} \Big\langle \Astat(x)  (\Astat)^{\dagger}(0)
              \Big\rangle_\mrm{stat} \\
   \label{e:meffstat}
    \meffstat_\mrm{AA}(x_0) &=& -\dzero\, \ln(\caastat(x_0))\,,
   \\ \label{e:fbstat}
   \quad
    \big[\Phi(\mu)\big]^2 
        &\equiv& \big|\langle B | \zastat\Astat |0 \rangle_\mrm{stat}\big|^2  
     \\ &=&
         2\lim_{x_0\to\infty} \exp(x_0\,\meffstat_\mrm{AA}(x_0)) (\zastat)^2\,\caastat(x_0)\,.
     \nonumber
\ees
The $\mu$-dependence of $\Phi$ results from the renormalization 
of the current in the effective theory,
\bes
  \zastat(g_0,\mu a) = 1+g_0^2\,[B_0-\gamma_0\ln(a\,\mu)]+\rmO(g_0^4)
\,.
\ees
Different renormalization schemes have different constants $B_0$.  
Alternatively one uses the renormalization group invariant operator $\Argi$.
We come to that shortly.

\section{Matching} \label{s:match}
The correlation function $\caa$ and the
matrix element $\Phiqcd$, \eq{e:fb},
are independent of any renormalization scale, due to the chiral symmetry of QCD
in the massless limit. But of course they depend
on the mass of the b-quark. 

In the effective 
theory we first renormalize in an arbitrary scheme, which we
do not need to specify for the following, resulting in 
a scale-dependent $\Phi(\mu)$. The two quantities are 
then related through the matching equation 
(without explicit superscripts ``QCD''
we refer to HQET quantities, here static), 
\be \label{e:match1}
  \Phiqcd(m) =\widetilde C_\mrm{match}(m,\mu)\times\Phi(\mu)
                                + \rmO(1/m)\,.
\ee
Somewhat symbolically the same equation could be written
for the current instead of its matrix element; we write ``symbolically'' 
since the two currents belong to theories with different field
contents. However, thinking in terms of the currents, it is
clear that \eq{e:match1} can be thought of as a change of renormalization
scheme in the effective theory, where the new renormalization scale is
$m=\mbeauty$ and the finite part is exactly fixed by \eq{e:match1}.
In fact, since at tree level we have constructed the effective
theory such that $\Phi=\Phiqcd$, the tree-level value for
$\widetilde{C}_\mrm{match}$ is one and we have a perturbative expansion
\be
 \label{e:c1_m_mu}
 \widetilde C_\mrm{match}(m,\mu) = 
 1 + c_1(m/\mu) \gbar^2(\mu) +\ldots
\ee
The finite renormalization factor $\widetilde C_\mrm{match}$
may be determined 
such that \eq{e:match1} holds for some particular matrix element
of the current and will then be valid {\em for all matrix elements}
or correlation functions.
Some aspects of the above equation still need explanation. 
The $\mu$-dependence is not present on the left-hand-side and
this should be made explicit also on the right-hand-side; further
one may wonder which 
definition of the quark mass and coupling constant one is to choose.

\subsection{One-loop}
Before coming to these issues, it is illustrative
to write down explicitly what \eq{e:match1} looks like
at 1-loop order. Ignore for now how we renormalized
the current in \eq{e:zastatdef}
and use instead lattice minimal subtraction,
\bes
   \zastat(g_0,\mu a) = 1-\gamma_0\ln(a\,\mu)\,g_0^2+\ldots\,,
\ees
in the static theory.

Instead of the decay constant, use as an observable
a perturbatively accessible quantity. We take\footnote{
The correlation functions $\fa$, $\fone$ are the 
relativistic versions of  $\fastat$, $\fonestat$.
} 
\bes
  \Phiqcd =Y_\mrm{R}^\mrm{QCD}(\theta,\mr,L) &\equiv&\lim_{a\to0} 
  \za(g_0)\,{\fa(L/2,\theta,\mr) \over \sqrt{\fone(\theta,\mr)}}  \,,
\ees
where for our one-loop discussion we do not need to specify
the normalization condition for the renormalized
heavy quark mass $\mbar=\mr$ and coupling $\gbar=\gr$.
The one-loop expansion of these functions has been
computed~\cite{zastat:pap2}, and the result can be summarized 
as
\bes
  \Phi=Y_\mrm{R}^\mrm{stat}(\theta,\mu,L) &=&\lim_{a\to0} 
  \zastat(g_0,\mu a)\,{\fastat(L/2,\theta) \over \sqrt{\fonestat(\theta)}}
  \\ &=& A(\theta)[1-\gamma_0\ln(\mu L)\,\gr^2] 
  +D(\theta) \gr^2 + \rmO(\gr^4)   
  \nonumber
\ees
in static approximation and
\bes
  \Phiqcd  &=& 
   A(\theta)[1+(D'-\gamma_0\ln(\mr L))\gr^2] + D(\theta) \gr^2 
   + \rmO(1/(\mr L)) +
   \rmO(\gr^4) \,
  \nonumber 
\ees
in QCD. 
From these expressions we can read off
\bes
  \label{e:c1}
  c_1(\mr/\mu) = \gamma_0\,\ln(\mu/\mr) + D'\,.
\ees
Furthermore, the fact that the same functions
$A(\theta),\,D(\theta)$ appear in the static theory and in
QCD is a (partial) confirmation that the static approximation is 
the effective theory for QCD. In particular the logarithmic
L-dependence in QCD matches the one
in the static theory. With \eq{e:c1},
the matching of QCD and static theory holds for all $\theta$,
and also for other matrix elements of $A_0$.

\subsection{Renormalization group invariants}\label{s:RGI}
Having seen how QCD and effective theory match at 
one-loop order, we now proceed to a general discussion of 
\eq{e:match1}, beyond one-loop.
Obviously, the $\mu$-dependence in \eq{e:match1} is artificial,
since we have a scale-independent
quantity in QCD. Only the mass-dependence is for real. We 
may then choose any value for $\mu$.
For convenience 
we set all renormalization scales equal to the mass itself~\footnote{
Note that $\mstar$ is implicitly defined through $\mstar = \mbar(\mstar)$.},
\bes
  \mu = \mstar &=& \mbar(\mstar)\,, \;  \gstar=\gbar(\mstar)\,, 
  \label{e:mstar}
\ees
where $\mbar(\mu),\gbar(\mu)$ are running mass and coupling in an 
unspecified massless
renormalization scheme.\footnote{In a massless renormalization
scheme, the renormalization factors do not depend on the masses.
Consequently the renormalization group functions do not depend
on the masses.} 
This simplifies the matching function to
\bes
  \widetilde C_\mrm{match}(\mstar,\mstar) &=&
  C_\mrm{match}(\gstar)=1 + c_1(1)\,\gstar^2+\ldots\,.
\ees
Further we want to eliminate the dependence on the 
renormalization scheme for $\mbar,\gbar,\Aren$. 
As a first step we change
from $\Phi(\mu)$ 
to the RGI matrix element 
\bes \label{e:phirgi}
   \PhiRGI &=& % \varphi_{\Astat}(\gbar(\mu))\, \Phi(\mu) =
    \exp \left\{-\int^{\gbar(\mu)} \rmd x\,
          \bfrac{\gamma(x)}{ \beta(x)}        \right\} \,\Phi(\mu)\,,
\ees
and arrive at the form
\bes 
   \Phiqcd &=& C_\mrm{match}(\gstar) \times\Phi(\mu) 
   = C_\mrm{match}(\gstar)  
       \exp \left\{\int^{\gstar} \rmd x \bfrac{\gamma(x)}{\beta(x)}\right\} 
       \PhiRGI 
   \\
   &\equiv&  \exp \left\{\int^{\gstar} \rmd x\, 
        \bfrac{\gamma_\mrm{match}(x)}{\beta(x)}\right\} 
        \PhiRGI \,.
   \label{e:defgammamatch}
\ees
Everywhere terms of order $1/m$ are dropped, 
since we are working to static order. \Eq{e:defgammamatch}
defines $\gamma_\mrm{match}$, which describes the physical mass dependence
via,
\bes 
  {\mstar \over  \Phiqcd} {\partial \Phiqcd \over \partial \mstar} 
  &=& \gamma_\mrm{match}(\gstar)  \label{e_RG_Phi}\,,
\ees
but it still depends on the chosen renormalization scheme
through the choice of $\mbar$ (the scheme, not the scale). 
We eliminate also this scheme dependence by switching to the RGI mass, $M$,
and the $\Lambda$-parameter,
\bes \label{e:lammu}
   {\Lambda \over \mu} &=& % \varphi_g(\gbar) = 
   \exp \left\{-\int^{\gbar(\mu)} \rmd x\;
          \bfrac{1}{ \beta(x)}          \right\} \,,
   \\ \label{e:mmbar}
   {M \over \mbar(\mu)} &=& % \varphi_m(\gbar) = 
   \exp \left\{-\int^{\gbar(\mu)} \rmd x\;
          \bfrac{\tau(x)}{ \beta(x)}          \right\} \,.
\ees
Exact expressions, defining the constant parts in these
equation, are given in the appendix.

Just based on dimensional analysis, we expect a relation
\bes
     \Phiqcd &=& C_\mrm{PS}(M/\Lambda) \times\PhiRGI \,  
\ees
to hold.
Indeed, remembering \eq{e:mstar}, $\mu=\mstar=\mbar$, we can
combine \eq{e:lammu} and \eq{e:mmbar} to 
\bes
{\Lambda \over M} &=&  \exp \left\{-\int^{\gstar(M/\Lambda)} \rmd x\;
          \bfrac{1-\tau(x)}{ \beta(x)}          \right\} \,,  
  \label{e:gstarM} 
\ees
from which $\gstar$ can be determined for any value of 
$M/\Lambda$; we write
$\gstar = \gstar(M/\Lambda)$.
It follows that 
\bes
   M {\partial \gstar(\mstar(M/\Lambda)) \over  \partial M} 
  = {\beta(\gstar) \over 1-\tau(\gstar)} \,,
\ees
and the matching function is 
\bes
  \label{e:cpsrgi}
  C_\mrm{PS}(M/\Lambda) = \exp \left\{\int^{\gstar(M/\Lambda)} \rmd x\; 
        \bfrac{\gamma_\mrm{match}(x)}{\beta(x)}\right\} \,.
\ees
We note that the dependence on $M$ is described by a function\footnote{
This is seen from
\bes
  {M \over  \Phi} {\partial \Phi \over \partial
    M}
  = \underbrace{{M \over  \mstar}{\partial \mstar \over  \partial M}
         }_{1\over 1-\tau(\gstar)} 
         \underbrace{{\mstar \over  \Phi}{\partial \Phi \over \partial \mstar}
         }_{\gamma_\mrm{match}(\gstar) }  
   =   { \gamma_\mrm{match}(\gstar)\over 1-\tau(\gstar)} \,,
\ees
where we used
\bes
  \mstar &=& M\; \exp \left\{\int^{\gstar} \rmd x
          \bfrac{\tau(x)}{ \beta(x)}          \right\} \\
  {\partial \mstar \over  \partial M} &=& 
   {\mstar \over  M} + \bfrac{\tau(\gstar)}{ \beta(\gstar)}
   {\partial \gstar \over  \partial M} 
          \mstar  
   = {\mstar \over  M} + \bfrac{\tau(\gstar)}{ \beta(\gstar)} 
       \,\beta(\gstar) {\partial \mstar \over  \partial M}\,,
\ees
which shows that
\bes
   {M \over  \mstar}{\partial \mstar \over  \partial M} = {1\over
     1-\tau(\gstar)} \,.
\ees
}
\bes
   \left.{M \over  \Phi} {\partial \Phi \over \partial M}\right|_\Lambda = 
   \left.{M \over C_\mrm{PS} } {\partial C_\mrm{PS} \over \partial M}
   \right|_\Lambda
   = { \gamma_\mrm{match}(\gstar)\over 1-\tau(\gstar)}\,,
          \quad \gstar=\gstar(M/\Lambda)\,.
   \label{e:Mdep}
\ees
With
\bes
  \gamma_\mrm{match}(\gstar) \simas{\gstar\to0} -\gamma_0\gstar^2 - 
   \gamma_1^\mrm{match}\gstar^4 + \ldots\,,\qquad
  \beta(\gbar) \simas{\gbar\to0} -b_0\gbar^3 + \ldots
\ees
we can now give  
the leading large mass behavior 
\bes
  \Cps &\simas{M\to\infty}& (2b_0\gstar^2)^{-\gamma_0/2b_0} \sim 
[\ln(M/\Lambda)]^{\gamma_0/2b_0}\,.
\ees
Functions such as $\Cps$ convert from the static RGI matrix elements to 
the QCD matrix element; we call them conversion functions.

An interesting application is the asymptotics of the decay constant
of a heavy-light pseudo-scalar (e.g. B):\footnote{Note the slow, logarithmic, 
decrease of the corrections in \eq{e:f_asymptotoics}. We will see below,
in the discussion of Figs. \ref{f:cps},\ref{f:ratios}, that the perturbative
evaluation of $\Cps(\Mbeauty/\Lambda)$ is somewhat problematic.}
\bes
  \label{e:f_asymptotoics}
  F_\mrm{PS} &\simas{M\to\infty}& 
  { [\ln(M/\Lambda)]^{\gamma_0/2b_0} \over \sqrt{m_\mrm{PS}}} \Phirgi\
  \times[1+\rmO([\ln(M/\Lambda)]^{-1}]\,.
\ees
At leading order in $\minv$ the conversion function
$\Cps$ contains the full (logarithmic) mass-dependence.
The non-perturbative effective theory matrix elements, $\PhiRGI$,
are mass independent numbers.
Conversion functions such as $\Cps$ are universal for all
(low energy) matrix elements of their associated operator. For example
\bea
    \caar(x_0)
                    &\simas{x_0 \gg 1/m }&
                    [\Cps(\frac{M}{\Lambda_\msbar})\,\zastatRGI]^2
                    \underbrace{\langle \Astat(x)^\dagger \Astat(0)\rangle}_{
                    \caastat(x_0)\;\;\mbox{(bare)}}
                    +\rmO(\frac1m)\,,
\eea
%%%%%%%%%%%%%%%%%%%%%%%%%%%%%%%%%%%%%%%%%%%%%%%%%%%%%%%%%%%%%%%%%%%%%%%%%%%%%%%%%%%%%%%%%%
\begin{figure}[tb]
\vspace{0pt}
\centerline{\epsfig{file=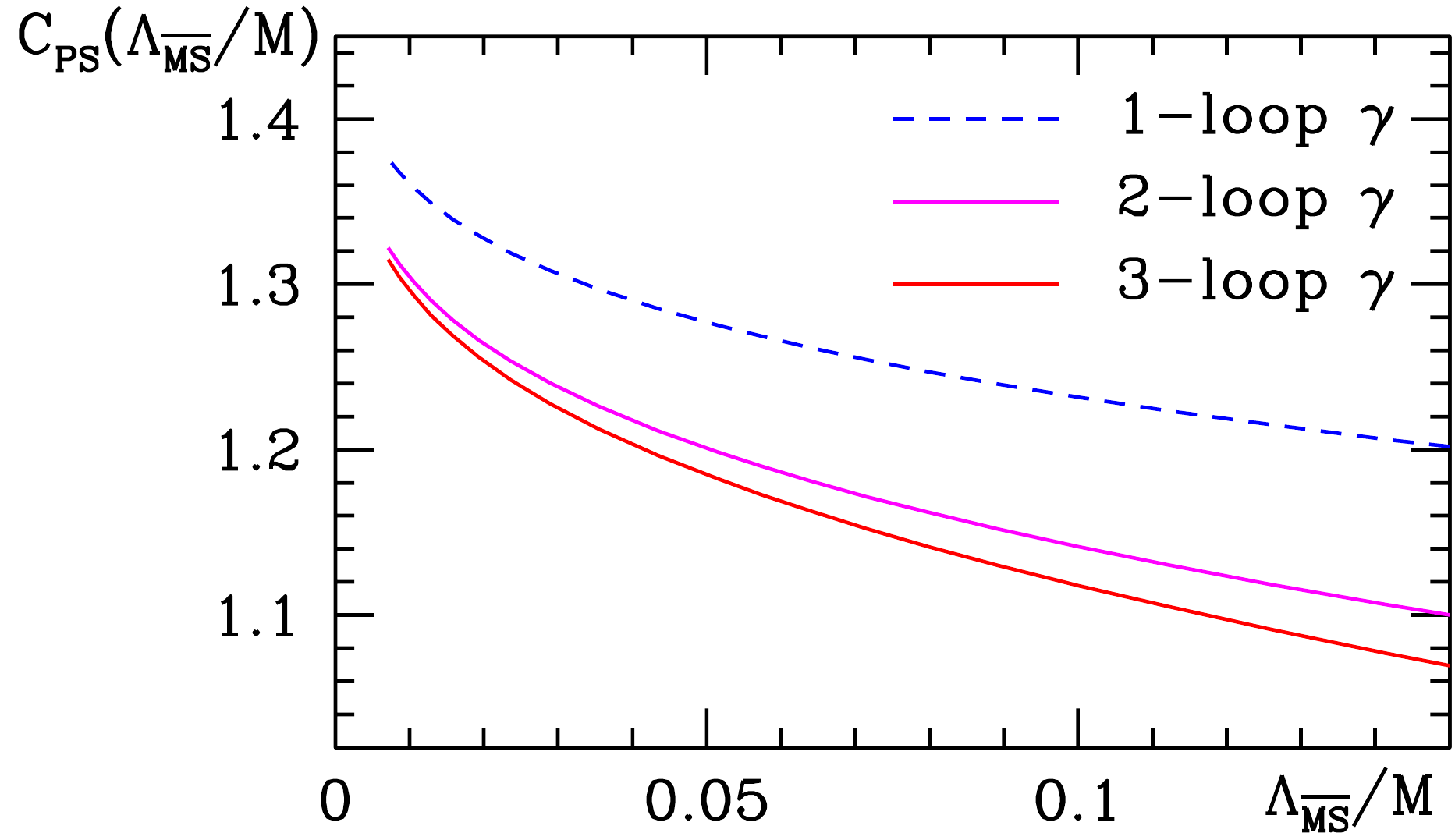,
       width=7.5cm} }
  \caption{\footnotesize $\Cps$ estimated in perturbation
        theory. For B-physics
        we have $\Lambda_\msbar/\Mbeauty\approx0.04$. Figure from \protect\cite{hqet:pap3}.
        }\label{f:cps}
\end{figure}
%%%%%%%%%%%%%%%%%%%%%%%%%%%%%%%%%%%%%%%%%%%%%%%%%%%%%%%%%%%%%%%%%%%%%%%%%%%%%%%%%%%%%%%%%%
is a straight forward generalization of \eq{e:match1}.

Analogous expressions for the conversion functions
are valid for the time component of the
axial current replaced by other composite fields, for example
the space components of the vector current.
Based on the work of \cite{BroadhGrozin,Shifman:1987sm,Politzer:1988wp}
and recent efforts
their perturbative expansion is known including the 3-loop anomalous dimension
$\gamma_\mrm{match}$ obtained from the
3-loop anomalous dimension  $\gamma$~\cite{ChetGrozin}
in the $\msbar$-scheme
and the 2-loop matching function $C_\mrm{match}$
\cite{Ji:1991pr,BroadhGrozin2,Gimenez:1992bf}.

\Fig{f:cps}
seems to indicate that the remaining $\rmO(\gbar^6(\mbeauty))$ errors in $\Cps$
are relatively small. However, as discussed in more detail in \app{s:conv},
such a conclusion is premature. By now  {\em ratios} of
conversion functions for different currents are known to even one
more order in perturbation theory~\cite{hqet:match3lp}.  We show an example
in the first column of \fig{f:ratios},
where the x-axis is approximately proportional to
$\gstar^2(M/\Lambda)$ and for B-physics one needs
$1/\ln(\Lambda_\msbar/\Mbeauty) \approx 0.3$.
For a quark mass around the mass of the b-quark and lower, 
the higher order contributions in perturbation theory do not 
decrease significantly and perturbation theory is not trustworthy. It 
seems impossible to estimate a realistic error of the 
perturbative expansion. Only for somewhat higher masses the 
expansion looks reasonable. 

Moreover, using the freedom to choose
the scale $\mu$ in \eq{e:c1_m_mu}, the $l$'th order coefficients 
(as far as they are known) can be brought down in magnitude below about $(4\pi)^{-l}$,
which means there is a fast decrease of terms in the perturbative series
once $\alpha(\mu)\lesssim 1/3$. This is shown in columns two and three 
of the figure.
Unfortunately, the required scale $\mu$ is around a factor 4 or more below 
the mass of the quark. For the b-quark, $\alpha$ is rather large 
at that scale and  the series is again unreliable. 
Only for even larger masses, say $\mstar>15\,\GeV$, the asymptotic convergence
of the series is noticeably better after adjusting the scale. 
More details are found in \app{s:conv}.
Unfortunately we see no way out of the conclusion that for B-physics 
with a trustworthy error budget aiming at the few percent level,
one needs a non-perturbative matching, {\em even in the 
static approximation}.

%%%%%%%%%%%%%%%%%%%%%%%%%%%%%%%%%%%%%%%%%%%%%%%%%%%%%%%%%%%%%%%%%%%%%%%%%%%%%%%%%%%%%%%%%%
\begin{figure}[tb]
\vspace{0pt}
\centerline{\epsfig{file=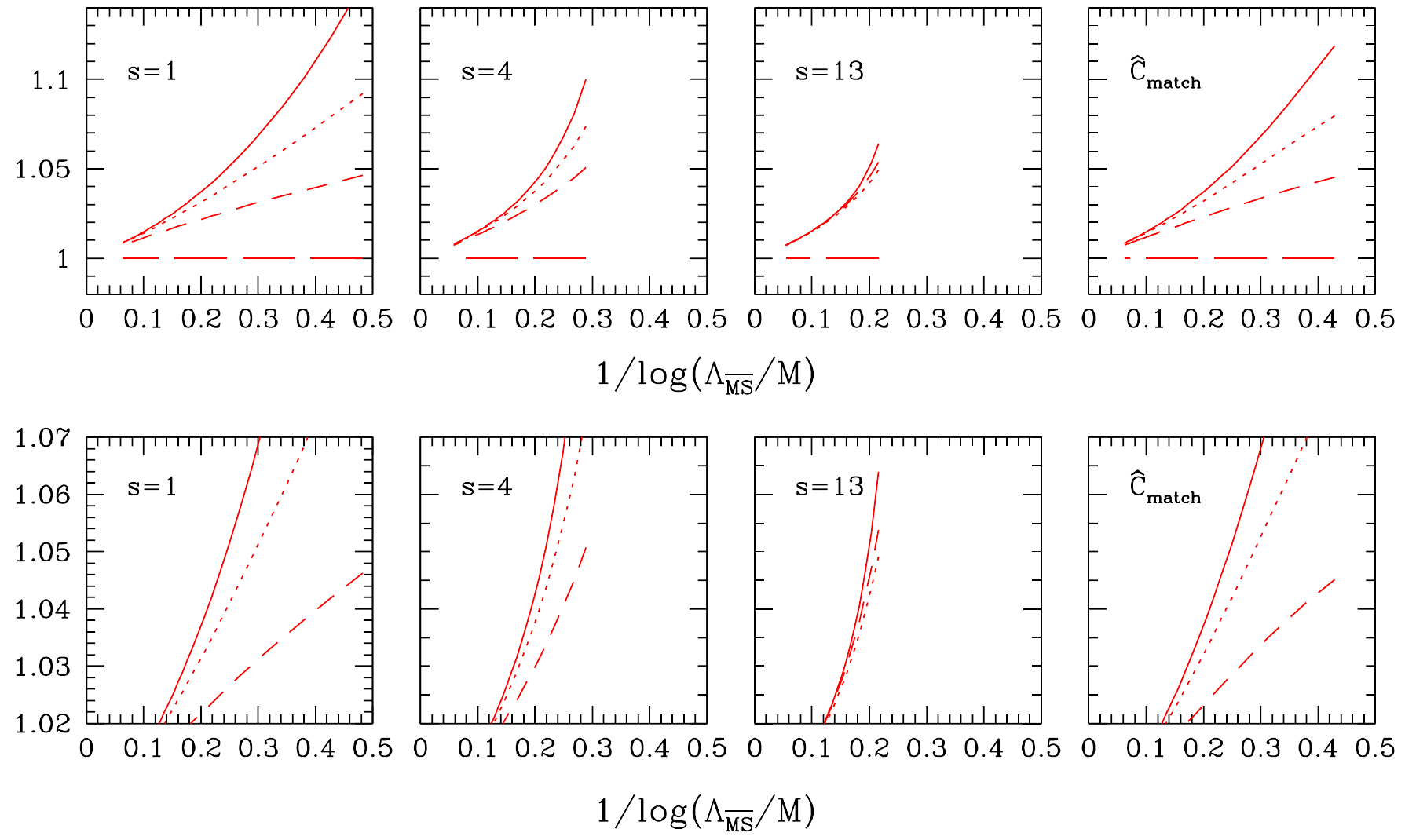,
       width=1.0\textwidth} }
  \caption{\footnotesize The ratio $\Cps/\Cv$, evaluated in the first column
        as described here. In columns two and three the expansion in $\gstar$ is generalized
        to an expansion in $\gbar(\mstar/s)$, see \app{s:numad}. The last column contains
        the conventionally used 
        $\hat{C}_\mrm{match}^\mrm{PS}(\mpole,\mpole,\mpole)/\hat{C}_\mrm{match}^\mrm{V}(\mpole,\mpole,\mpole)$, 
        see \app{s:conv}. For B-physics
        we have $\Lambda_\msbar/\Mbeauty\approx0.04$ and 
        $1/\ln(\Lambda_\msbar/\Mbeauty) \approx 0.3$. 
        The loop order changes from one-loop  
        (long-dashes) up to 4-loop (full line) anomalous dimension.
        }\label{f:ratios}
\end{figure}
%%%%%%%%%%%%%%%%%%%%%%%%%%%%%%%%%%%%%%%%%%%%%%%%%%%%%%%%%%%%%%%%%%%%%%%%%%%%%%%%%%%%%%%%%
We return to the full set of heavy-light flavor currents of \sect{s:statfields}.
The bare fields satisfy the symmetry relations \eq{e:ddvk}. The same
is then true for the RGI fields
in static approximation.  It follows that {\em in static
approximation} the effective currents are given by
\bes
  \Ahqet &=& C_{\rm PS}(\Mbeauty/\Lambda_\msbar)\, \zastatrgi(g_0)\,\Astat\,, \\
  \Vkhqet &=& C_{\rm V}(\Mbeauty/\Lambda_\msbar)\, \zastatrgi(g_0)\,\,
         \Vkstat\,,\\
  \Vhqet &=& C_{\rm PS}(\Mbeauty/\Lambda_\msbar)\, \zastatrgi(g_0)\,Z^\mrm{stat}_{\rm V/A}(g_0)\,
         \Vstat\,,\\
  \Akhqet &=& C_{\rm V}(\Mbeauty/\Lambda_\msbar)\, \zastatrgi(g_0)Z^\mrm{stat}_{\rm V/A}(g_0)
         \,\Akstat\,.
\ees
The factor $\zastatrgi(g_0)$ is known as discussed in the 
previous lecture.
Note that $\zastatrgi(g_0)$ 
is common to all (components of the) currents. Due to the 
HQET symmetries, there is one single anomalous dimension.
A dependence on the different fields comes in only through matching,
i.e. through the QCD matrix elements. In the above equations, chiral
symmetry (of the continuum theory),   \eq{e_var_quarks},
has been used to relate 
conversion functions of axial and vector currents.

\mynotes{We finally note that the bare pseudo scalar density, $P$, 
and the scalar density, $S$,
are  identical to $-\Astat$ and $\Vstat$, respectively. 
This then also holds for
the RGI fields in HQET; they are given by  
\bes
 \Prgi=-\zastatrgi(g_0) \Astat \,,\quad
 \Scalarrgi=\zastatrgi(g_0)\,Z^\mrm{stat}_{\rm V/A}(g_0)\Vstat
\ees

For the renormalized $\Phqet$ and $\Scalarhqet$, there is in principle an 
arbitrariness, since in QCD they depend on a scale and a scheme.
This arbitrariness is fixed by considering the RGI fields
in QCD. We leave it for an exercise to show
that
\bes
  \Phqet &=& -C_{\rm PS}(\Mbeauty/\Lambda_\msbar)\, {\mb \over \Mbeauty}\,
  \zastatrgi(g_0) \Astat\, = P^\mrm{QCD}_\mrm{RGI} +\rmO(1/m) \\
  \Scalarhqet &=& C_{\rm V}(\Mbeauty/\Lambda_\msbar)\, {\mb \over \Mbeauty}\, 
  Z^\mrm{stat}_{\rm V/A}(g_0)
                  \,\zastatrgi(g_0)\,\Vstat\, =\, S^\mrm{QCD}_\mrm{RGI}
  + \rmO(1/m)
\ees

}
\exercis{ Pseudo-scalar and Scalar densities}{
Start from the PCAC, PCVC relations in QCD
\bes
  \partial_\mu (\ar)_\mu 
  &=& (\mbar_\beauty(\mu)+\mbar_\mrm{l}(\mu)) P_\mrm{R}(\mu) \,, \\
  \partial_\mu (\vr)_\mu 
  &=& (\mbar_\beauty(\mu)-\mbar_\mrm{l}(\mu)) S_\mrm{R}(\mu) \,.  
\ees
Replace all quantities by their RGI's. Take the matrix elements between
vacuum and a suitable B-meson state to show that
\bes
  \Phqet &=& -C_{\rm PS}(\Mbeauty/\Lambda_\msbar)\, 
              {\mb \over \Mbeauty}\, \zastatrgi(g_0) \Astat\,, \\
  \Scalarhqet &=& C_{\rm V}(\Mbeauty/\Lambda_\msbar)\, 
              {\mb \over \Mbeauty}\, Z^\mrm{stat}_{\rm V/A}(g_0)
                  \,\zastatrgi(g_0)\,\Vstat\,,
\ees
is valid up to terms of order $\minv$. What happens if you choose 
a different matrix element?
}

\subsection{Applications}
As an application, we can now modify the scaling
law for the decay constant to include renormalization and matching
effects
\bes \label{e:fbscal}
   &&{\Fb \sqrt{\mB} \over C_{\rm PS}(\Mbeauty/\Lambda_\msbar)} 
   =  \Phirgi + \rmO(1/m) \\
   && \quad {\fb \over \fd} \approx 
  {\sqrt{\md} \; C_{\rm PS}(\Mbeauty/\Lambda_\msbar)\over 
   \sqrt{\mb} \; C_{\rm PS}(\Mcharm/\Lambda_\msbar)}\,, 
\ees
where the latter equation is maybe stretching
the applicability domain of HQET.
 
\begin{figure}[tb]
\vspace{0pt}
\centerline{\epsfig{file=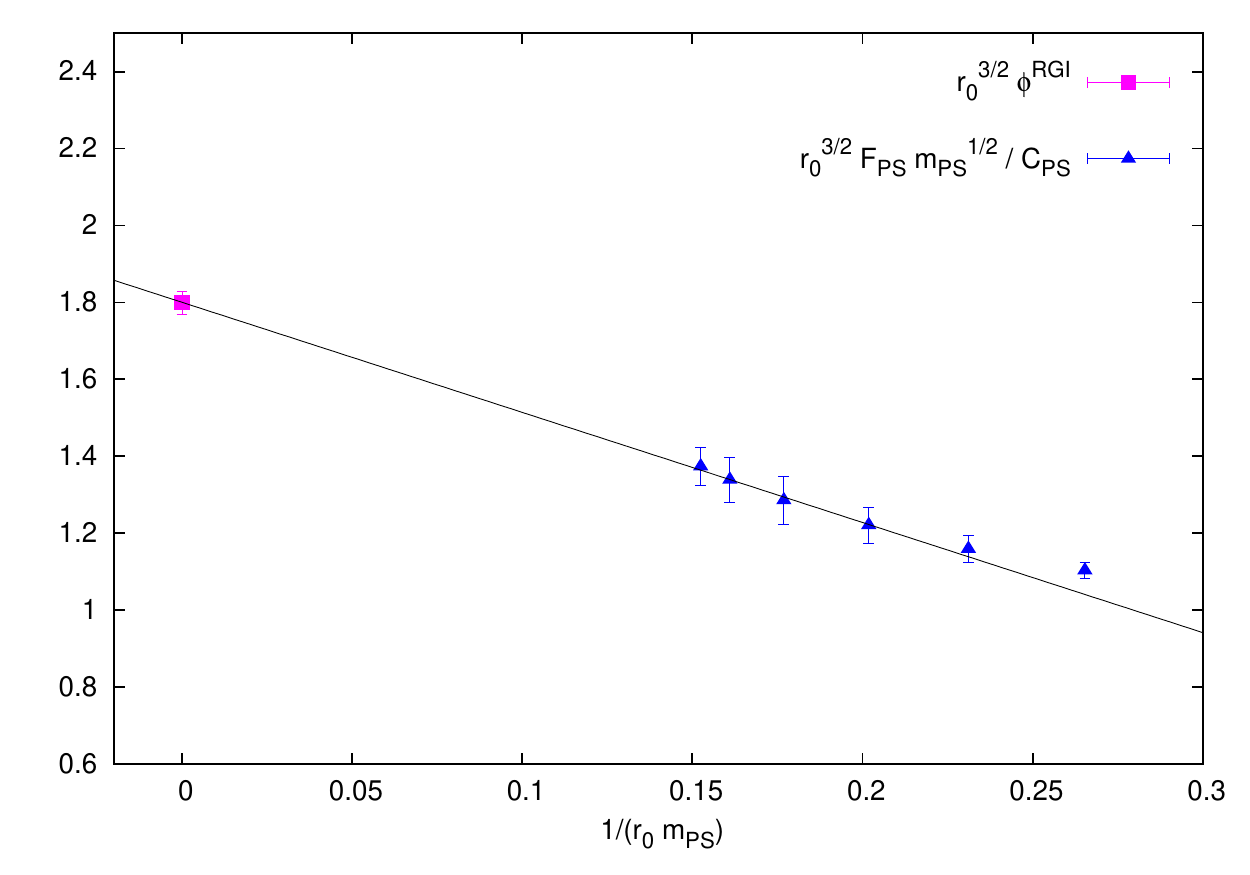,width=0.68\textwidth} }
  \caption{\footnotesize Example of an interpolation between a
        static result and results with $m_\mrm{h} < \mbeauty$.
        The function $\Cps$ is estimated at three-loop order.
        Continuum extrapolations are done before the
        interpolation \protect\cite{hqet:first3}.
        The point at $1/r_0\mp = 0$ is given by $r_0^{3/2}\Phi_\mrm{RGI}$.
        This quenched computation is done for validating and demonstrating
        the applicability of HQET.
        }\label{f:fbinterpol}
\end{figure}

Despite the discussion above, let us assume 
that the conversion functions $C$ are known with
reasonably small errors from perturbation theory.  
In this case, the knowledge of the leading term in expansions
such as \eq{e:fbscal}
is very useful to constrain the large mass behavior
of QCD observables, computed on the lattice with unphysical
quark masses $m_\mrm{h} < \mbeauty$, typically
$m_\mrm{h} \approx m_\mrm{charm}$. (Such a calculation is done
with a relativistic (Wilson, tmQCD, \ldots) formulation,
extrapolating $am_\mrm{h}\to 0$ at fixed $m_\mrm{h}$.)
As illustrated in \fig{f:fbinterpol},
one can then, with a reasonable smoothness assumption,
interpolate to the physical point. 

Given the unclear precision of the perturbative predictions,
the above interpolation method has to be taken with care. 
The inherent perturbative error remains to
be estimated.

%%%%%%%%%%%%%%%%%%%%%%%%%%%%%%%%%%%%%%%%%%%%%%%%%%%%%%%%%%%%%%%%%%%%%%%%%%%%%%%%%%%%%%%%%%
%%%%%%%%%%%%%%%%%%%%%%%%%%%%%%%%%%%%%%%%%%%%%%%%%%%%%%%%%%%%%%%%%%%%%%%%%%%%%%%%%%%%%%%%%%

The relation
between the RGI fields and the bare fields has also  been obtained
for the two parity violating $\Delta B=2$ four fermion operators
        \cite{stat:zbb_pert,stat:zbb_nf0} for $\nf=0$  and $\nf=2$
        \cite{hqet:4ferm_nf2}.
Their matrix elements,
evaluated in twisted mass QCD will give the standard model B-parameter
for B-$\rm\overline{B}$ mixing.

We now turn to the natural question whether one can directly compute the 
$1/m$ corrections in HQET, which will lead us again to the necessity of
performing a non-perturbative matching between HQET and QCD.

\exercis{Anomalous dimension $\gamma_\mrm{match}$}{
Show that
\bes \label{e:gammamatch}
   \gamma_\mrm{match} = -  \gamma_0\gstar^2 - [\gamma_1 + 2b_0 c_1(1)]\gstar^4
   + \ldots \,.
\ees
where $c_1$ is the 1-loop matching coefficient in the same scheme
as $\gamma_1$.
}

\chapter{Renormalization and matching at order 1/m}  
\label{s:renorm}

\section{Including $\minv$ corrections} \label{s:first}

We here work directly in lattice regularization. 
The continuum formulae are completely analogous. 
The expressions for $\Okin,\Ospin$ 
are discretized in a straight forward way, 
\bes
D_kD_k \to \nabstar{k}\nab{k}\,, \quad
F_{kl} \to \widehat{F}_{kl} 
\ees
with the latter given by the clover leaf representation,
defined e.g. in \cite{impr:pap1}. Of course other discretizations of these
composite fields are possible.

Apart from the terms in  the classical Lagrangian,
renormalization can in principle introduce new local fields  
compatible with the symmetries (but not necessarily the
heavy quark symmetries which are broken by $\Ospin,\Okin$)
and with dimension  $d_\mrm{op}\leq5$. Also the field equations
can be used to eliminate terms. With these rules one finds 
that no new terms
are needed and it suffices to treat the coefficients of
$\Ospin,\Okin$ as free parameters which depend on the bare coupling of the 
theory and on $m$.

The $\minv$ Lagrangian then reads
\be
  \Lh^{(1)}(x) = -(\omegakin\,\Okin(x) + \omegaspin\,\Ospin(x)) \,.
\ee
Since these terms are composite fields of dimension five, the theory defined
with a path integral weight ($\lag{light}$ collects all contributions of
QCD with the heavy quark(s) dropped)
\bes
  W_\mrm{NRQCD} &\propto&
  \exp(-a^4\sum_x[\lag{light}(x)+\Lh^\mrm{stat}(x)+\Lh^{(1)}(x)])
\ees
is {\em not} renormalizable. In perturbation theory,
new divergences will occur at each order in the loop expansion, 
which necessitate to introduce
new counter-terms. The continuum limit of the lattice theory will
then not exist\cite{nrqcd:first}.  Since the effective theory is ``only'' 
supposed to reproduce the $\minv$ expansion of the observables
order by order in $\minv$, 
we instead expand the weight $W$ in $\minv$, counting 
$\omegakin=\rmO(\minv)=\omegaspin$,
\bes 
   W_\mrm{NRQCD} \to W_\mrm{HQET} &\equiv & \exp(-a^4\sum_x[\lag{light}(x)+\Lh^\mrm{stat}(x)]) \left\{
   1 - a^4\sum_x \Lh^\mrm{(1)}(x)\right\} \,.
   \nonumber
\ees
This rule is part of the definition of HQET, just like the same step is part of
Symanzik's effective theory discussed by Peter Weisz.

Let us remark here on the difference to chiral perturbation theory.
In chiral perturbation theory one computes the asymptotic expansion 
in powers of $p^2$.
Each term in the expansion requires a finite number
of counter terms, since there are only a finite number of (pion) loops.
The theory is thus renormalizable order by order in the expansion.
In NRQCD and HQET one expands in $\minv$. At each order of the
expansion an arbitrary number of loops remain, coming from the gluons
and light quarks. In fact, we are even
interested in more than an arbitrary number of loops: in non-perturbative results 
in $\alpha$.

NRQCD can then only be formulated with a cutoff and results depend
on how the cutoff is introduced and on it's value. 
On the lattice, the cutoff is identified with the one present 
for the other fields, $\Lambda_\mrm{cut}\sim 1/a$. 
Instead of taking a continuum limit, one then relies on physics results
not depending on the lattice spacing (the cutoff) within  
a window \cite{nrqcd:first}
\bes
   1/m \ll a \ll \Lambda_\mrm{QCD} \,.\qquad\mbox{[in NRQCD]}
\ees
In HQET the discussion is rather simple, since the static theory is (believed to be)
renormalizable; we will come to the renormalization of the
insertion of $\Lh^\mrm{(1)}$ shortly.

Up to and including $\rmO(\minv)$, 
expectation values in HQET are therefore defined as 
\bes
  \langle \op{} \rangle &=& 
         \langle  \op{}  \rangle_\mrm{stat} 
        + \omegakin  a^4\sum_x \langle  \op{} \Okin(x) \rangle_\mrm{stat}
        + \omegaspin a^4\sum_x \langle \op{} \Ospin(x) \rangle_\mrm{stat} 
        \nonumber \\
  &\equiv&  \langle  \op{}  \rangle_\mrm{stat} 
        + \omegakin\langle  \op{}  \rangle_\mrm{kin} 
        + \omegaspin\langle  \op{}  \rangle_\mrm{spin} \,,
        \label{e:exp}
\ees
where the path integral average
\be
  \label{e:expval}
         \langle  \op{}  \rangle_\mrm{stat} = {1 \over \cal Z} \int_\mrm{fields} 
          \op{} \exp(-a^4\sum_x[\lag{light}(x)+\Lh^\mrm{stat}(x)])\, 
\ee
is taken  with respect to the lowest order action.
The integral extends over 
all fields and the normalization $\cal Z$ is
fixed by $\langle 1  \rangle_\mrm{stat} = 1$.\footnote{
A straight expansion gives e.g.
  $\omegakin  a^4\sum_x \langle  \op{}
  [ \Okin(x)- \langle \Okin(x)\rangle_\mrm{stat} \rangle_\mrm{stat}$,
  but this just corresponds to an irrelevant shift of $\Okin(x)$ 
  etc. by a constant.
} 

In order to compute matrix elements or correlation functions in the 
effective theory, we also need the effective composite fields. 
At the classical level they can again be obtained from the 
Fouldy-Wouthuysen rotation. In the quantum theory one
adds all local fields with the proper quantum numbers and 
dimensions. For example the effective axial current (time component) 
is given by
\bes
  \label{e:ahqet}
 \Ahqet(x)&=& \zahqet\,[\Astat(x)+  \sum_{i=1}^2\cah{i}\Ah{i}(x)]\,, \\
 \Ah{1}(x) &=& \lightb(x){1\over2}
            \gamma_5\gamma_i(\nabsym{i}-\lnabsym{i})\heavy(x)\,,
\\
 \Ah{2}(x) &=& -\drvsym{i}\,\Ahi\,,
\ees
where all derivatives are symmetric,
\bes
  \drvsym{i} = \frac12(\drv{i}+\drvstar{i})\,,\quad
  \lnabsym{i} = \frac12(\lnab{i}+\lnabstar{i})\,,\quad
  \nabsym{i} = \frac12(\nab{i}+\nabstar{i})\,,
\ees
and we recall $\Ahi(x)=\lightb(x)\gamma_i\gamma_5\heavy(x)$.
One arrives at these currents, writing down all dimension four operators 
with the right flavor structure and transformation under spatial lattice rotations
and parity. The equations of motion of the light and static quarks are used to
eliminate terms but 
heavy quark symmetries (spin and local flavor) can't be used 
since they are broken at order $\minv$.\footnote{An operator ${m_l \over m}\Astat $ is
included as a corresponding mass-dependence of $\zahqet$. In practice, since
${m_l \over \mbeauty} \lll 1$, and this term appears only at one-loop order, this dependence
on the light quark mass can be neglected.}

For completeness let us write down the other HQET currents:
\bes
  \label{e:vhqet}
 \Akhqet(x)&=& \zakhqet\,[\Akstat(x)+ \sum_{i=3}^6\cah{i}\Akh{i}(x)]\,, \\
% \Vhqet(x)&=& \zvhqet\,[\Vstat(x)+  \sum_{i=1}^2\cvh{;}\Vh{i}(x)]\,, \\
% \Vkhqet(x)&=& \zvkhqet\,[\Vkstat(x)+ \sum_{i=3}^6\cvh{i}\Vkh{i}(x)]\,, \\
\nonumber
 \Akh{3}(x) &=& \lightb(x){1\over2}
           \gamma_k\gamma_5\gamma_i (\nabsym{i}-\lnabsym{i})\heavy(x)\,,
\quad
 \Akh{4}(x) = \lightb(x){1\over2}
            (\nabsym{k}-\lnabsym{k})\gamma_5\heavy(x)\,,
\\
\nonumber
 \Akh{5}(x) &=& \drvsym{i}\,\left(\lightb(x)\gamma_k\gamma_5\gamma_i\heavy(x)\right)\,,
\quad 
 \Akh{6}(x) = \drvsym{k}\,\Astat\,.
%\\ \Vkh{3}(x) &=& -\lightb(x){1\over2}
%            (\lnabsym{i}-\nabsym{i})\gamma_i\gamma_k\heavy(x)\,,
%\\ \Vkh{4}(x) &=& -\drvsym{i}\,\lightb(x)\gamma_i\gamma_k\heavy(x)\,,
%\\ \Vkh{5}(x) &=& \lightb(x){1\over2}
%            (\lnabsym{k}-\nabsym{k})\heavy(x)\,,
%\\ \Vkh{6}(x) &=& \drvsym{k}\,\lightb(x)\heavy(x)\,,
\ees
The vector current components are just obtained by dropping $\gamma_5$ in these
expressions and changing $\cah{i}\to\cvh{i}$. The classical values of the coefficients are
$\cah{1}=\cah{2}=\cah{3}=\cah{5}=-{1 \over 2m}$, while $\cah{4}=\cah{6}=0$.
We note that with periodic boundary conditions in space we have
\bes
   \label{e:ah2_p0}
   a^3\sum_\vecx \Aone(x) &=& a^3\sum_\vecx  \lightb(x)
            \lnabsym{i}\gamma_i\gamma_5\heavy(x)\,, \quad 
   a^3\sum_\vecx \Ah{2}(x) = 0\,, 
\ees
which for instance may be used in the determination of the 
B decay constant.

Before entering into details of the 
renormalization, we show some examples how the $\minv$-expansion works.

\section{$\minv$-expansion of correlation functions and matrix elements}
\label{s:corr}

For now we assume that the coefficients
\bes
  \rmO(1)    \,:&& \dmstat\,,\; \zahqet\,,  \nonumber \\[-1ex]
  \label{e:counting} \\[-1ex] \nonumber
  \rmO(\minv)\,:&& \omegakin\,, \; \omegaspin\,, \; \cahqet\,, \;
\ees
are known as a function of the bare coupling $g_0$ and the quark mass $m$.
Their non-perturbative determination will be discussed later.

The rules of the $\minv$-expansion are illustrated on the example
of $\caar(x_0)$, \eq{e:caa}. One uses \eq{e:exp} and the HQET representation
of the composite field \eq{e:ahqet}. Then the expectation value is expanded
consistently in $\minv$, counting powers of $\minv$ as in \eq{e:counting}. 
At order $\minv$, terms proportional to $\omegakin \times \cahqet$ etc. 
are to be dropped.
As a last step, we have to take the energy shift between 
HQET and QCD into account.
Therefore correlation functions with a time separation $x_0$ 
obtain an extra factor $\exp(-x_0\,m)$,
where the scheme dependence of $m$ is compensated by a
corresponding one in $\dmstat$. Dropping all terms $\rmO(\minv^2)$ without 
further notice, one arrives at the expansion 
\bes
   \caa(x_0) &=& \label{e:caahqet}
           \rme^{-m x_0} (\zahqet)^2 \,\Big[
                 \caastat(x_0) + \cahqet\,\cdaa^\mrm{stat}(x_0) 
           \\ && \qquad   
           +\,\omegakin\,\caakin(x_0)+\omegaspin\,\caaspin(x_0)  
                 \Big] 
           \nonumber \\
             &\equiv& \rme^{-m x_0} (\zahqet)^2 \,\caastat(x_0) 
                 \Big[1 + \cahqet\,\rda^\mrm{stat}(x_0) 
           \\ && \qquad 
           +\,\omegakin\,\raa^\mrm{kin}(x_0)+\omegaspin\,\raa^\mrm{spin}(x_0)  
                 \Big] \nonumber
\ees
with (remember the definitions in \eq{e:exp})
\bes
   \cdaa^\mrm{stat}(x_0) &=& 
        a^3\sum_\vecx \langle 
  \Astat(x) (\Ah{1}(0))^\dagger  \rangle_\mrm{stat}
  \,+\, a^3\sum_\vecx
  \langle \Ah{1}(x) (\Astat(0))^\dagger  \rangle_\mrm{stat} \,,
        \nonumber \\
   \caakin(x_0) &=& a^3\sum_\vecx \langle  
   \Astat(x) (\Astat(0))^\dagger  \rangle_\mrm{kin} 
   %     \, =\, \,  a^7\sum_\vecx\sum_y \langle  
   % \Astat(x) (\Astat(0))^\dagger  \Okin(y)\rangle_\mrm{stat}\,, 
   \nonumber\\
   % && \mbox{ a graph} \\
   \caaspin(x_0) &=& a^3\sum_\vecx 
   \langle  \Astat(x) (\Astat(0))^\dagger  \rangle_\mrm{spin}
        \,. \nonumber
\ees
The contribution of $\Ah{2}$ vanishes due to \eq{e:ah2_p0}.
It is now a straight forward exercise to 
obtain the expansion of the B-meson mass\footnote{It follows  from the 
simple form of
the static propagator that there is no dependence on $\dmstat$ 
except for the explicitly shown energy shift $\dmstathat$.} 
\bes
    \mB &=& - \lim_{x_0\to\infty} \dzero \ln\caa(x_0) \\
        &=& \mhbare - \lim_{x_0\to\infty} \dzero
         \big[\,\ln\caastat(x_0) +   \cahqet\,\rda^\mrm{stat}(x_0) + \\
        && \qquad\quad\qquad\quad  \nonumber
           +\,\omegakin\,\raa^\mrm{kin}(x_0)+\omegaspin\,\raa^\mrm{spin}(x_0) \,\big]_{\dmstat=0}
    \\
        &=& \mhbare + \Estat+ \omegakin \Ekin + \omegaspin \Espin \label{e:mBexp}
     \,,\\
    \Estat&=&  \left.- \lim_{x_0\to\infty} \dzero\,\ln\caastat(x_0)
               \right|_{\dmstat=0}\,, 
    \label{e:estat}\\
    \Ekin &=& - \lim_{x_0\to\infty} \dzero\, \raa^\mrm{kin}(x_0)\,,\quad
    \Espin = - \lim_{x_0\to\infty} \dzero\, \raa^\mrm{spin}(x_0)\,.
\ees
Again we have made the dependence on $\dmstat$ explicit through 
$\mhbare = \mbeauty+\dmstathat$
and then quantities in the theory with $\dmstat=0$ appear. Note that the ratios
$\raa^{x}$ (and therefore $\Ekin,\Espin$) do not
depend on $\dmstat$; the quantities $\Ekin,\Espin$ have mass dimension two 
and we have already anticipated  \eq{e:eda}.

The expansion for the decay constant is
\bes 
 \fb\sqrt{\mB}&=& \lim_{x_0\to\infty} \big\{2\exp(\mB x_0)\,\caa(x_0)\big\}^{1/2} 
  \\
  &=& \zahqet\, \Fhatstat \, 
       \lim_{x_0\to\infty} \big\{1 + \frac12 x_0 \big[ \omegakin \Ekin + \omegaspin\Espin\big]
  \nonumber \\
  && + \frac12 \cahqet \rda^\mrm{stat}(x_0) 
                          + \frac12 \omegakin \raa^\mrm{kin}(x_0)
                          + \frac12 \omegaspin \raa^\mrm{spin}(x_0) 
        \big\}\,, \label{e:fBexp}
  \\ 
         \Fhatstat &=& \lim_{x_0\to\infty} 
	               \big\{2\exp(\Estat x_0)\,\caastat(x_0)\big\}^{1/2}
	\,.
	\nonumber
\ees
Using the transfer matrix formalism 
(with normalization $\langle B | B \rangle = 2L^3$), 
one further observes that (do it as an exercise)
\bes
 \label{e:ekin}
    \Ekin &=& - {1\over 2L^3}\langle B | a^3\sum_{\vecz} \Okin(0,\vecz)
      | B \rangle_\mrm{stat}   
     = - {1\over 2}\langle B | \Okin(0)| B \rangle_\mrm{stat}
   \\ \label{e:espin}
    \Espin &=& - {1\over 2}\langle B | \Ospin(0)| B \rangle_\mrm{stat} \,, \\
     0 &=&  \lim_{x_0\to\infty} \dzero \rda^\mrm{stat}(x_0)\,.
     \label{e:eda}
\ees
As expected, only the parameters of the action are relevant
in the expansion of hadron masses. 

A correct split of the terms in \eq{e:mBexp} and  \eq{e:fBexp} into leading 
order and next to leading order pieces which are separately renormalized and
which hence {\em separately have a continuum limit} requires
more thought on the renormalization of the $\minv$-expansion. We 
turn to this now. 

\section{Renormalization beyond leading order} \label{s:renorm1}

For illustration we check the self consistency of \eq{e:caahqet}. 
The relevant question concerns
renormalization: are the ``free'' parameters 
$\dmstat \ldots \cahqet$ sufficient to absorb
all divergences on the r.h.s.? We consider the term 
$\propto\caakin(x_0)$ since its renormalization displays all subtleties.
As a first step we rewrite $\omegakin \Okin=\frac1{2\mr}\Okinr$ in terms
of a renormalized mass and the renormalized operator 
\bes \label{e:okinr}
  \big(\Okin\big)_\mrm{R}(z) &=& Z_{\Okin} \big( \Okin(z) + {c_1\over a}\, \heavyb(z) D_0 \heavy(z) + 
  {c_2\over a^2}\, \heavyb(z)\heavy(z) \big)\,. 
\ees
The latter involves a subtraction of lower dimensional ones 
with dimensionless coefficients $c_i(g_0)$.
The renormalization scheme for $\mr$ is irrelevant,  as any change of
scheme can be compensated by  $Z_{\Okin},c_i$ whose finite parts need
to be fixed by matching to QCD. We further expand 
\bes
  \label{e:expandzahqet}
  (\zahqet)^2  &=& (\zastat)^2 + 2 \zastat \za^{\first} +\rmO(\minv^2)\,
\ees
which we will discuss more below. With these rules
we then have 
\bes \label{e:caakinr}
   \big(\zastat\big)^2\,\omegakin \caakin(x_0) = \frac1{2\mr} a^7 \sum_{\vecx,\, z} G(x,z)
    + \text{ subtraction terms}\,,
\ees
where 
\bes
\label{e:3ptfct} G(x,z) = 
 \Big\langle 
  [\Astat]_\mrm{R}(x)\, ([\Astat]_\mrm{R}(0))^\dagger  \,\big(\Okin\big)_\mrm{R}(z)\Big\rangle_\mrm{stat}\,.
\ees
The subtraction terms are due to the lower dimensional operators with coefficients
$c_1$ and $c_2$.  
Since we are interested in on-shell observables
($x_0>0$ in \eq{e:caahqet}), 
we may use the equation of motion
$D_0 \heavy(z)=0$ to see that the $c_1$-term does not contribute, 
while  ${c_2\over a^2} \heavyb(z)\heavy(z)$, is equivalent
to a mass shift. In the full correlation function \eq{e:caahqet} it 
hence contributes to $\dmstat$ which becomes quadratically
divergent when the $\minv$ terms are included. 

While $G(x,z)$ is a renormalized correlation function 
for all physical separations, its integral over $z$ (or on the lattice the continuum
limit of the sum over $z$)
does not exist due to singularities  
at $z\to 0$ and as $z\to x$. These contact term singularities can be analyzed by the 
operator product expansion. 
We discuss them first in the continuum and regulate the short distance region by
just integrating for $z^2\geq r^2$ with some small $r$. The operator product expansion then yields
\bes
    &&\int_{z^2 \geq r^2}\rmd^4 z\,G(x,z) \\
    &&\simas{r\to0} 
    \Big\langle   [\Astat]_\mrm{R}(x)\,   
    [d''_1{1\over r} \;(\Astat(0))^\dagger + d''_2 (\Ah{1}(0))^\dagger + d''_3 (\Ah{2}(0))^\dagger ]
    \Big\rangle_\mrm{stat}  
    \nonumber 
\ees
up to terms which are finite as $r\to0$. 
The coefficients $d''_i$ in the operator product expansion 
have a further logarithmic dependence on $r$.\footnote{
We have written down the integrated version, since then a smaller number of operators
can appear and we are ultimately interested in the integral.}
For (the continuum version of) \eq{e:3ptfct} we need $r\to0$. In this limit
short distance divergences
emerge which have to be subtracted by counter-terms. In the lattice
    regularization, 
short distance singularities are regulated 
by the lattice spacing $a$ and we have in full analogy
\bes
    \label{e:contterms}
    &&\,\Big\langle 
    [\Astat]_\mrm{R}(x)\, 
    \big[a^4  \sum_{z}\,([\Astat]_\mrm{R}(0))^\dagger  \,\big(\Okin\big)_\mrm{R}(z)\big]
    \Big\rangle_\mrm{stat}  \\
    &&\simas{a\to0} 
    \Big\langle   [\Astat]_\mrm{R}(x)\,   
    [d'_1{1\over a} \;(\Astat(0))^\dagger + d'_2 (\Ah{1}(0))^\dagger + d'_3 (\Ah{2}(0))^\dagger ]
    \Big\rangle_\mrm{stat}  
    \nonumber 
\ees
up to terms which have a continuum limit $a \to 0$ and up to the singular terms originating
from $z\approx x$ . The coefficients $d_i$ contain a logarithmic dependence on $a$. 
Treating the singular terms at $z\approx x$ in the same way and noting that the term with 
$\Ah{2}(0)$ vanishes upon summation over $\vecx$ we find 
\bes
  \zastat\, [\, d_1{1\over a} \,\caastat(x_0) + d_2\,\cda^\mrm{stat}(x_0) \,]\,
\ees
for the contact term singularities
in \eq{e:caakinr}.
These are absorbed in \eq{e:caahqet} through counter-terms contained in
$\zahqet$ and $\cah{1}$,
\bes
  2 \za^{\first} = - {d_1 \over 2a\mr} + \ldots\,,\quad 
  \cah{1} = - {d_2 \over 2\mr\zastat} + \ldots \;.
\ees
The change from $d'_i$ to $d_i$ is due to the use of the equation of motion 
above. This step is valid only up to contact terms, resulting in the 
shift $d'\to d$.
The ellipses contain the physical, finite $\minv$ terms. 

We now comment further on the expansion  
\eq{e:expandzahqet}.
Our discussion shows that the quadratic term $(\za^{\first})^2$ in \eq{e:expandzahqet} 
{\em must be  dropped};
otherwise an uncanceled $1/(a^2m^2)$ divergence remains. As we have seen there is
no $1/(a^2m^2)$ in $\caakin(x_0)$ and the other pieces in \eq{e:caahqet}
are less singular. This is just a manifestation of the general rule
of an effective field theory that all quantities are to be expanded 
in $\minv$ whether they are divergent
or not. With this rule the various HQET parameters can be determined
such that they absorb all divergences.
\footnote{
It is convenient to avoid the multiplication of $\minv$ terms 
explicitly by a choice of observables, for example
\bes
  \nonumber
  \tilde \Phi &=& \ln(\fb\sqrt{\mB}) =
  \ln(\zahqet) + \ln(\Fhatstat) + \lim_{x_0\to\infty}
  \big\{ 
  \frac12 x_0  \omegakin \Ekin + \frac12 \omegakin \raa^\mrm{kin}(x_0) 
  +\ldots
  \big\}\,,\\
  && \ln(\zahqet) = \ln(\zastat) + {\za^\first \over \zastat} \equiv 
     \ln(\zastat) + [\ln(\za)]^\minv
  \nonumber
\ees
In this convention all $\minv$-terms appear linearly. 
}

The lesson of our discussion is that counter-terms with the correct structure are automatically
present because in
the effective theory all the relevant local composite fields are included
with free coefficients.
These free parameters may thus be chosen such that the continuum
limit of the HQET correlation functions exists. Finally, their finite parts are to
be determined such that the effective theory yields the $\minv$ expansion of the QCD
observables.

\off{
\section{The flavor currents in the effective theory\label{s:fields}}

Following our general rules 
for finding the HQET fields which represent the QCD ones we find
\bes
 \Ahqet(x)&=& \zahqet\,[\Astat(x)+ \cahqet\Ah{1}(x)]\,, \\ 
 \Vhqet(x)&=& \zvhqet\,[\Vstat(x) + \cvhqet\delta\Vstat(x)]\,,\\
  \label{e:vhqet}
 \Vkhqet(x)&=& \zvkhqet\,[\Vkstat(x) + \cvkhqet\delta\Vkstat(x)]\,,\\
 \Akhqet(x)&=& \zakhqet\,[\Akstat(x) + \cakhqet\delta\Akstat(x)]\,.
\ees
We note that in the static approximation
we have
\bes
  \zahqet &=& \Cps(M/\Lambda) \zastatrgi(g_0) \,.
\ees
}

\section{The need for non-perturbative
        conversion functions} \label{s:need}

An important step remains to be explained: 
the determination of the HQET parameters.
As discussed in \sect{s:mass} at the leading order in $\minv$, this
can be done with the help of perturbation theory for conversion functions
such as  $\Cps$. However, as soon as a $\minv$ correction is to be 
included, the leading
order conversion functions have to be known
non-perturbatively. This general feature in the
determination of power corrections in QCD is seen in the following
way. Consider
the error made in \eq{e:match1},
when the
anomalous dimension has been computed at $l$ loops and
$C_\mrm{match}$ at $l-1$ loop order. The conversion
function 
\bes 
\Cps &=& 
    \exp \left\{ -\int^{\gstar} \rmd x 
        \bfrac{\gamma_0 x^2 + \ldots + \gamma_{l-1}^\mrm{match}
        x^{2l}}{\beta(x)}\right\}\,
     + \Delta(\Cps)
\ees
is then known up to a relative {\em error}
\bes    
     {\Delta(\Cps) \over \Cps} &\propto&  [\gbar^2(m)]^{l} \sim
        \left\{{1 \over 2b_0\ln(m/\Lambda_\mrm{QCD})}\right\}^{l}
        \ggas{m\to\infty}\; {\Lambda_\mrm{QCD} \over m} \,.
    \label{e:deltacps}
\ees
As $m$ is made large, this perturbative error becomes
dominant over the power correction one wants to determine.
Taking a
perturbative conversion function and adding power corrections
to the leading order effective theory is thus
a phenomenological approach, where one assumes
that for example at the b-quark mass,
the coefficient of the $[\gbar^2(\mbeauty)]^{l}$ term
(as well as higher order ones)
is small, such that the $\Lambda/\mbeauty$ corrections dominate. 
In such a phenomenological
determination of a power correction, its size depends on the
order of perturbation theory considered.
A theoretically
consistent evaluation of power corrections requires
a fully non-perturbative formulation of the theory including
a non-perturbative matching to QCD. 
Note that the essential point of \Eq{e:deltacps} is not the 
expected factorial growth of the coefficients of the perturbative
expansion. 
Rather it is due to the truncation of perturbation theory
as such.  
Of course a renormalon-like growth of the coefficients
does not help.

The foregoing discussion is completely generic, applying to any
regularization. 
When we define the theory on the lattice, there are in addition  
power divergences, e.g. in \eq{e:contterms}. It is well known
that they have to be subtracted non-perturbatively if one wants
the continuum limit to exist.

\section{Splitting leading order (LO) and next to leading order (NLO)}

We just learned that the very definition of a NLO correction
to $\fb$ means to take \eq{e:fBexp} with all coefficients $\zahqet \ldots \cahqet$
determined non-perturbatively. We want to briefly explain that, as a 
consequence, the split
between  LO and NLO is not unique. This is
fully analogous to the case of standard perturbation theory in $\alpha$, where the
split between different orders depends on the renormalization scheme used, and
on the experimental
observable used to determine $\alpha$ in the first place.

Consider the lowest order. The only coefficient needed in \eq{e:fBexp} is
then $ \zahqet = \Cps\zastatrgi$.
It has to be fixed by matching some matrix element of $\Astat$ to the
matrix element of $A_0$ in QCD. For example one may choose $\langle B'|A_0^\dagger|0\rangle$,
with $|B'\rangle$ denoting some other state such as an excited  pseudo-scalar state.  Or one
may take a finite volume matrix element defined through the \SF as we
will do later. Since the matching involves the QCD matrix element,
there are higher order in $\minv$ ``pieces'' in these equations. There is no reason
for them to be independent of the particular matrix element. So from
matching condition to matching condition,
$\Cps\zastatrgi$ determined at the leading order
in $\minv$ differs by $\rmO(\Lambda_\mrm{QCD}/\mbeauty)$ terms.

The matrix element $\fb$ in static approximation inherits
this $\rmO(\Lambda_\mrm{QCD}/\mbeauty)$ ambiguity.
These corrections are hence not unique.
Fixing a matching condition, the leading order $\fb$ as well as the one including
the corrections can be computed and have a continuum limit.  Their difference
can be defined as the $\minv$ correction.  However, what matters is not the
ambiguous NLO term, but the fact that the uncertainty is reduced
from $\rmO(\Lambda_\mrm{QCD}/\mbeauty)$ in the LO term
to $\rmO(\Lambda^2_\mrm{QCD}/\mbeauty^2)$ in the sum.

The following table illustrates the point explicitly.\\[0.5ex]
\begin{center}
\begin{tabular}{lccccllllllll}
 \mbox{Observables} & $\langle B|A_0^\dagger|0\rangle$ 
   & $\langle B'|A_0^\dagger|0\rangle$ & $\langle B''|A_0^\dagger|0\rangle$ \\[1ex]
  \mbox{matching condition} &  * \\
  \mbox{error in HQET result} & 0 &\rmO($\Lambda/\mbeauty$)& 
                                   \rmO($\Lambda/\mbeauty$)\\[1ex]
  \mbox{matching condition} &  & * \\
  \mbox{error in HQET result} & \rmO($\Lambda/\mbeauty$)& 0 &
                                   \rmO($\Lambda/\mbeauty$)\\[1ex]
  \mbox{matching condition} &  & & * \\
  \mbox{error in HQET result} & \rmO($\Lambda/\mbeauty$)& 
                                   \rmO($\Lambda/\mbeauty$) & 0\\[1.5ex]
\end{tabular}
\end{center}
As a consequence, there is no strict meaning to the statement 
``{\em the} $\minv$ correction to $\Fb$ is 10\%''. 

%%%%%%%%%%%%%%%%%%%%%%%%%%%%%%%%%%%%%%%%%%%%%%%%%%%%%%%%%%%%%%%%%%%%%%%%%%%%%%%%%%%%%%%%%%
\section{Mass formulae}

Often cited mass formulae are
\bes
     {\mbav} &\equiv& {1\over 4} [\mB+3\mBstar] = \mbeauty + \overline{\Lambda}
    + {1\over 2\mbeauty} \lambda_1 +\rmO(1/\mbeauty^2)
  \\
     {\mbsplitt} &\equiv& \mBstar-\mB =  -{2\over \mbeauty} \lambda_2
     +\rmO(1/\mbeauty^2) 
\ees
with (ignoring renormalization)
\bes
   \lambda_1 = \langle B | \Okin | B \rangle \,,\quad 
   \lambda_2 = \frac13 \langle B | \Ospin | B \rangle \,.
\ees
The quantity $\overline{\Lambda}$ is termed ``static binding energy''.
Also here, depending on how one formulates the matching condition
which determines $\mbeauty$, one changes $\overline{\Lambda}$ by a term  
of order $\Lambda_\mrm{QCD}$, e.g. one may define  $\overline{\Lambda}=0$. 
Similarly, the kinetic term $\lambda_1/ (2\mbeauty)$ has a non-perturbative 
matching scheme
dependence of order $\Lambda_\mrm{QCD}$ and thus $\lambda_1$ itself 
has a matching scheme dependence of order $\mbeauty$.
The situation for $\overline{\Lambda}$ is similar to the gluon ``condensate''.
The non-perturbative scheme dependence has the same size as
the gluon ``condensate'' itself.
In contrast, $\lambda_2$ is the leading term in the $\minv$ expansion
and does not have such an ambiguity. We refer also to the more detailed
discussion in \cite{nara:rainer}.

\section{Non-perturbative determination of HQET parameters}
\label{s:param1}

We close our theoretical discussion of HQET by stating the 
correct procedure to determine the $\Nhqet$ parameters in the effective
theory at a certain order in $\minv$.  One requires
\be \label{e:matchnpgen}
  \Phiqcd_i(m) = \Phihqet_i(m,a) \,,\quad i=1\ldots\Nhqet\,,
\ee
where the $m$-dependence on the r.h.s. is entirely inside the 
HQET parameters. On the l.h.s. the continuum limit in QCD
is assumed to have been taken, but the r.h.s. refers to a given lattice spacing
where it defines the bare parameters of the theory at that 
value of $a$. We emphasize that as this matching has to be
invoked by numerical data, it is done at a given finite value
of $\minv$. Carrying it out with just the static parameters
defines the static approximation etc.

As simple as it is written down, it is non-trivial
to implement \eq{e:matchnpgen} in practice such that 
\bi
  \item[1)] the HQET expansion is accurate and one may thus truncate
  at a given order,
  \item[2)] the numerical precision is sufficient,
  \item[3)] lattice spacings are available 
  for which large volume computations of physical
  matrix elements can be performed.
\ei
In the following section we explain how these criteria can be satisfied 
using \SF correlation functions and a step scaling method. 
The first part will be a test of HQET on some selected 
correlation functions. This establishes how 1) and 2) can be met.
We can then explain the complete strategy which also achieves 3). 

\section{Relation to RGI matrix elements and conversion functions}

The matching equations \eq{e:matchnpgen} provide a definition of 
all HQET parameters, in principle at any given order
in the expansion. If considered at the static order,
it also provides the renormalization of the static axial current,
which we discussed at length in \sect{s:mass}. The relation between 
the two ways of parametrizing the current in static approximation 
are 
\bes
   \zahqet = \zastatrgi \, \Cps(M/\Lambda) + \rmO(\minv,a)\,.
\ees 
While \eq{e:matchnpgen} is a matching equation determining directly the product 
$\zastatrgi \, \Cps$,  the r.h.s. separates the problem 
into a pure HQET problem, the determination of the 
RGI operator, and a pure QCD problem, the 
the ``anomalous dimension'' $\gamma_\mrm{match}$, see
\eq{e_RG_Phi}. Note that in
this simple form, such a separation is only possible at the lowest
order in $\minv$. 

Since the breaking of spin symmetry is due to a single operator
at order $\minv$, there is also an analogous representation of
$\omegaspin$. We refer the interested reader to 
\cite{hqet:zospin}.

\chapter{Non-perturbative HQET}\label{s:nphqet}

After our long discussion of the theoretical issues in the 
renormalization of HQET, we turn to a complete strategy for 
the non-perturbative implementation. To this end the three criteria
in \sect{s:param1} have to be fulfilled. Establishing 1) is equivalent 
to testing HQET. We therefore start with such a test. Item 2)
has to do with finding matching conditions sensitive to the 
$\minv$-suppressed contributions. For this purpose we then expand a 
little on correlation functions in the \SF before coming to
a full description of the matching strategy.

%%%%%%%%%%%%%%%%%%%%%%%%%%%%%%
\begin{figure}[tb]
%\centering
  \includegraphics[width=0.49\textwidth]{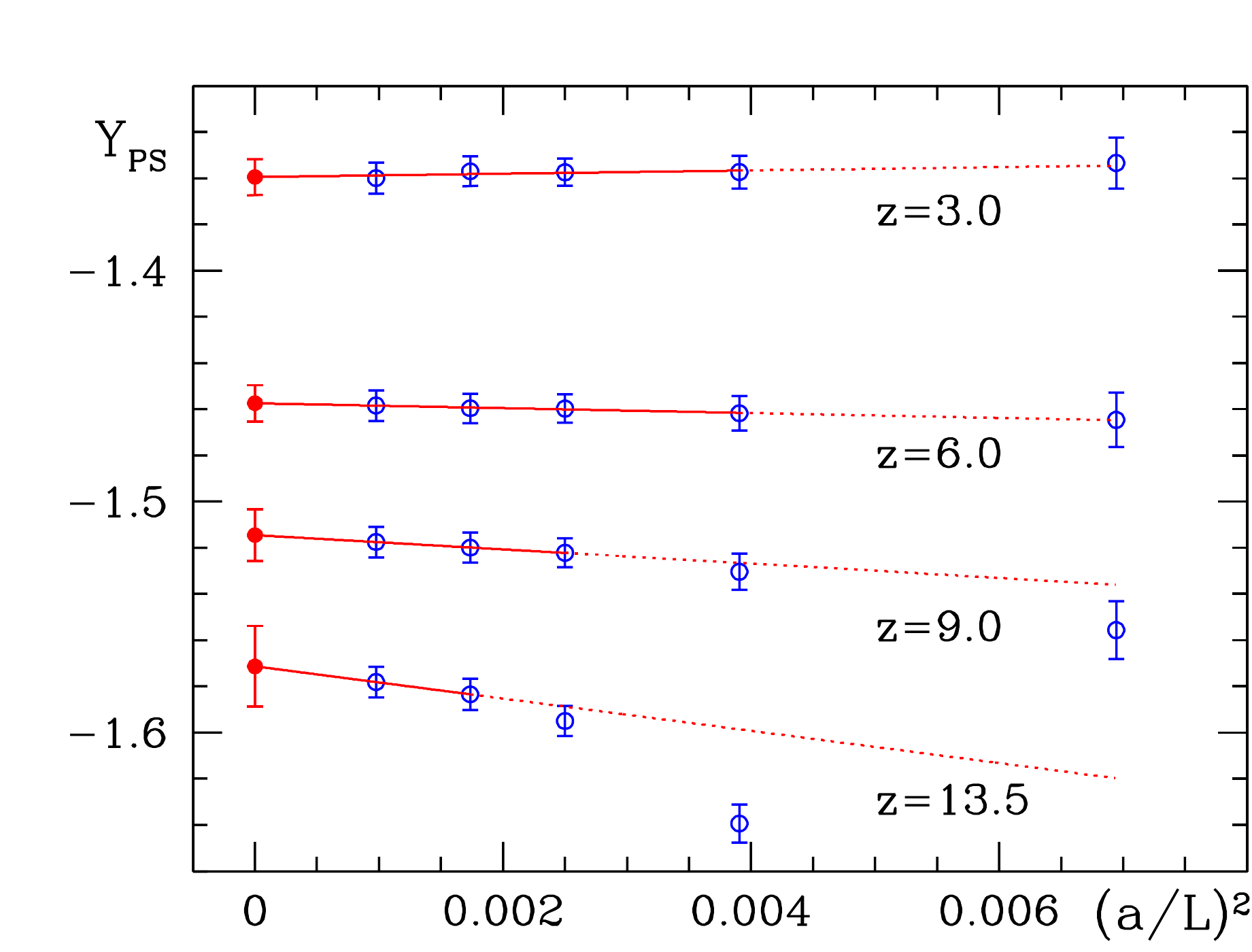}\hfill
  \includegraphics[width=0.49\textwidth]{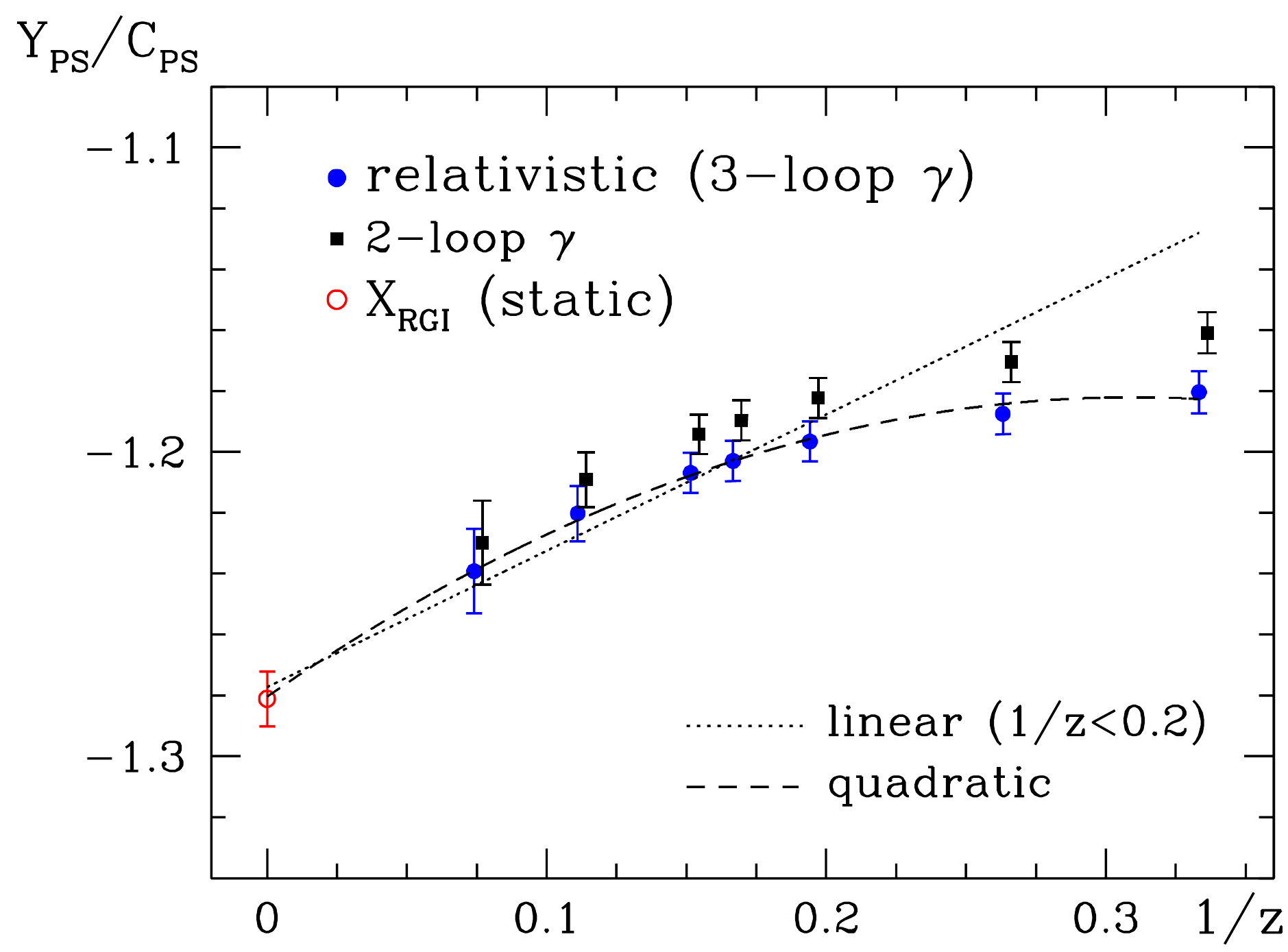}
\caption{ \footnotesize
Testing \eq{e:yrequiv} through numerical simulations in
the quenched approximation and for $L\approx0.2\,\fm$ \protect\cite{hqet:pap3}.
The graph uses notation $Y_\mrm{R}^\mrm{QCD}\equiv Y_\mrm{PS}$).
The physical mass of the b-quark corresponds to $z\approx6$.
Two different orders of perturbation theory for $\Cps$ are shown.  
}\label{f:yrmatch}
\end{figure}
%%%%%%%%%%%%%%%%%%%%%%%%%%%%%%

\section{Non-perturbative tests of HQET}\label{s:tests}

Although it is generally accepted that HQET is an effective theory of
QCD, tests of this equivalence
are rare and mostly based on phenomenological analysis of experimental
results. A pure theory test can be performed if
QCD including a heavy enough quark can be simulated on the lattice
at lattice spacings which are small enough to be able to take the continuum limit.
This has been done in the last few years~\cite{hqet:pap3,lat08:patrick}
and will be summarized below.

We start with the QCD side of such a test.
Lattice spacings such that $a \mbeauty \ll 1$ can be reached if
one puts the theory in a finite volume, $L^3 \times T$ with $L,T$ not
too large. We shall use $T=L$.
For various practical reasons,
\SF boundary conditions are chosen. Equivalent
boundary conditions are imposed in the effective theory.
As in \sect{s:match} we consider the ratio 
$Y_\mrm{R}^\mrm{QCD}(\theta,\mr,L)$ built from the
correlation functions $\fa$ and $\fone$.

It can be written as
\bes
Y_\mrm{R}^\mrm{QCD}(\theta,\mr,L) &=&
            {\langle \Omega(L)| A_0 |B(L)\rangle
              \over
              || \,| \Omega(L)\rangle \,||\; ||\, | B(L)\rangle \, ||}
            ,\\[1ex] \nonumber
            |B(L)\rangle &=& \rme^{ -L \ham/2 } |\varphi_{\rm B}(L)\rangle \,,\;
            |\Omega(L)\rangle =\rme^{ -L \ham/2 } |\varphi_{0}(L)\rangle \,,
\ees
in terms of the boundary states $|\varphi_{\rm B}(L)\rangle\,,\;|\varphi_{0}(L)\rangle$.
Expanded in  energy eigenstates with energies $E_n\geq \mb$ in
the B-sector and energies $\tilde E_n$ in the vacuum sector,
we have
\bes
   |B(L)\rangle &=& \sum_n \rme^{ -L E_n/2 } \langle n,B|\varphi_{\rm B}(L)\rangle
                  \;| n, B \rangle \\
                &\sim& \sum_{n\;|\; E_n-\mb < k/L} \rme^{ -L E_n/2 } 
               \langle n,B|\varphi_{\rm B}(L)\rangle
                  \;| n, B \rangle  +\rmO(\rme^{-k/2}) \,,\\
   |\Omega(L)\rangle &=& \sum_n \rme^{ -L \tilde E_n/2 } 
               \langle n,0|\varphi_{0}(L)\rangle
                  \;| n, 0 \rangle \,,\\
               &\sim& \sum_{n\;|\; \tilde E_n < k/L}  \rme^{ -L \tilde E_n/2 }
                  \langle n,0|\varphi_{0}(L)\rangle
                  \;| n, 0 \rangle +\rmO(\rme^{-k/2}) \,,
\ees
which shows that only energy eigenstates with $E_n-E_0 = \rmO(1/L)$ contribute
significantly. For $z=L \Mbeauty \gg 1$, HQET will thus describe the
correlation functions and the ratio $Y_\mrm{R}^\mrm{QCD}$.
We come to the conclusion that 
\bes \label{e:yrequiv}
        Y_\mrm{R}^\mrm{QCD}(\theta,\mr,L)  &=& \Cps(\Mbeauty/\Lambda)\, \XRGI + \rmO(1/z)\,,
        \quad z=\Mbeauty L\,,
        \\
        \XRGI &=& \lim_{a\to0} \zastatrgi(g_0){\fastat(L/2,\theta) \over \sqrt{\fonestat(\theta)}}
\ees
and similarly for other observables.
Note that one could also just argue that the only relevant scales are $L,\Lambda,\mbeauty$.
Therefore with $L\approx1/\Lambda$ there is a $\Lambda/\mbeauty \sim 1/z$ expansion.

Of course relations such as \eq{e:yrequiv} are expected after the continuum limit
of both sides has been taken separately. For the case of
$Y_\mrm{R}^\mrm{QCD}$, this is done by the following steps:
\begin{itemize}
\item Fix a value $u_0$ for the renormalized coupling
        $\gbar^2(L)$ (in the \SF scheme) at
        vanishing quark mass. In \cite{hqet:pap3}
        $u_0$ was chosen such that
        $L\approx0.2\,\fm$.
\item For a given resolution $L/a$, determine the bare coupling
        from the condition $\gbar^2(L)=u_0$. This step is well known 
        by now~\cite{mbar:pap1}.
\item Fix the bare quark mass $\mq$ of the heavy quark such that
        $LM=z$  using the known renormalization
        factors $\zM,Z$ in $M =\zM Z\, (1+a\bm \mq)\,\mq$,
        where $Z,\zM,\bm$ are all known non-perturbatively
        \cite{impr:babp,lat07:jochen}.
\item Evaluate $Y_\mrm{R}^\mrm{QCD}$ and repeat for better resolution $a/L$.
\item Extrapolate to the continuum as shown in \fig{f:yrmatch}, left.
\end{itemize}

\begin{figure}
  \centerline{\includegraphics[width=7.5cm]{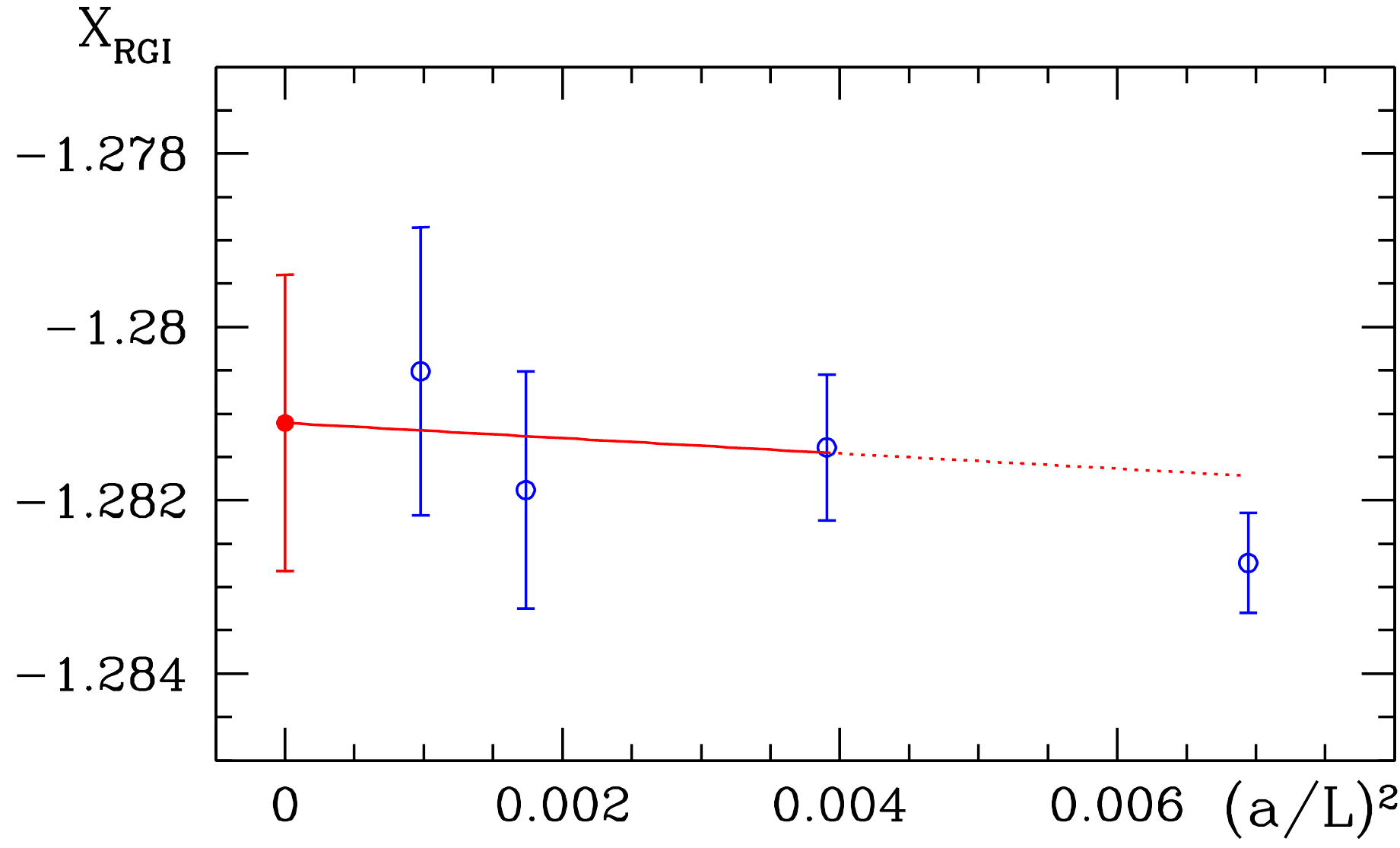}}
\caption{
\footnotesize
Continuum extrapolation of $\XRGI$ \protect\cite{hqet:pap3}.
}
\label{f:XRGI}
%\end{figure} %%%%%%%%%%%%%%%%%%%%%%%%%%%%%%%%%%%%%%%%%%%%%
\end{figure}

In the effective theory the same steps are followed.
As a simplification, no quark mass needs to be fixed and
the continuum extrapolation is much easier
as illustrated in \fig{f:XRGI}.
%%%%%% figure: XRGI %%%%%%%%%%%%%%%%%%%%%%%%%%%%%%%%%%%%%%%%%%%%%
%

The comparison of the static result and the relativistic theory,
\fig{f:yrmatch}, looks rather convincing,\footnote{Note that the comparison
\fig{f:yrmatch} has to be taken with a grain of salt
due to the perturbative uncertainty in $C_\mrm{PS}$
discussed in \sect{s:RGI}.} 
but we note that the b-quark mass point is
$1/z=1/z_\mrm{b}\approx 0.17$, where $1/z^2$ terms are
not completely negligible. The displayed fit has a 8\% contribution
by the $1/z$ term and a 2\% $1/z^2$ piece.

\begin{figure}[htb!]
  \centerline{\includegraphics[width=0.6\textwidth]{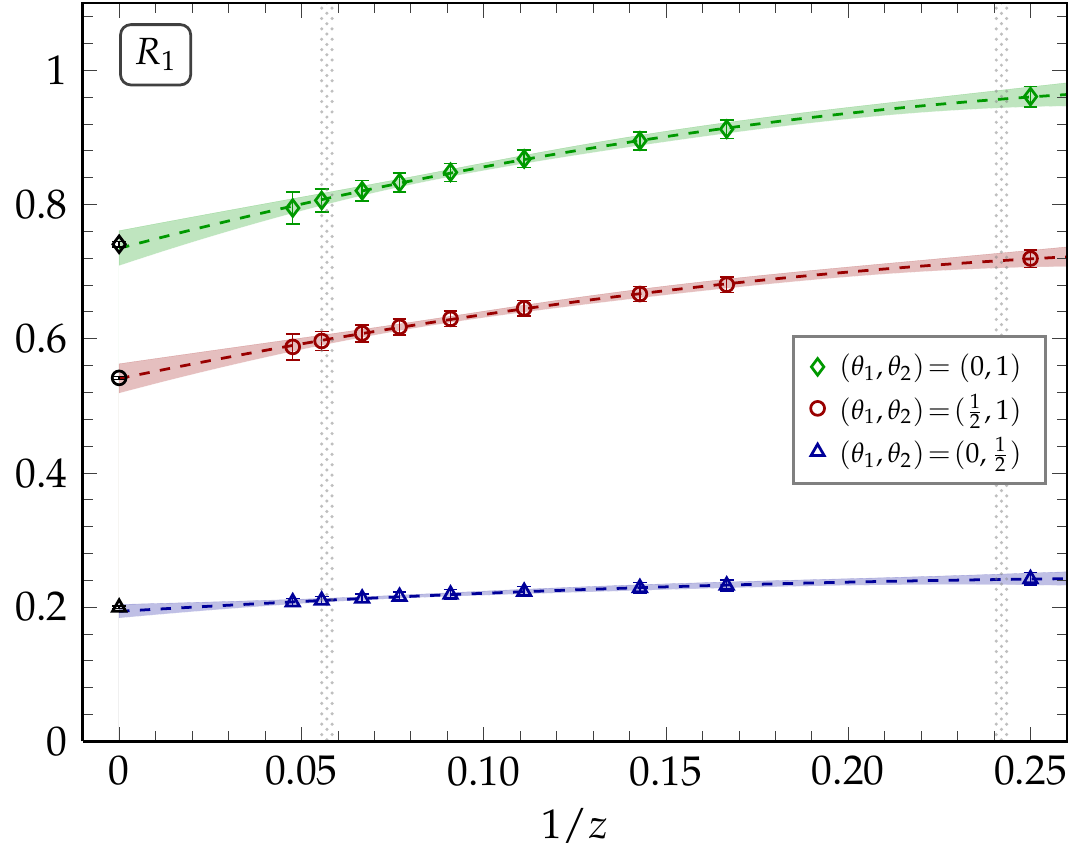}}
\caption{
\footnotesize
The logarithmic ratio $R_{1}$ for different pairs $(\theta_1,\theta_2)$ with $\nf=2$ 
flavors and for $L=T\approx 0.5\,\fm$
\protect\cite{lat08:patrick} with $\nf=2$. The value of $1/z$ for charm and bottom quarks are
indicated by the vertical bands.
}
\label{f:rone}
\end{figure}

\begin{figure}[htb!]
  \centerline{\includegraphics[width=0.6\textwidth]{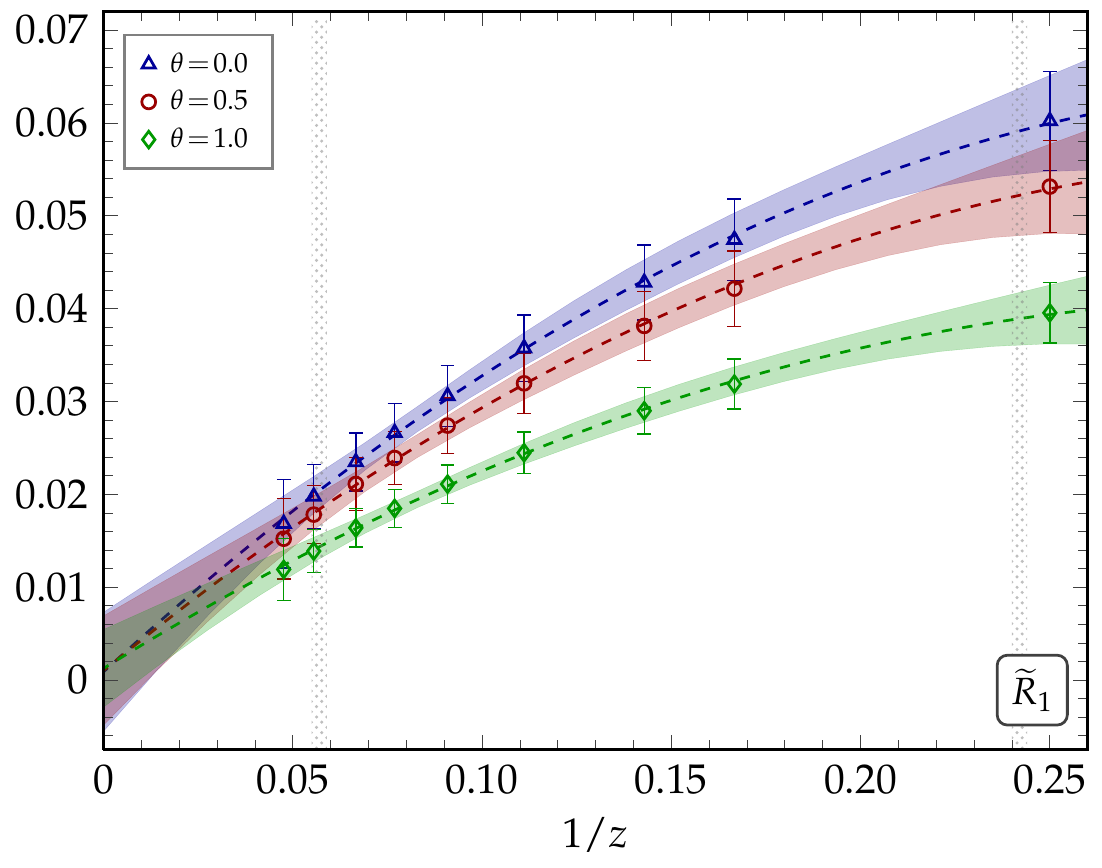}}
\caption{
\footnotesize
The logarithmic ratio $\widetilde{R}_{1}$ for different values $\theta_0$ with $\nf=2$ 
flavors and for $L=T\approx 0.5\,\fm$
\protect\cite{lat08:patrick} with $\nf=2$.
}
\label{f:rspin}
\end{figure}

For a precision application (\sect{s:strat}) it is thus
safer to have $L\grtsim0.4\,\fm$ instead of the  
$L=0.2\,\fm$ chosen in the first test, reducing $1/z^2$ by a factor four. 
For $L\approx0.5\, \fm$ we show two different examples, \fig{f:rone},
\fig{f:rspin} which involve
\bes
  \kone(\theta) &=&
  -{a^{12} \over 6L^6}\sum_{\vecu,\vecv,\vecy,\vecz,k}
  \left\langle
  \zetalbprime(\vecu)\gamma_k\zzetaprime_{\rm b}(\vecv)\,
  \zetabar_{\rm b}(\vecy)\gamma_k\zetal(\vecz)
  \right\rangle\,
\ees
in addition to the previously introduced correlation functions. The considered
combinations are
\bes
  \rone &=& \frac14\left(
\ln\left(\fone(\theta_1)\kone(\theta_1)^3\over\fone(\theta_2)\kone(\theta_2)^3\right)
\right) \label{e:rone}
\\
 \widetilde{R}_1  &=& {3\over4}\ln\left({\fone\over\kone}\right)  \,. \label{e:ronespi}
\ees
Their HQET expansion contains no conversion functions at leading order
and they are thus free of the associated perturbative uncertainty. 
While $\rone$ has a finite static limit, $\widetilde{R}_1$ vanishes as 
$z\to\infty$ due to the spin symmetry. The expected HQET behavior 
is confirmed with surprisingly small $1/z^2$ corrections for a charm quark.
The quadratic fits in $1/z$ displayed in the figures are not constrained
to pass through the separately displayed static limit.

%%%%%% figure: R1 %%%%%%%%%%%%%%%%%%%%%%%%%%%%%%%%%%%%%%%%%%%%%
%
%\end{figure} %%%%%%%%%%%%%%%%%%%%%%%%%%%%%%%%%%%%%%%%%%%%%

%%%%%% figure: R1 %%%%%%%%%%%%%%%%%%%%%%%%%%%%%%%%%%%%%%%%%%%%%
%
%\end{figure} %%%%%%%%%%%%%%%%%%%%%%%%%%%%%%%%%%%%%%%%%%%%%

\section{HQET expansion of \SF correlation functions} \label{s:SF1}

In complete analogy to the case of a manifold without boundary we can write
down the 
expansions of the \SF correlation functions to first order in $\minv$:
\bes
  \left[\fa\right]_\mrm{R} &=& \zahqet \zzetah\zzeta \rme^{-\mhbare x_0}
        \left\{ \fastat + \cahqet \fdeltaastat + \omegakin \fakin
                + \omegaspin \faspin
        \right\}\,, \label{e:faexp} \\
  \label{e:foneexp}
  \left[\fone\right]_\mrm{R} &=& \zzetah^2\zzeta^2 \rme^{-\mhbare T}
        \left\{ \fonestat + \omegakin \fonekin
                + \omegaspin \fonespin
        \right\}\,, \\
  \label{e:koneexp}
  \left[\kone\right]_\mrm{R} &=& \zzetah^2\zzeta^2 \rme^{-\mhbare T}
        \left\{ \fonestat + \omegakin \fonekin
                -\frac13 \omegaspin \fonespin
        \right\}\,.
\ees
Apart from
\bes
  \fdeltaastat(x_0,\theta) =  -{a^6 \over 2}\sum_{\vecy,\vecz}\,
  \left\langle
  \Ah{1}(x)\,\zetabar_{\rm h}(\vecy)\gamma_5\zeta_{\rm l}(\vecz)
  \right\rangle
\ees
the labeling of the different terms follows directly the one
introduced in \eq{e:exp}. The relation between the $\minv$ terms
in $\fone$ and $\kone$ is
a simple consequence of the spin symmetry of the static action,
valid at any lattice spacing.
A further simplicity is that no $\minv$ boundary corrections are present.
Potential such terms have dimension four. After using the equations of motion, 
only one candidate remains, which however does
not contribute to any correlation function.\footnote{In the notation of 
\cite{impr:pap1} 
it reads $\rhobar_\mrm{h}(\vecx) \gamma_k D_k \rho_\mrm{h}(\vecx)$ at $x_0=0$.
Such a term does not contribute to any correlation function due to the form of 
the static propagator.} 

\section{Strategy for non-perturbative matching}\label{s:strat}

After the tests of HQET described above,
it is clear how one can non-perturbatively match 
HQET to QCD. Consider the action as well as $A_0$ (just at $\vecp=0$)
and denote the free parameters of the effective 
theory by $\omega_i\,,\,i=1\ldots \Nhqet$. 
In static approximation we then have 
\bes
  \omega^\mrm{stat} &=& (\,\mhbare^\stat\,,\, [\ln(\za)]^\stat\,)^t \,, \quad 
  \Nhqet=2
  \label{e:paramstat}
\ees
and including the first order terms
in $1/m$ together with the static ones, the HQET parameters are
\bes
  \omega^\mrm{HQET} &=& (\,\mhbare\,,\,\, \ln(\zahqet)\,,\,\cahqet\,,\,\omegakin\,,\,\omegaspin\,)^t \, \quad 
  \Nhqet=5\,.
  \label{e:paramhqet}
\ees
The pure $\minv$ parameters may be defined as 
$\omega^\first=\omega^\mrm{HQET}-\omega^\mrm{stat}$,
with all of them, e.g. also $\mhbare^\first$, non-zero. In fact our discussion
of renormalization of the $\minv$ terms shows that $\mhbare^\first$ diverges
as $1/(a^2 m)$.

With suitable observables 
\bes
  \Phi_i(L_1,M,a)\,,\;i= 1\ldots \Nhqet\,, \nonumber
\ees
in a \SF with $L=T=L_1\approx 0.5\,\fm$,
we then require matching\footnote{Recall that observables
without a superscript refer to HQET.}
\bes
  \label{e:matchnp}
  \Phi_i(L_1,M,a)=\Phi_i^\mrm{QCD}(L_1,M,0)\,,\;i= 1\ldots \Nhqet\,.
\ees
Note that the continuum limit is taken in QCD, while in HQET we want to
extract the bare parameters of the theory from the matching equation
and thus have a finite value of $a$.
It is convenient to pick observables with HQET expansions linear in $\omega_i$,
\bes
  \label{e:phiexp}
  \Phi(L,M,a) &=& \phistat(L,a) + \phimat(L,a)\,\omega(M,a)\,,
\ees
in terms of a $\Nhqet\times\Nhqet$ coefficient matrix $\phimat$.
A natural choice for the first two observables is
 \bes
  \Phi_1 &=& L\meffp \equiv -L\dzero \ln(-\fa(x_0))_{x_0=L/2} \;\simas{L\to\infty}\; 
             L \mB \\
  \Phi_2 &=& \ln(\za{-\fa\over\sqrt{\fone}}) \;\simas{L\to\infty}\; 
             L^{3/2}\fb\sqrt{\mB/2}\,,
\ees
since in static approximation these determine directly
$\omega_1$ and $\omega_2$. We will introduce the other $\Phi_i$ later.
The explicit form of $\phistat,\phimat$ is
\bes
  \phistat &=&\pmat{\meffstat  \\
                   \zetaa \\ \ldots}
           \,,\quad
   \phimat = \pmat{ L & 0 &  \ldots\\
                      0 & 1 & \ldots\\ \ldots} \,
  \label{e:phiexp2}
\ees
with
\bes
  \meffstat = -L\dzero \ln(\fastat(x_0))_{x_0=L/2}\,,\quad
     \zetaa = \ln({-\fastat\over\sqrt{\fonestat}})\,.
   \label{e:phiexp3}
\ees
In static approximation, the structure of the matrix $\phimat$ is perfect:
one observable determines one parameter. This is possible since there 
is no (non-trivial) mixing at that order. 

Having specified the matching conditions, 
the HQET parameters $\omega_i(M,a)$ 
can be obtained from eqs.(\ref{e:matchnp},\ref{e:phiexp}), but only for 
rather small lattice
spacings since a reasonable suppression
of lattice artifacts requires $L_1/a=\rmO(10)$ and thus $a=\rmO(0.05\,\fm)$.

\begin{figure}
  \centerline{\includegraphics[width=0.95\textwidth]{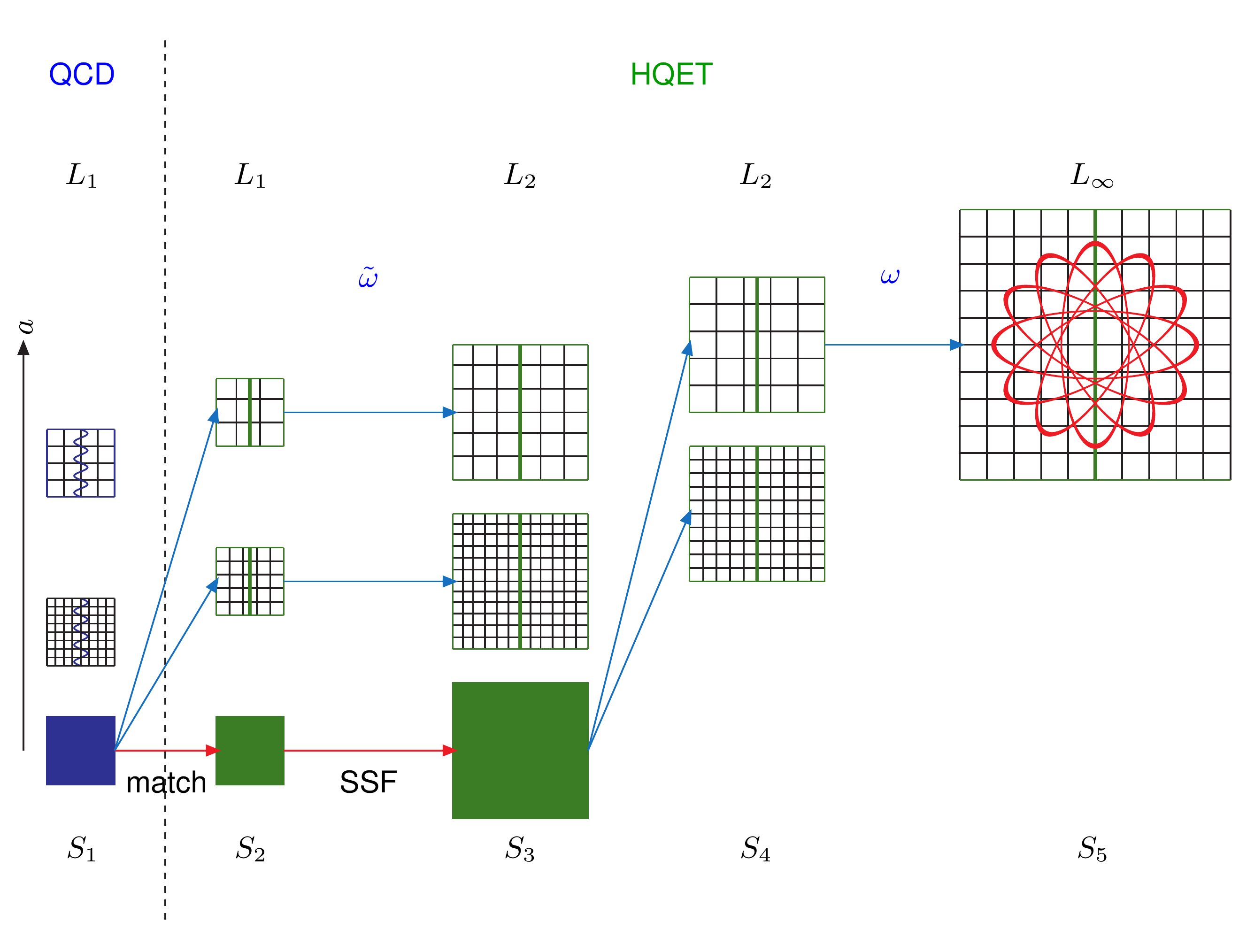}}
\caption{
\footnotesize
Strategy for non-perturbative HQET \protect\cite{hqet:first1}.
Note that in the realistic implementation\protect\cite{hqet:first1} finer resolutions are used.
}
\label{f:strat}
%\end{figure} %%%%%%%%%%%%%%%%%%%%%%%%%%%%%%%%%%%%%%%%%%%%%
\end{figure}

Larger lattice spacings as needed in large volume,
can be reached by adding a step scaling strategy,  
illustrated in \fig{f:strat}.
%%%%%% figure: hqetstrategy %%%%%%%%%%%%%%%%%%%%%%%%%%%%%%%%%%%%%%%%%%%%%
%
Let us now go through the various steps of this strategy.
\bi
\item[(1)] Take the continuum limit
\bes
  \label{e:phiqcd}
 \Phiqcd_i(L_1,M,0)=\lim_{a/L_1\to 0} \Phi^\mrm{QCD}_i(L_1,M,a)\,.
\ees
This is similar to the HQET tests and as we saw there, 
it requires $L_1/a = 20 \ldots 40$\,, or
$a=0.025\,\fm \ldots 0.012\,\fm$.
\item[(2a)] Set the HQET observables equal to the QCD ones,
\eq{e:matchnp} and extract the parameters
\bes
  \label{e:match}
  \tilde\omega(M,a) &\equiv& \phimat^{-1}(L_1,a)\,
  [\Phi(L_1,M,0)-\phistat(L_1,a)]\, \\
  &=& \pmat{L_1^{-1}  \Phi_1(L_1,M,0) - \meffstat(L_1,a) \\
                    \Phi_2(L_1,M,0) - \zetaa(L_1,a) \\
                    \ldots } \,.
\ees
The only restriction here is $L_1/a\gg1$, so one can use 
$L_1/a = 10 \ldots 20$\,, which means
$a=0.05\,\fm \ldots 0.025\,\fm$.
\item[(2b.)] Insert $\tilde\omega$ into $\Phi(L_2,M,a)$:
\bes
  \Phi(L_2,M,0)&=&  \lim_{a/L_2\to0}
  \left\{ \phistat(L_2,a) + \phimat(L_2,a)\,\tilde\omega(M,a)  \right\}\,
  \\ \nonumber&=&  \lim_{a/L_2\to0} 
  \pmat{L_2\meffstat(L_2,a) + {L_2\over L_1} \Phi_1(L_1,M,0) - L_2\meffstat(L_1,a) 
        \\
        \zetaa(L_2,a) + \Phi_2(L_1,M,0) - \zetaa(L_1,a) \\
                    \ldots }
  \\ \nonumber&=&  \lim_{a/L_2\to0} 
  \underbrace{\pmat{L_2[\meffstat(L_2,a) - \meffstat(L_1,a)] \\
        \zetaa(L_2,a) - \zetaa(L_1,a) \\
                    \ldots }}_{\mbox{finite HQET SSF's}} 
        + 
  \underbrace{\pmat{{L_2\over L_1} \Phi_1(L_1,M,0) \\
        \Phi_2(L_1,M,0) \\
                    \ldots }}_{\mbox{QCD, mass dependence}} .
\ees
In the last line we have identified pieces which are separately finite.
This step can be done as long as the lattice spacing is common to
the $n_2=L_2/a$ and $n_1=L_1/a$-lattices and
\bes
   s = L_2/L_1 =  n_2/n_1
\ees
is kept at a fixed, small, ratio.\footnote{
A fixed ratio $s$ ensures that 
the cutoff effects are a smooth function of $a/L_i$.
}
\item[(3.)]
Repeat (2a.) for $L_1\to L_2$:
\bes
  \label{e:fin}
  \omega(M,a) \equiv \phimat^{-1}(L_2,a)\,
  [\Phi(L_2,M,0)-\phistat(L_2,a)]\,.
\ees
With the same resolutions $L_2/a = 10 \ldots 20$ one has now reached
$a=0.1\,\fm \ldots 0.05\,\fm$. 
\item[(4.)]
Finally insert $\omega$ into the expansion of large volume
observables, e.g. 
\bes
   \mB =& \omega_1 + \Estat\,.
\ees
\ei
In the chosen example the result is the relation between the RGI b-quark mass 
and the B-meson mass $\mB$. It is illustrative to put 
the different steps into one equation,
\bes
\label{e:mbmB}
 \\[-3ex]
\begin{aligned}
  \nonumber
   \mB &= 
         \\ &
         \lim_{a \to 0} [\Estat - \meffstat(L_2,a)]  \quad &a=&0.1\fm\ldots0.05\fm&[S_4,S_5&]
         \\ & + \lim_{a \to 0} [\meffstat(L_2,a) - \meffstat(L_1,a)] 
                \quad &a=&0.05\fm\ldots0.025\fm&[S_2,S_3&]
         \\ & +  {1 \over L_1} \lim_{a \to 0}\Phi_1(L_1,\Mbeauty,a) 
               \quad &a=&0.025\,\fm \ldots 0.012\,\fm \,&[S_1&]\;.
\end{aligned}
\ees 
 We have indicated the lattices drawn in
\fig{f:strat} and the typical lattice spacings of these lattices.
The explicit expression for the decay constant in static approximation 
is even more simple; write it down as an exercise!

So far we have spelled out only those observables which are needed
in the static approximation. The following heuristics helps to 
find observables suitable for the determination of the $\minv$-terms. 
Recall that $\theta\ne0$ means 
$\frac12(\nab{j}+\nabstar{j})\sim i\theta/L$ (acting onto 
a quark field) 
when the gauge fields are weak, as is the case in small volume.
Hence, expanding in $\minv$ 
\bes
  \Phi_3(L,M,a)={\fa(\theta_1) \over \fa(\theta_2)} \sim \ldots +
   \cahqet\,[\theta_2-\theta_1]/L  \,
\ees
for weakly coupled quarks. In the same way the combination (recall \eq{e:rone})
\bes
  \Phi_4(L,M,a) &=&  R_1 =  \ronestat 
  + \omegakin \,{\ronekin} \nonumber
\ees
has a sensitivity to $\omegakin$ of $\ronekin\propto\; \theta_1^2-\theta_2^2$
while in the specific linear combination of $\fone$ and $\kone$ which form
$R_1$ the parameter $\omegaspin$ drops out.
Finally the choice
\bes
  \Phi_5(L,M,a) &=& \widetilde{R}_1  = \omegaspin \ronespin \, 
\ees
allows for a direct determination of $\omegaspin$.
These choices leave relatively many zeros in the matrix $\phimat$,
which has a block structure,
\bes
  \label{e:AB}
  \phimat = \pmat{C & B \\ 0 & A}\,,\quad \phimat^{-1}=
            \pmat{C^{-1} & -C^{-1}BA^{-1} \\ 0 & A^{-1}}\,,\quad
     C = \pmat{ L & 0 \\
                0 & 1  } \,.
\ees
The listed observables $\Phi_i$ have been shown to work in practice, i.e.
in a numerical application \cite{hqet:first1}.

\section{Numerical computations in the effective theory}

Before showing some  results, we should briefly mention that it is not entirely
straight forward to obtain precise numerical results in the effective theory. The reason 
is a generically rather strong growth of statistical errors as a function of
the Euclidean time separation of the correlation functions. Two ideas help to
overcome this problem. We sketch them here; more details are available in the
cited literature.

\subsection{The static action}
Consider a typical two-point function, for example \eq{e:caastat}. At large time it decays 
exponentially and so does the variance. Setting $\dmstat=0$ the decay of the
signal is 
\bes
  C(x_0) \sim \rme^{- E_{\rm stat} \, x_0 }\,,
\ees
while the variance decays with an exponential rate given
by the pion mass. Thus the noise-to-signal ratio for the B-meson correlation
function behaves as
\be
R_{\rm NS} \propto e^{[E_{\rm stat}-m_{\pi}/2]\,x_0} \;.
\label{Lepf}
\ee
The self energy of
a static quark is power divergent, in particular 
in perturbation theory 
\bes \label{e:r1}
 \Estat &\sim& \left({{1}\over{a}} r^{(1)} + {\rm O}(a^0)\right) \; g_0^2+ \rmO(g_0^4) \,.
\ees
This divergence yields the leading behavior of \eq{Lepf}
for small $a$. It is potentially dangerous since we are interested 
in the continuum limit. The scale of the problem
can be reduced considerably by the replacement
\bes
  \label{e:HYP}
  U(x,0) \;\to\; W_\mrm{HYPi}(x,0) \,,
\ees
in the covariant derivative $\nabstar{0}$ in the static action. Here $W_\mrm{HYPi}$
is a so-called HYP-smeared link. \Tab{t:deltam} shows how the self energy is
reduced for two choices of $W_\mrm{HYPi}$.
%%%%%%%%%%%%%%%%%%%%%%%%%%%%%%%%%%%TABLE%%%%%%%%%%%%%%%%%%%%%%%%%%%%%%%
%
\begin{table}[h!]
\centering
\begin{tabular}{c|ccc}
\hline\\[-2.0ex]
 $S_{\rm h}^{\rm W}$ & $r^{(1)}$ & $aE_{\rm stat}$& \\[.5
ex]
\hline\\[-2.ex]
$S_{\rm h}^{\rm EH}$   & 0.16845(2)    & 0.68(9)  \\[.7ex]
$S_{\rm h}^{\rm HYP1}$ & 0.04844(1)    & 0.44(2)   \\[.7ex]
$S_{\rm h}^{\rm HYP2}$ & 0.03523(1)    & 0.41(1)   \\[.7ex]
\hline
\end{tabular}\\
\caption{\footnotesize{
One loop coefficients $r^{(1)}$, 
\eq{e:r1} and
non-perturbative values for $aE_{\rm stat}$  at $\beta=6/g_0^2=6$
and a (quenched) light quark with the mass of the strange quark.
``EH'' refers to Eichten-Hill, i.e. $W(x,0) = U(x,0)$,
while ``HYP1,HYP2'' are two versions of HYP-smearing
\protect\cite{HYP,stat:actpaper}
}}\label{t:deltam}
\end{table}
%%%%%%%%%%%%%%%%%%%%%%%%%%%%%%%%%%%%%%%%%%%%%%%%%%%%%%%%%%%%%%%%%%

It is mandatory to check that such a change of action does not introduce
large cutoff effects. This was done for single smearing
in \cite{stat:actpaper}: the points with smallest error bars in
\fig{f:scal} are for these actions. We expect that large cutoff effects would
however appear if smearing was repeated several times.

\subsection{Generalized Eigenvalue Method \label{s:gevp}}

For the numerical evaluation of matrix elements 
such as $\Phistat$, \eq{e:Phistat},
or of energy levels it is advisable
to use an improvement over the straight forward formula 
\eq{e:fbstat}. The reason is as follows.
Let us label the energies in the sector contributing to a given 
correlation function by $E_n, \; n=1,2,3$. Then
there are corrections to the desired ground state matrix element
due to excited state contaminations 
of order $\rme^{-x_0 \Delta}$ and $\Delta=E_2-E_1$. From an investigation of the 
spectrum in the B-meson sector one finds numerically   
$\Delta \approx600  \,\MeV$ and thus $\Delta\,x_0 \approx 3 x_0 / \fm$.
The suppression of excited state contaminations 
is then not necessarily small enough for $x_0\sim1\fm$\,
but using \eq{e:fbstat} beyond $x_0\sim1\fm$ is very difficult 
because statistical errors grow quite rapidly with $x_0$.

A considerable improvement is achieved if one 
considers the generalized eigenvalue problem 
(GEVP)\cite{gevp:michael,phaseshifts:LW,gevp:pap}.
It uses additional information in the form of
a matrix correlation function formed from 
$N$ different interpolating fields on one time slice and the same 
interpolating fields on another time slice. When this 
matrix correlation function is analyzed in a specific way, 
described in \cite{gevp:pap}, 
one can prove that a much larger gap, $\Delta=E_{N+1}-E_1$ appears for
the  dominating correction terms
due to excited states. 
These then disappear much more quickly with growing time.

The GEVP is straight forwardly applicable to HQET, order by order
in $\minv$.
The precision of the numerical results that we show below
is largely due to this method, together with the use of HYP1/2 actions.

\section{Examples of results} \label{s:res}

%%%%%%%%%%%%%%%%%%%%%%%%%%%%%%%%%%%%%%%%%%%%%%%%%%%%%%%%%%%%%%%%%%%%%%%%%%%%%%%%%%%%%%%%%%
\begin{figure}[t!]
\vspace{0pt}
\centerline{\epsfig{file=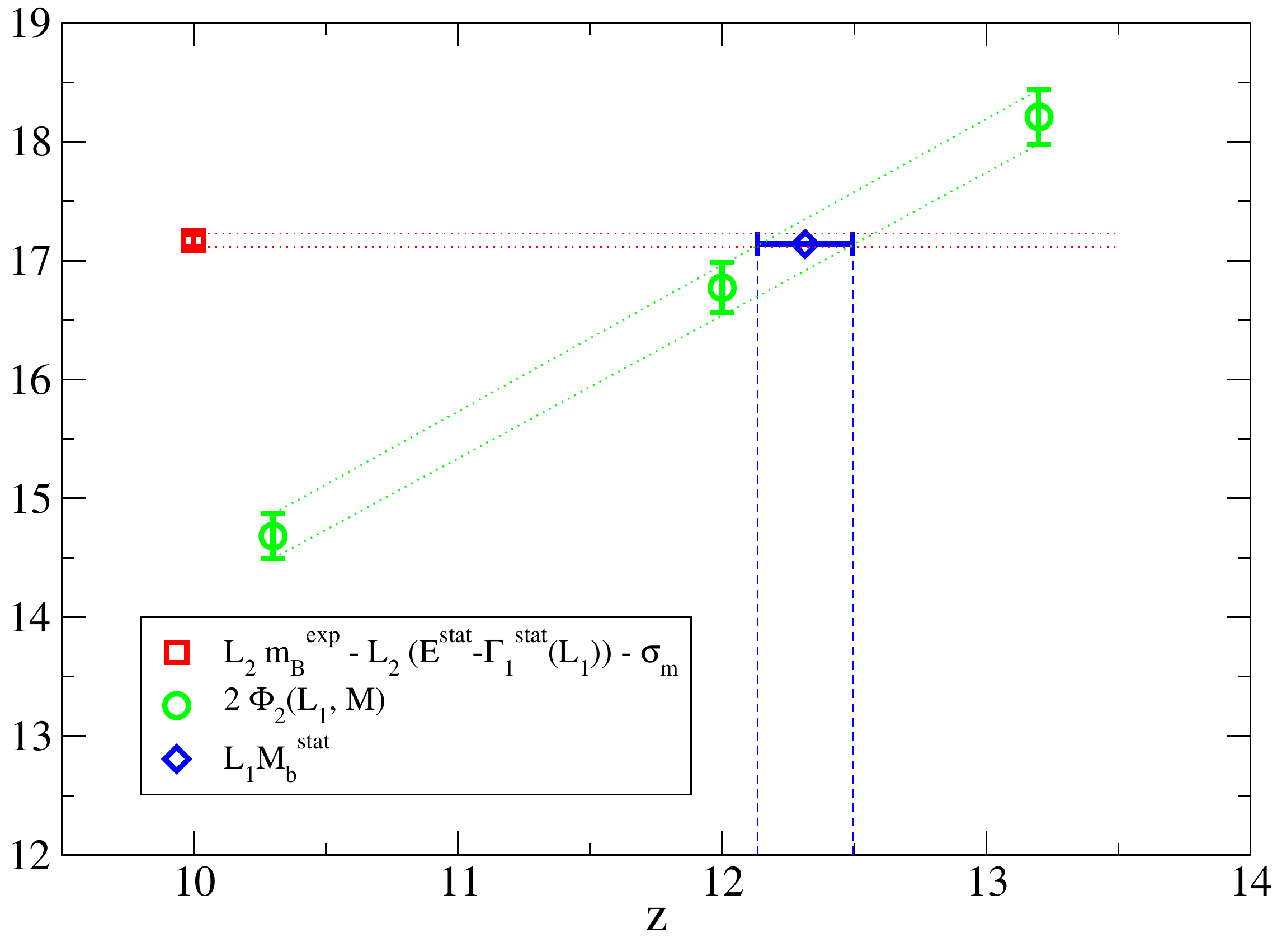,
       width=0.68\textwidth} }
  \vspace*{-5mm}
  \caption{\footnotesize 
        Numerical solution of the equation for $\Mbeauty$
        \protect\cite{hqet:first1} made dimensionless by
        multiplication with $L_2$. The figure uses a notation
        $\sigma_\mrm{m}=\lim_{a \to 0} L_2\,[\meffstat(L_2,a) - \meffstat(L_1,a)]$
        and $\Phi_2$ in the figure is $\Phi_1$ in our notation.
        }\label{f:mbstat}
\end{figure}
%%%%%%%%%%%%%%%%%%%%%%%%%%%%%%%%%%%%%%%%%%%%%%%%%%%%%%%%%%%%%%%%%%%%%%%%%%%%%%%%%%%%%%%%%%

 \begin{table}[!p] 
 \hspace{-1.cm} 
 \begin{center} 
 \begin{tabular}{||c||c|ccc||}
 \hline
 \hline
 & LO (static) & \multicolumn{3}{c||}{NLO (static + $\rmO(1/m)$) }  \\
 \hline
 \hline
 & & $(\theta_1, \theta_2) = (0,0.5)$ & $(\theta_1, \theta_2) = (0.5,1)$ & $(\theta_1, \theta_2) = (0,1)$ \\
 \hline
$\theta_0 = 0  $&$  17.1 \pm 0.2  $&$   17.1 \pm 0.2  $&$ 17.1 \pm 0.2 $&$  17.1 \pm 0.2 $ \\
$\theta_0 = 0.5$&$  17.2 \pm 0.2  $&$   17.2 \pm 0.2  $&$ 17.2 \pm 0.2 $&$  17.1 \pm 0.2 $ \\
$\theta_0 = 1  $&$  17.2 \pm 0.2  $&$   17.3 \pm 0.3  $&$ 17.3 \pm 0.3 $&$  17.3 \pm 0.3 $ \\
 \hline \hline
 \end{tabular} 
 \end{center} 
 \caption[ ]{Dimensionless b-quark mass, $r_0 M_{\rm b}$, obtained from the $B_{\rm s}$ meson mass, 
for different values of $\theta_i$.}
 \label{table_mb_ps} 
 \end{table}

%%%%%%%%%%%%%%%%%%%%%%%%%%%%%%%%%%%%%%%%%%%%%%%%%%%%%%%%%%%%%%%%%%%%%%%%%%%%%%%%%%%%%%%%%%
\begin{figure}[!p]
\vspace{0pt}
\centerline{\epsfig{file=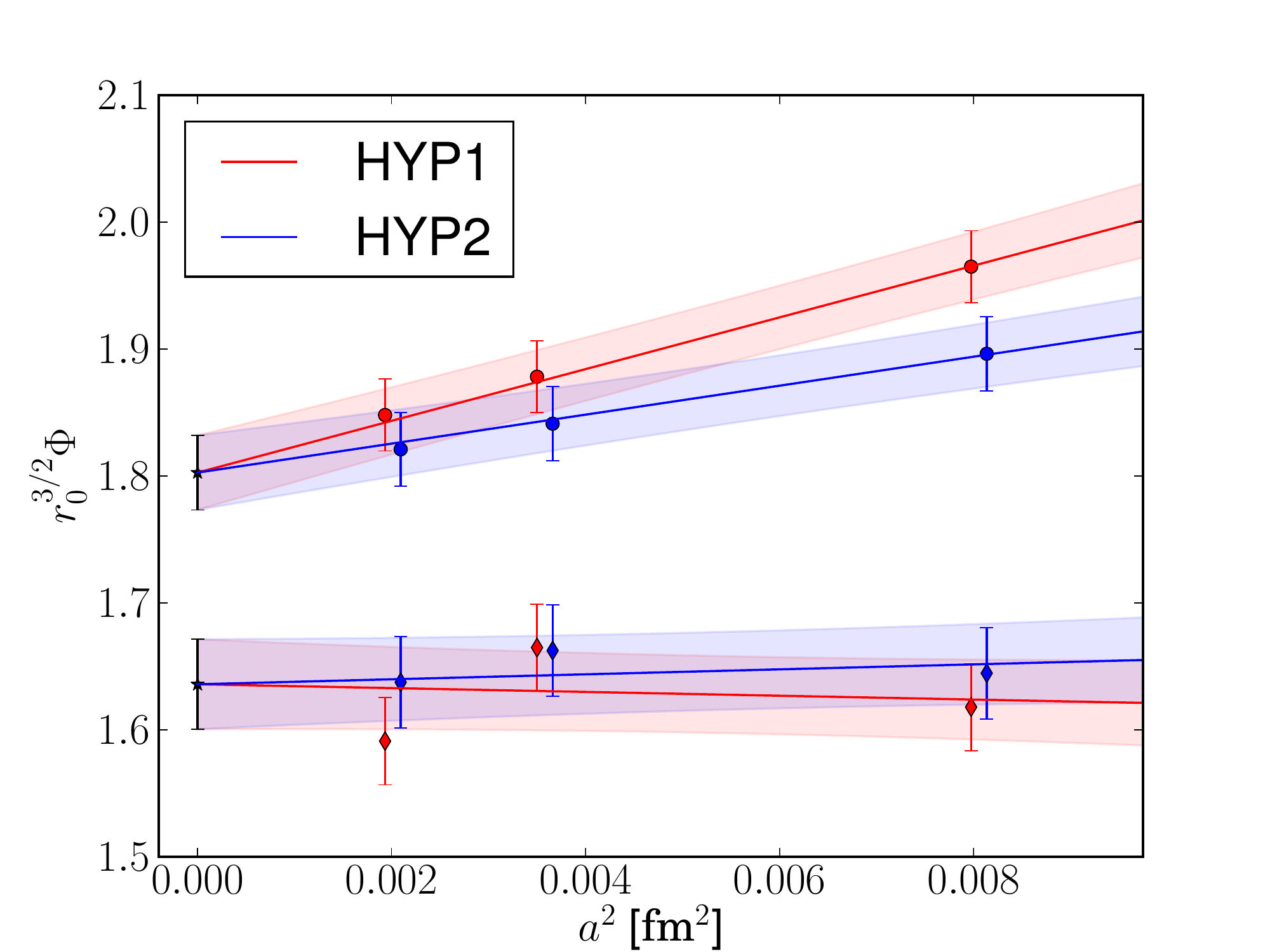,
       width=0.50\textwidth} \hfill 
       \epsfig{file=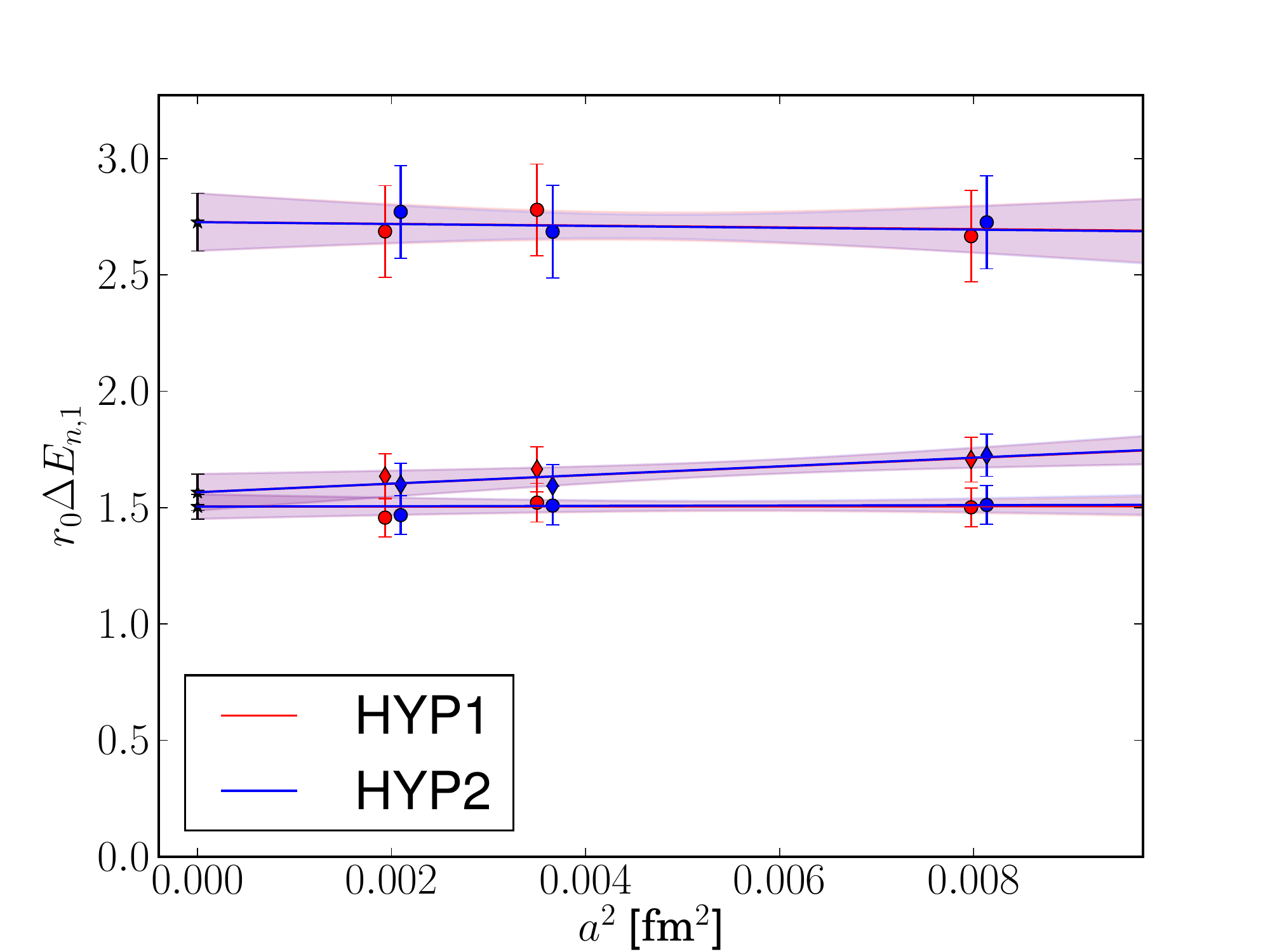,
       width=0.50\textwidth} \vspace*{-3mm}}
  \caption{\footnotesize 
        Continuum extrapolations in HQET. Left: $\Phihqet=\fbs\sqrt{\mbs}/\Cps$ (diamonds)
        in HQET with $\minv$ corrections included\protect\cite{hqet:first3}
        and its static limit $\Phirgi$ (circles). The value of $\Cps$ does
        not depend on the lattice spacing. It renders the two quantities 
        directly comparable. Right: pseudo scalar energy levels
        \protect\cite{hqet:first2}. From bottom to top:
        2s -- 1s splitting static, 
        2s -- 1s splitting static + $\minv$, 
        3s -- 1s splitting static.
        }\label{f:fbhqetcl}
\end{figure}
%%%%%%%%%%%%%%%%%%%%%%%%%%%%%%%%%%%%%%%%%%%%%%%%%%%%%%%%%%%%%%%%%%%%%%%%%%%%%%%%%%%%%%%%%%
%%%%%%%%%%%%%%%%%%%%%%%%%%%%%%%%%%%%%%%%%%%%%%%%%%%%%%%%%%%%%%%%%%%%%%%%%%%%%%%%%%%%%%%%%%
\begin{figure}[!p]
\vspace{0pt}
\centerline{\epsfig{file=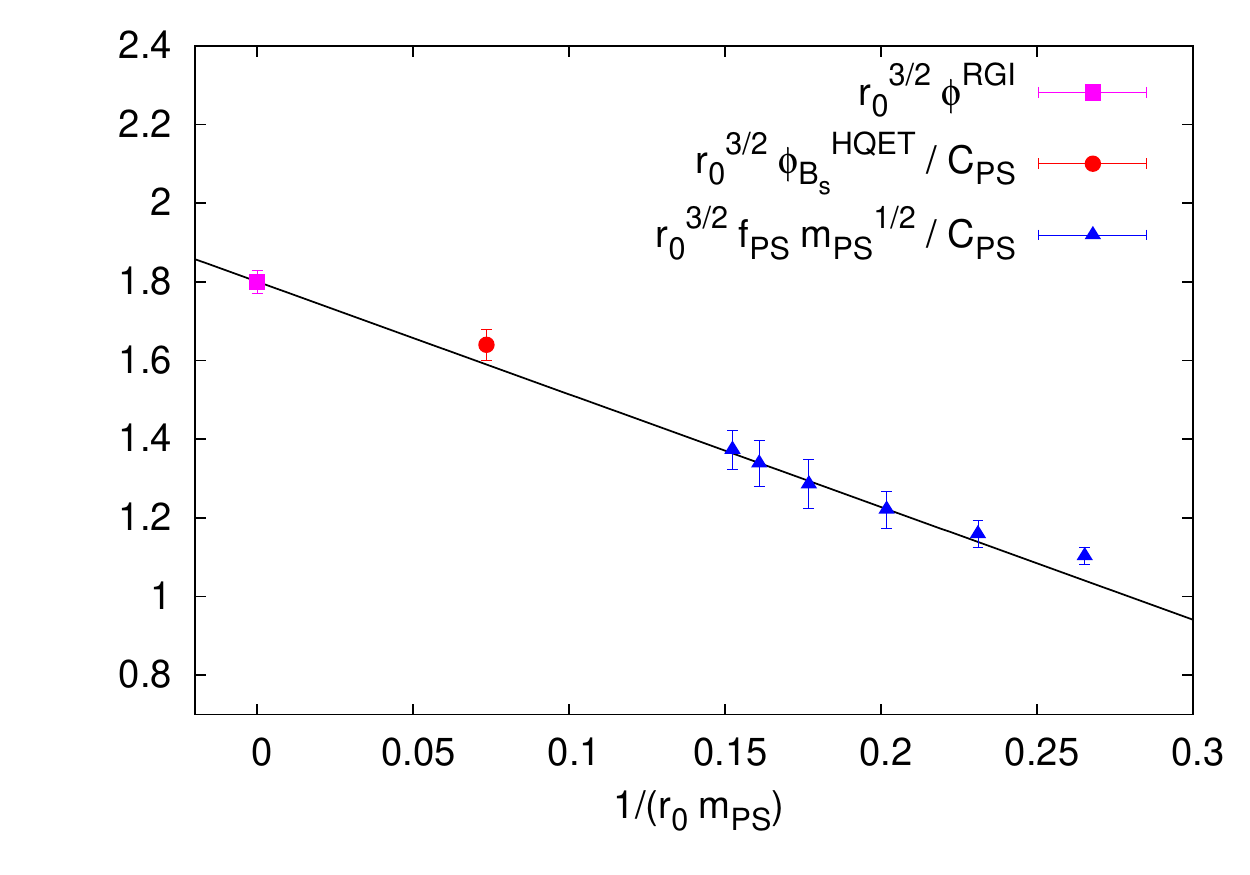,
       width=0.68\textwidth \vspace*{-3mm}} }
  \caption{\footnotesize 
        Static results together with results with $m_\mrm{h} < \mbeauty$
        and an HQET computation with $\minv$ corrections included.
        Continuum extrapolations are done before the
        interpolation \protect\cite{hqet:first3}. $\Cps$ is evaluated with
        the three-loop approximation of $\gamma_\mrm{match}$.
        }\label{f:fbhqet}
\end{figure}
%%%%%%%%%%%%%%%%%%%%%%%%%%%%%%%%%%%%%%%%%%%%%%%%%%%%%%%%%%%%%%%%%%%%%%%%%%%%%%%%%%%%%%%%%%

We now discuss a few numerical results \cite{hqet:first1,hqet:first2,hqet:first3}
in order to give an
indication of what can be done at present. 
The graphs and numbers are for the quenched approximation
(the light quark is a strange quark) but these computations are also on the 
way for dynamical fermions. The statistics employed in the quenched
approximation is rather modest: only 100 configurations were analyzed.
One can easily use a larger number, 
even with dynamical fermions. We skip numerical details in the following
discussion.

As a first step, one wants to fix the b-quark mass. 
This is done through \eq{e:mbmB} and its $1/m$ corrections. 
Its graphical solution 
is illustrated in \fig{f:mbstat} where all plotted numbers
originate from prior continuum extrapolations.
The resulting mass of the b-quark is displayed in \tab{table_mb_ps}.
Observe that it depends very little on the matching condition,
i.e. the choice of $\theta_0,\theta_1,\theta_2$ and moreover
the $1/m$ corrections are small.

Next we look at  the lattice spacing dependence 
of the decay constant. For the results including $\minv$ corrections
no significant
dependence on $a$ is seen in \fig{f:fbhqetcl} 
despite a good precision of about 2\%. In static approximation,
discretization errors are visible but small. 
\Tab{table_fb} lists the $B_\mrm{s}$ decay constant using
$r_0=0.5\fm$ to convert to MeV  for illustration. 
The actual number is affected by an unknown ``quenching effect''
and thus not so important. It is more relevant to observe the precision
that can be reached with just 100 configurations and 
how the spread in the numbers in static approximation is reduced when the 
$\minv$ corrections are included.

 \begin{table}[!htb] 
 \hspace{-1.cm} 
 \begin{center} 
 \begin{tabular}{||c||c|ccc||}
 \hline
 \hline
 & LO (static) & \multicolumn{3}{c||}{NLO (static + $\rmO(1/m)$) }  \\
 \hline
 \hline
 & & $(\theta_1, \theta_2) = (0,0.5)$ & $(\theta_1, \theta_2) = (0.5,1)$ & $(\theta_1, \theta_2) = (0,1)$ \\
 \hline
$\theta_0 = 0  $&$ 233 \pm 6 $&$ 220 \pm 9 $&$ 218 \pm 9 $&$ 218 \pm 9 $\\
$\theta_0 = 0.5$&$ 229 \pm 7 $&$ 221 \pm 9 $&$ 219 \pm 8 $&$ 219 \pm 9 $\\
$\theta_0 = 1  $&$ 219 \pm 6 $&$ 223 \pm 9 $&$ 221 \pm 8 $&$ 222 \pm 8 $\\
 \hline \hline
 \end{tabular} 
 \end{center} 
 \caption[ ]{Pseudo-scalar heavy-light decay constant $f_{{\rm B}_{\rm s}}$ in MeV, 
for different
values of $\theta_i$.}
 \label{table_fb} 
 \end{table}

Further, the comparison with results in the charm mass region, 
\fig{f:fbhqet}, seems to indicate that the $\minv$ expansion works very 
well even for charm quarks. This is a bit surprising and certainly requires 
further confirmation. Note also that this comparison makes use of
the perturbatively evaluated $\Cps$ whose 
intrinsic uncertainty due to perturbation theory is
difficult to evaluate. Of course this uncertainty does neither 
affect the non-perturbatively computed static value at $1/(r_0\mp)=0$,
nor $\fbs\sqrt{\mbs}$ computed with $\minv$ corrections at the mass
of the b-quark, corresponding to $1/(r_0\mp)\approx0.07$. It only
affects the comparison to the results for $1/(r_0\mp) \grtsim 0.15$
since only for the purpose of this comparison the logarithmic mass dependence
described by $\Cps$ has to be divided out.

Finally we show some results concerning the spectrum. The
splitting between 
radial excitations in the pseudo-scalar sector is displayed in the right part of
\fig{f:fbhqetcl}.
As throughout in our results, the $\minv$-correction is
rather small.

\section{Perspectives}

Meanwhile it has been established
that HQET with non-perturbatively determined
parameters is a precision tool. However, we are still 
at the beginning concerning applications. Results for the quantities shown
here will be available for
$\nf=2$ dynamical fermions rather soon. 
But there are many more applications which remain unexplored and
open interesting avenues of research for the future.

\appendix
\chapter{} %\label{s:notation}

\section{Notation} \label{s:notation}

\subsection{Index conventions}

Lorentz indices $\mu,\nu,\ldots$ are taken from the middle of the 
Greek alphabet and run from 0 to 3. Latin indices $k,l,\ldots$
run from 1 to 3 and 
are used to label the components of spatial vectors.  
For the Dirac indices capital letters $A,B,\ldots$ from the 
beginning of the alphabet are taken. They run from 1 to 4.
Color vectors in the fundamental representation of SU($N$)
carry indices $\alpha,\beta,\ldots$ ranging from 1 to $N$, 
while for vectors in the adjoint representation, Latin indices
$a,b,\ldots$ running from 1 to $N^2-1$ are employed.

Repeated indices are always summed over unless otherwise
stated and scalar products are taken with Euclidean metric.

\subsection{Dirac matrices}

In the {\em chiral representation} for 
the Dirac matrices, we have
\be
  \dirac{\mu}=\pmat{0                 & e_{\mu}  \\
                    e_{\mu}^{\dagger} & 0      }. 
\ee
The $2\times2$ matrices $e_{\mu}$ are taken to be 
\be
  e_0=-1,\qquad e_k=-i\sigma_k,
\ee
with $\sigma_k$ the Pauli matrices.
It is then easy to check that 
\be
  \diracstar{\mu}{\dagger}=\dirac{\mu},\qquad
  \{\dirac{\mu},\dirac{\nu}\}=2\delta_{\mu\nu}.
\ee
Furthermore, if we define
$\dirac{5}=\dirac{0}\dirac{1}\dirac{2}\dirac{3}$, we have
\be
  \dirac{5}=
  \pmat{1&0\\ 0&-1} .
\ee
In particular, $\dirac{5}=\diracstar{5}{\dagger}$
and $\diracstar{5}{2}=1$.
The hermitian matrices
\be
  \sigma_{\mu\nu}={i\over2}\left[\dirac{\mu},\dirac{\nu}\right]
  %\eqno\enum
\ee
are explicitly given by ($\sigma_i\sigma_j=i\epsilon_{ijk}\sigma_k$)
\be
  \sigma_{0k}=\pmat{\sigma_k&0\\ 0&-\sigma_k},
  \qquad
  \sigma_{ij}=-\epsilon_{ijk}\pmat{\sigma_k&0\\ 0&\sigma_k} \equiv -\epsilon_{ijk}\sigma_k,
  %\eqno\enum
\ee  
where $\epsilon_{ijk}$ is the totally anti-symmetric tensor with
$\epsilon_{123}=1$.
\\[2ex]
In the {\em Dirac representation}  we have
\bes
  \dirac{k}&=&\pmat{0                 &  -i\sigma_k \\
                   i\sigma_k & 0      }\,,
  \quad
  \dirac{0}=\pmat{ 1               &  0 \\
                   0                     & -1     }\,,
  \\
  \dirac{5}&=&
  \pmat{0&1\\ 1&0}\,, \qquad
  \sigma_{ij}=-\epsilon_{ijk}\pmat{\sigma_k&0\\ 0&\sigma_k} = \sigma_k,
\ees

\subsection{Lattice conventions}

Ordinary forward and backward lattice derivatives 
act on color singlet functions 
$f(x)$ and are defined through
\bes
  \drv{\mu}f(x)&=&{1\over a}\bigl[f(x+a\hat{\mu})-f(x)\bigr], \nonumber
  \\
  \drvstar{\mu}f(x)&=&{1\over a}\bigl[f(x)-f(x-a\hat{\mu})\bigr],
  \label{e_deriv}
\ees  
where $\hat{\mu}$ denotes the unit vector in direction $\mu$. We also use the 
symmetric derivative
\bes
  \widetilde{\partial}_{\mu} = \frac12 (\drv{\mu}+\drvstar{\mu})\,.
\ees
The gauge covariant derivative operators,
acting on a quark field $\psi(x)$, are given by
\bes
  \nab{\mu}\psi(x)&=&
  {1\over a}\bigl[\lambda_{\mu}U(x,\mu)\psi(x+a\hat{\mu})-\psi(x)\bigr],
  \\
  \nabstar{\mu}\psi(x)&=&
  {1\over a}\bigl[\psi(x)-\lambda_{\mu}^{-1}U(x-a\hat{\mu},\mu)^{-1}
  \psi(x-a\hat{\mu})\bigr]\,,
\ees
with the constant phase factors 
\be
 \label{e_lambdamu}
  \lambda_{\mu}=\rme^{ia\theta_{\mu}/L},\qquad
  \theta_{0}=0,\quad -\pi<\theta_k\leq\pi\,,
\ee
explained in \sect{s:SF}.
The left action of the lattice derivative operators
is defined by
\bes
  \psibar(x)\lvec{\nab{\mu}}&=&
  {1\over a}\left[\,
  \psibar(x+a\hat{\mu})U(x,\mu)^{-1}\lambda_{\mu}^{-1}
  -\psibar(x)\,\right],
  \\
  \psibar(x)\lvec{\nabstar{\mu}}&=&
  {1\over a}\left[\,
  \psibar(x)-\psibar(x-a\hat{\mu})U(x-a\hat{\mu},\mu)\lambda_{\mu}
  \,\right].
\ees  
Our lattice version of $\delta$-functions are 
\be
 \label{e_delta}
 \delta(x_\mu) = a^{-1} \delta_{x_\mu 0}\,,\quad
 \delta(\vecx) = \prod_{k=1}^3 \delta(x_k) \,,\quad
 \delta(x) = \prod_{\mu=0}^3 \delta(x_\mu)  
\ee
and we use
\bes
 \label{e_theta}
  \theta(x_\mu) &=& 1 \hbox{ for } x_\mu\geq0 \\
  \theta(x_\mu) &=& 0 \hbox{ otherwise }      \nonumber
\ees
Fields in momentum space are introduced by the Fourier transformation
\bes
  \tilde f(p) = a^4 \sum_x \rme^{-ipx}f(x) \Leftrightarrow \left\{
  \barr{ll}  f(x) = {1 \over L^3 T } \sum_p  \rme^{ipx}\tilde f(p) 
      & \quad\mbox{in a $T\times L^3$ volume} \\[1ex]
        f(x) = \int_{-\pi/a}^{\pi/a}{\rmd^4 p \over (2\pi)^4}  \rme^{ipx}\tilde f(p) 
      & \quad\mbox{in infinite volume}
  \earr
  \right.
\ees

\off{\\
\subsubsection{Lattice momenta}

\bes
 \phat_\mu &=& {2\over a} \sin\left({ap_\mu\over 2}\right) \\
 \ptil_\mu &=& {1\over a} \sin(ap_\mu) = \pcirc_\mu
 \momp{p}{\mu} &=&\\
 \momm{p}{\mu} &=&\\
 \mompm{p}{\mu} &=&\\
 \smomp{p}{\mu} &=&\\
 \smomm{p}{\mu} &=&\\
 \smompm{p}{\mu} &=&
\ees
}

\subsection{Continuum gauge fields}

An SU($N$) gauge potential in the continuum theory 
is a vector field $A_{\mu}(x)$
with values in the Lie algebra su($N$). It may thus be written as
\be
  A_{\mu}(x)=A_{\mu}^a(x)T^a
\ee
with real components $A_{\mu}^a(x)$ and
\be
  (T^a)^\dagger = - T^a,\quad \tr\{T^aT^b\}=-\frac{1}{2}\delta^{ab}.
\ee 
The associated field tensor,
\be
  F_{\mu\nu}(x)=
  \partial_{\mu}A_{\nu}(x)-\partial_{\nu}A_{\mu}(x)
  +[A_{\mu}(x),A_{\nu}(x)],
\ee
may be decomposed similarly and 
the right and left action of the covariant derivative 
$D_{\mu}$ is defined by
\bes
  D_{\mu}\psi(x)&=&
  (\drv{\mu}+A_{\mu})\psi(x),
   \\
  \psibar(x)\lvec{D_{\mu}}&=&
  \psibar(x)(\lvec{\drv{\mu}}-A_{\mu}).
\ees  
We note that periodic boundary conditions up to a phase $\theta_\mu$
are equivalent to adding a constant
abelian gauge field $i\theta_{\mu}/L$: in the above we replace 
$A_{\mu}\to A_{\mu}+i\theta_{\mu}/L$.

\subsection{Lattice action}
Let us first assume that the theory is defined on an infinite
lattice. 
A gauge field $U$ on the lattice is an assignment of a matrix 
$U(x,\mu)\in\SUn$ to every lattice point $x$ and direction 
$\mu=0,1,2,3$.
Quark and anti-quark fields, $\psi(x)$ and $\psibar(x)$,
reside on the lattice sites and  
carry Dirac, colour and flavour indices.
The (unimproved) lattice action is of the form
\be
  S[U,\psibar,\psi\,]=\Sg[U]+\Sf[U,\psibar,\psi\,],
\ee
where $\Sg$ denotes 
the usual Wilson plaquette action and $\Sf$ the Wilson quark action.
Explicitly we have
\bes
  \Sg[U]&=&{1\over g_0^2}\sum_p\tr\{1-U(p)\} = 
         {1\over g_0^2}\sum_x\sum_{\mu,\nu}P_{\mu\nu}(x)\,,\\
  &&P_{\mu\nu}(x)= U(x,\mu)\, U(x+a\hat\mu,\nu)\, 
                   U(x+a\hat\nu,\mu)^{-1}\, U(x,\nu)^{-1}
  %\eqno\enum
\ees
with $g_0$ being the bare gauge coupling and 
$U(p)$ the parallel transporter around the plaquette $p$.
The sum runs over all {\it oriented}\/ plaquettes $p$
on the lattice, i.e. independently over $\mu,\nu$. 
The quark action,
\be
  \label{e_Sf}
  \Sf[U,\psibar,\psi\,]=a^4\sum_{x}\psibar(x)(\DW+m_0)\psi(x),
\ee
is defined in terms of the 
Wilson-Dirac operator
\be
  \label{e_dirop}
  \DW=\frac{1}{2}\left\{
  \dirac{\mu}(\nabstar{\mu}+\nab{\mu})-a\nabstar{\mu}\nab{\mu}\right\},
\ee
which involves the gauge covariant lattice derivatives 
$\nab{\mu}$ and $\nabstar{\mu}$, \eq{e_deriv}, and 
the bare quark mass {\em matrix}, $m_0 = \mrm{diag}(m_{0\up},m_{0\down},\ldots)$ .

\off{
\subsection{Advanced Lattice QCD}

The improved action is given by \cite{impr:sw,impr:pap1}
\bes
  \label{e_simpr}
    \Simpr[U,\psibar,\psi]&=&S[U,\psibar,\psi]+\delta S[U,\psibar,\psi],
  \\
  \delta S[U,\psibar,\psi]&=&
  a^5\sum_{x}
  \csw\,
  \psibar(x)\frac{i}{4}\sigma_{\mu\nu}\widehat{F}_{\mu\nu}(x)\psi(x),
\ees
with
\bes
    \widehat{F}_{\mu\nu}(x)&=&{1\over8a^2}\left\{
  Q_{\mu\nu}(x)-Q_{\nu\mu}(x)\right\},
  \\
  Q_{\mu\nu}(x)
  &=&\,U(x,\mu)U(x+a\hat{\mu},\nu)
    U(x+a\hat{\nu},\mu)^{-1} U(x,\nu)^{-1}
  \\
  && +\,U(x,\nu)U(x-a\hat{\mu}+a\hat{\nu},\mu)^{-1}
    U(x-a\hat{\mu},\nu)^{-1}U(x-a\hat{\mu},\mu)
  \nonumber \\ && 
  +\,U(x-a\hat{\mu},\mu)^{-1}U(x-a\hat{\mu}-a\hat{\nu},\nu)^{-1}
  \nonumber \\ && \qquad\qquad\qquad\qquad\qquad\quad
   \times U(x-a\hat{\mu}-a\hat{\nu},\mu)U(x-a\hat{\nu},\nu)
  \nonumber \\ && 
  +\,U(x-a\hat{\nu},\nu)^{-1}U(x-a\hat{\nu},\mu)
    U(x+a\hat{\mu}-a\hat{\nu},\nu)U(x,\mu)^{-1}.\nonumber 
\ees

\subsection{\SF}
Turning now to the case of \SF boundary conditions,
the pure gauge action on the lattice is \cite{SF:LNWW}
\be  
  \Sg[U]={1\over g_0^2}\sum_p w(p)\tr\{1-U(p)\},
\ee
with a weight $w(p)$ equal to 1 for all $p$ except for
the spatial plaquettes at $x_0=0$ and
$x_0=T$, where $w(p)= \frac{1}{2}$.

The quark action in the \SF can be written exactly as in \eq{e_Sf} when
the conventions
\be
   \psi(x)=0\quad\hbox{if $x_0<0$ or $x_0>T$}
\ee
and
\be
    P_{-}\psi(x)|_{x_0=0}=
  P_{+}\psi(x)|_{x_0=T}=0
\ee
are adopted.
}

\off{
\vfill\eject
\subsection{Chiral symmetry and mass factors in the effective fields.}
This is just an incomplete argument suggesting that mass dependences
of the type $\mstrange/\mbeauty$ are not forbidden by chiral symmetry.
We consider the strange quark mass dependence of an observable 
$\Phi$ by taking a derivative of $\Phi$ with respect to the mass
and then setting the mass to zero:
\bes
   \Phi = \langle \op{} \rangle\,,
   \quad \partial_{\mstrange} \Phi =  \langle \op{}\; \bar s s  \rangle_{\rm conn}\,.
\ees
In QCD with $\mstrange=0$ and $\mbeauty=0$, we have a non-anomalous symmetry
\bes
   \pmat{ s \\ b} &\to& \rme^{i \tau^3 \gamma_5 \pi/2} \pmat{ s \\ b} 
   = \pmat{ i\gamma_5 s \\ -i\gamma_5 b}\\
   \pmat{ \bar s & \bar b} &\to&  \pmat{ \bar s & \bar b} \rme^{i \tau^3 \gamma_5 \pi/2} 
   = \pmat{ \bar s i\gamma_5 & -i \bar b \gamma_5}\,.
\ees
Some field transformations are 
\bes
  \bar b \gamma_0 \gamma_5 s \to - \bar b \gamma_0 \gamma_5 s\,,\quad
  \bar s \gamma_0 \gamma_5 b \to - \bar s \gamma_0 \gamma_5 b\,,\quad
  \bar s  s \to - \bar s s\,.
\ees
From this we see immediately that
\bes
&&\langle \bar b \gamma_0 \gamma_5 s\;\; \bar s \gamma_0 \gamma_5 b \;\; 
  \bar s s  \rangle_{\rm conn, {\mstrange=\mbeauty=0}} =0 
\\
&\to& \partial_{\mstrange} \Phi|_{\mstrange=\mbeauty=0} = 0 \quad \mbox{for} \quad 
     \Phi =  \langle \bar b \gamma_0 \gamma_5 s\; \bar s \gamma_0 \gamma_5 b \rangle  \,.
\ees
This excludes linear terms in $\mstrange$.

Of course, $b$ can be any flavor here, but we note that it has to be a massless flavor.
Otherwise either the consider transformation is not a symmetry
or the currents do not transform into themselves. As soon as we make the 
b-quark mass finite (and large) chiral symmetry does not seem to forbid
terms linear in $\mstrange$ any more. Terms such as $\mstrange\mbeauty$ or
$\mstrange/\mbeauty$ can appear.
}

\subsection{Renormalization group functions and invariants}
Our RG functions are defined through 
\bes
  \mu {\partial \bar g \over \partial \mu} &=& \beta(\bar g) \enspace ,
     \label{e_RG} \\
  {\mu \over \mbar} {\partial \mbar \over \partial \mu} &=& \tau(\bar g) \enspace ,
     \label{e_RG_m}
  \\
  {\mu \over  \Phi} {\partial \Phi \over \partial\mu} &=& \gamma(\gbar)  \label{e_RG_Phione}
\ees
in terms of running coupling and running quark mass as well as some
matrix element $\Phi$ of a (multiplicatively renormalizable) composite field.
They have asymptotic expansions
\bes
 \beta(\bar g) & \buildrel {\bar g}\rightarrow0\over\sim &
 -{\bar g}^3 \left\{ b_0 + {\bar g}^{2}  b_1 + \ldots \right\}
                      \enspace ,  \label{e_RGpert} \\ \nonumber
 &&b_0=\frac{1}{(4\pi)^2}\bigl(11-\frac{2}{3}\nf\bigr)
                      \enspace ,\quad
   b_1=\frac{1}{(4\pi)^4}\bigl(102-\frac{38}{3}\nf\bigr) \enspace ,
 \\
 \tau(\bar g) & \buildrel {\bar g}\rightarrow0\over\sim &
 -{\bar g}^2 \left\{ d_0 + {\bar g}^{2}  d_1 + \ldots \right\}
                      \, , \qquad
 d_0={8}/{(4\pi)^2}
 \enspace ,  \label{e_RGpert_m}
 \\
  \gamma(\gbar)& \buildrel {\bar g}\rightarrow0\over\sim &
  - \gbar^2 \left\{ \gamma_0 + {\bar g}^{2}  \gamma_1 + \ldots \right\} 
  \label{e_RGpert_Phi}
\ees
The integration constants of the solutions to the 
RGEs define the RG invariants
\bes
 \Lambda &=&\mu \left(b_0\gbar^2\right)^{-b_1/(2b_0^2)} \rme^{-1/(2b_0\gbar^2)}%%\\         %% && \times
           \exp \left\{-\int_0^{\gbar} \rmd x
          \left[\frac{1}{ \beta(x)}+\frac{1}{b_0x^3}-\frac{b_1}{b_0^2x}
          \right]
          \right\} \enspace , \label{e_lambdapar}
  \\
  M &=& \mbar\,(2 b_0\gbar^2)^{-d_0/2b_0}
   \exp \left\{- \int_0^{\gbar} \rmd x \left[{\tau(x) \over \beta(x)}
     - {d_0 \over b_0 g} \right] \right\}  \enspace .
  \\
  \PhiRGI &=& \Phi \left[\,2b_0 \gbar^2\,\right]^{-\gamma_0/2b_0}
                   \exp\left\{-\int_0^{\gbar} \rmd x
                     \left[\,{ \gamma(x) \over\beta(x)}
                           -{\gamma_0 \over b_0 x}\,\right]
                     \right\} \,
\ees
where $\gbar\equiv\gbar(\mu)$ ... $\Phi\equiv\Phi(\mu)$. We will also
us the shorthand notation
\bes
   {\Lambda \over \mu} &=& \varphi_g(\gbar) = \exp \left\{-\int^{\gbar} \rmd x
          \frac{1}{ \beta(x)}          \right\} \,,
   \\
   {M \over \mbar} &=& \varphi_m(\gbar) = \exp \left\{-\int^{\gbar} \rmd x
          \frac{\tau(x)}{ \beta(x)}          \right\} \,,
   \\
   {\PhiRGI \over \Phi} &=& \varphi_\Phi(\gbar) 
    = \exp \left\{-\int^{\gbar} \rmd x
          \frac{\gamma(x)}{ \beta(x)}        \right\} \,,
\ees
with the constants exactly as defined above.
\section{Conversion functions and anomalous dimensions
} \label{s:conv}
\def\vkin{{\cal V}_\mrm{kin}}
\def\gammastat{\gamma^\mrm{stat}}
\def\gammamatch{\gamma^\mrm{match}}
\newcommand{\abar}{\bar{a}}

Conversion functions and the anomalous dimensions 
$\gamma_\mrm{match}$ are not part of the standard phenomenology
literature. For completeness we give the explicit
relations to the matching coefficients found directly in the literature and
discuss the accuracy of their perturbative expansion.

\subsection{Matching coefficients and anomalous dimension}
We here describe the result \cite{hqet:match3lp} and its
relation to the anomalous dimension.
We denote a matrix element of some heavy-light quark bilinear 
$\psibar \Gamma \heavy$
in the effective theory by $\Phi(\mu)$. 
The Dirac structure $\Gamma$ is left implicit.\footnote{
The notation $C_{\tilde\Gamma}$ of 
\cite{BroadhGrozin2} translates to our $\Gamma$ as
$
  \tilde\Gamma=(1,\,\gamma_0,\,\gamma_1,\,\gamma_0\gamma_1)
  \to \Gamma=(\gamma_5,\,\gamma_0\gamma_5,\,\gamma_k,\,\gamma_0\gamma_k)
$
and \cite{hqet:match3lp} uses the notation of \cite{BroadhGrozin2}
when one sets 
$v_\mu\gamma_\mu=\gamma_0$, $\gamma_\perp=\gamma_k$ 
as it is the case in the rest frame. We will also refer to 
the bilinears as (PS,\,$A_0$,\,$V_k$,\,T).
In comparison to \cite{hqet:match3lp} we add a subscript Q
to the pole quark mass and a bar to the running
mass ($m\to\mpole,\;m(\mu)\to\mbar(\mu)$) for clarity. 
}

All quantities
are renormalized in the $\msbar$-scheme, with a scale $\mu_o$
for the QCD bilinear and a scale $\mu$ in HQET. Choosing the 
pole quark mass $\mpole$,\footnote{
While in the complete, non-perturbative theory,
the pole mass is ill-defined,
in perturbation theory it exists order by order
in the expansion. We use it here, because the formulae in
the literature are written in terms of it. It will be eliminated
in the final formulae.
}
the matrix element is then
(without explicit superscripts ``QCD''
we refer to HQET quantities, in the static approximation), 
\be \label{ea:match2}
  \Phiqcd(\mpole,\mu_o;\vkin) 
  =\widehat C_\mrm{match}(\mpole,\mu_o,\mu)\times\Phi(\mu;\vkin)
                                + \rmO(1/m)\,.
\ee
The kinematical variables entering the matrix element $\Phi$
are denoted by $\vkin$.
For the (partially) conserved currents
$V_\mu,A_\mu$ there is no  $\mu_o$-dependence on the l.h.s. of \eq{ea:match2},
($\partial_{\mu_o}\,\Phiqcd(\mpole,{\mu_o})=0$), while in general
we have 
\be
   {\mu\over \Phiqcd(\mpole,\mu_o)}
   {\partial \Phiqcd(\mpole,\mu_o) \over \partial \mu_o } =
   {\partial \ln(\Phiqcd(\mpole,\mu_o)) \over \partial \ln(\mu_o) } 
   \equiv \gamma_o(\gbar(\mu_o))\,.
\ee
We pass to the RGI matrix element in QCD via ($\rmO(1/m)$ is dropped without
notice)
\bes
  \Phiqcd_\mrm{RGI} &=&   
          \exp \left\{-\int^{\gbar(\mu_o)} \rmd x
          \bfrac{\gamma_o(x)}{ \beta(x)}        \right\}\, \Phiqcd(\mpole,\mu_o;\vkin)
     \\ &=&  \exp \left\{-\int^{\gbar(\mu_o)} \rmd x
          \bfrac{\gamma_o(x)}{ \beta(x)}\,  \right\}\,
         \widehat C_\mrm{match}(\mpole,\mu_o,\mu)\times\Phi(\mu;\vkin)\,. 
\ees
($\rmO(1/m)$ is dropped without notice). It depends on the quark mass
but not on a renormalization scale. The physical anomalous dimension
is given by
\bes 
  \gamma_\mrm{match}(\gstar) = {\rmd \ln(\mpole) \over  \rmd \ln(\mstar) }
          {\partial \ln(\widehat C_\mrm{match}(\mpole,\mu_o,\mu)) 
           \over \partial \ln(\mpole) }\,,
\ees
where the first factor is computed from the expansion
\cite{pert:mpole2lp,pert:mpole2lpb,pert:mpole3lp,hqet:match3lp}
\bes
   \label{ea:mpolestar}  
   \mpole &=& \mstar\,[  1+\sum_{l\geq 1} k_l [\abar(\mstar)]^l ]\,,
       \,,\quad \abar(\mu)={\gbar^2(\mu)\over 4\pi^2}  \\
       &&k_1= 4/3\,,\quad k_2=-1.0414 (\nf-1) + 13.4434\,,
       \nonumber \\
       &&k_3= 0.6527 (\nf-1)^2 - 26.655 (\nf-1) + 190.595\,.\nonumber
\ees
The authors of \Ref{hqet:match3lp} set $\mu_o=\mu$. Building on
\cite{Ji:1991pr,BroadhGrozin2,Gimenez:1992bf},
they give the perturbative expansion
\bes
 \widehat C_\mrm{match}(\mpole,\mu,\mu) &=& 
 1 + \sum_{l\geq 1}\,\sum_{k=0}^l L_{lk}[\ln(\mpole^2/\mu^2)]^k\, [\abar(\mpole)]^l \,,
 \label{ea:CGamma}
\ees
with coefficients $L_{lk}$ depending on the Dirac-structure, $\Gamma$.
%, but again this dependence is suppressed in our notation. 

Independence of the l.h.s. of \eq{ea:match2} of $\mu$  yields
\bes
   {\partial \ln(\widehat C_\mrm{match}(\mpole,\mu_o,\mu)) \over  \partial \ln(\mu) }
   &=&   \
   - {\partial \ln(\Phi(\mu)) \over \partial \ln(\mu) } = -\gammastat(\gbar(\mu))\,,
\ees 
and with ${\partial \ln(\widehat C_\mrm{match}(\mpole,\mu_o,\mu)) \over \partial \ln(\mu_o)} 
= \gamma_o(\gbar(\mu_o))$
we have
\bes
   \label{ea:derivrel}
   &&{\rmd \ln(\widehat C_\mrm{match}(\mpole,\mpole,\mpole)) \over \rmd \ln(\mpole)
   }  \\\nonumber
   &&\qquad=    
   \left.{\partial \ln((\widehat C_\mrm{match}(\mpole,\mu_o,\mu)) 
   \over  \partial \ln(\mpole) }\right|_{\mu_o=\mu=\mpole} 
   + \gamma_o(\gbar(\mpole))  -\gamma_\mrm{stat}(\gbar(\mpole)) \,.
\ees
From these equations $\gamma_\mrm{match}(\gstar)$ can be determined 
up to three-loop order and the differences 
$\gamma_\mrm{match}^{\Gamma'}(\gstar)-\gamma_\mrm{match}^{\Gamma}(\gstar)$
up to four-loop order.

\subsection{Numerical results and the behavior 
of perturbation theory} \label{s:numad} 

Let us now look at the numerical size of the perturbative coefficients
of the RG functions. The following
table lists results for $\nf=3$. This is enough to understand
the general picture since for smaller $\nf$ the higher order coefficients are 
generically somewhat
larger, but not by much. 
\begin{center}\begin{tabular}{lrrrccc}
  coefficient & $i=1$ &$i=2$ &$i=3$ &$i=4$ 
\\  \hline \\[-1ex]
  $(4\pi)^i b_{i-1}$ & 0.71620 & 0.40529 & 0.32445 & 0.47367 \\
  $(4\pi)^i d_{i-1}$          & 0.63662 & 0.76835 & 0.80114 & 0.90881\\
  $(4\pi)^i\,\gamma_{\mrm{stat},i-1}$   & -0.31831 & -0.26613 & -0.25917
\\  \hline \\[-1ex]
  $(4\pi)^i\,\gamma_{\mrm{match},i-1}^{\gamma_0\gamma_5} $  & -0.31831 & -0.57010 & -0.94645 \\
  $(4\pi)^i\,\gamma_{\mrm{match},i-1}^{\gamma_k}$   & -0.31831 & -0.87406 &  -3.12585 \\
  $(4\pi)^i\,[\gamma_{\mrm{match},i-1}^{\gamma_0\gamma_5}-\gamma_{\mrm{match},i-1}^{\gamma_k}] $  
 & 0 &  0.30396 &  2.17939 &  14.803 \\
  $(4\pi)^i\,[\gamma_{\mrm{match},i-1}^{\gamma_0\gamma_5}-\gamma_{\mrm{match},i-1}^{\gamma_5}] $  
  \\
\end{tabular}\end{center}
The normalization $(4\pi)^i$ has been inserted such that the series
is well behaved for $\alpha\lesssim 1/3$ if the coefficients
are order one. Indeed this is the magnitude of
the coefficients in the first three rows
which show as a comparison the beta-function, mass anomalous 
dimension and the anomalous dimension of the static-light bilinears
(all in the $\msbar$--scheme).
In contrast in the physical anomalous dimension of the 
vector current $\gamma_{\mrm{match}}^{\gamma_k}$, the 3-loop coefficient
is rather big and the difference 
$(4\pi)^i\,[\gamma_{\mrm{match},3}^{\gamma_0\gamma_5}-\gamma_{\mrm{match},3}^{\gamma_k}]
$ is even above ten. Perturbation theory is then useful only at rather
small $\alpha$; in particular not really for the b-quark.

An attempt to improve the perturbative series 
is to re-expand $\gamma_{\mrm{match}}$ in the coupling at a different
scale, adjusting the scale to obtain smaller coefficients. In fact, 
since the effective theory is valid at energy scales below the mass of the
quark, it is plausible that scales smaller than $\mstar$ are more 
suitable. So we choose a coupling
$  \hat g^2 = \gbar^2(s^{-1}\mstar) = \sigma(\gstar^2,s)$
and
\bes
  \hat\gamma_\mrm{match}(\hat g) &=& \gamma_\mrm{match}([\sigma(\hat g^2,1/s)]^{1/2}) 
\label{eaa:gammahat}\,,
\ees
which is of course expanded order by order,
\bes
  \gstar^2 = \sigma(\hat g^2,1/s) &=& \hat g^2  - 2b_0 \ln(s)\,\hat g^4 + \ldots \,.
\ees
The conversion functions are then expressed as
\bes
  C_\mrm{PS}(M/\Lambda) &=& \exp \left\{\int^{\hat g} \rmd x 
        \bfrac{\hat\gamma_\mrm{match}(x)}{\beta(x)}\right\} \,.
\ees
The difference comes from truncating \eq{eaa:gammahat} as a series in
$\hat g^2$. The argument above suggests $s>1$.
The perturbative coefficients are listed in the
following table for a few choices of $s$, for example the one
which brings the two-loop coefficient $\gamma_1$ to zero.
\begin{center}\begin{tabular}{lrrrccc}
  coefficient & $i=1$ &$i=2$ &$i=3$ &$i=4$ & $s$
\\  \hline
  $(4\pi)^i\,\gamma_{\mrm{match},i-1}^{\gamma_0\gamma_5} $   & -0.31831 & -0.57010 & -0.94645 &&1\\
                    & -0.31831 & 0        &  0.39720 && 3.4916 \\[0.5ex]
  $(4\pi)^i\,\gamma_{\mrm{match},i-1}^{\gamma_k}$   & -0.31831 & -0.87406 &  -3.12585 &&1\\
                    & -0.31831 & 0 & -0.231121 &&    6.8007 
  \\ \hline \\[-1ex]
  $(4\pi)^i\,[\gamma_{\mrm{match},i-1}^{\gamma_0\gamma_5}-\gamma_{\mrm{match},i-1}^{\gamma_k}] $ 
                     & 0 &  0.30396 &  2.17939 &  14.803 &1\\
                     & 0 &  0.30396 &  0.972221 & 4.733 &  4 \\
                     & 0 &  0.30396 & -0.05414 &   1.82678 &  13 \\
                     & 0 &  0.30396 & -0.23495 &   1.85344 &  16
\end{tabular}\end{center}
The higher order coefficients can indeed be reduced significantly 
but $s \grtsim 4 $ is required. For B-physics
$\alpha(m_{\star\beauty}/s)$ is then not small and there is no really
useful improvement for phenomenology, see \fig{f:ratios}. We emphasize, however,
that with $s \approx 4 $ the series is much better behaved for masses that are a
factor two or more higher than the b-quark mass. 
The pattern visible in in the tables reflects itself in \fig{f:ratios}.

Let us finally mention that the 
same behavior is found for
$\hat{C}_\mrm{match}(\mpole,\mpole,\mpole)$ for all Dirac structures of the 
currents. Their perturbative expansion in a coupling
$\gbar(\mpole/s)$ is better behaved for $s \grtsim 4 $ than for $s=1$.

\bibliographystyle{OUPnamed}  
\bibliography{latticen,books1,HQET,chir_pt}

\end{document}